\documentclass{aa}  

\usepackage{multirow}
\usepackage{graphicx}
\usepackage{txfonts}
\usepackage{subcaption}
\usepackage{lscape}
\usepackage[export]{adjustbox}

\newcommand{\msun}{\mbox{$M_{\odot}$}}

\begin{document} 

\titlerunning{THROES: PACS spectroscopy}
\authorrunning{J. Ramos-Medina et al.}

   \title{THROES: a caTalogue of HeRschel Observations of Evolved Stars}

   \subtitle{I. PACS range spectroscopy}

   \author{J. Ramos-Medina
          \inst{1},
          C. Sánchez Contreras\inst{1}, P. García-Lario\inst{2}, C. Rodrigo\inst{1}, J. da Silva Santos\inst{1} \and E. Solano\inst{1}
          }

   \institute{Department of Astrophysics, Astrobiology Center (CSIC-INTA), ESAC campus, PO Box 78, 28691 Villanueva de la Cañada, Madrid, Spain\\
              \email{jramos@cab.inta-csic.es}
         \and
             European Space Astronomy Centre, European Space Agency, PO Box 78, 28691, Villanueva de la Cañada, Madrid, Spain\\
            % \email{Pedro.Garcia-Lario@sciops.esa.int}
            %\and 
            %Centro de Astrobiología (CAB) \email{csanchez@cab.inta-csic.es}           
            % \thanks{The university of heaven temporarily does not
            %         accept e-mails}
             }

   \date{Received September 13, 2017; accepted October 27, 2017}

% \abstract{}{}{}{}{} 
% 5 {} token are mandatory

  \abstract{This is the first of a series of papers presenting the THROES 
  (A caTalogue of HeRschel Observations of Evolved Stars) project, 
  intended to provide a comprehensive overview of the spectroscopic 
    results obtained in the far-infrared (55-670 $\mu$m) with the \textit{Herschel} 
    space observatory on low-to-intermediate mass evolved stars in our Galaxy. Here we introduce the
    catalogue of interactively reprocessed PACS (Photoconductor Array Camera and Spectrometer)
    spectra covering the 55-200
    $\mu$m range for 114 stars in this category for which PACS range
    spectroscopic data is available in the \textit{Herschel} Science Archive (HSA).  Our
    sample includes objects spanning a range of evolutionary stages,
    from the asymptotic giant branch to the planetary nebula phase, displaying
    a wide variety of chemical and physical properties. The THROES/PACS
    catalogue is accessible via a dedicated web-based interface
    ({\tt https://throes.cab.inta-csic.es/}) and includes not only the
    science-ready Herschel spectroscopic data for each source, but also 
    complementary photometric and spectroscopic data from other infrared     
    observatories, namely IRAS (Infrared Astronomical Satellite), ISO (Infrared Space Observatory) or 
    AKARI, at overlapping wavelengths. Our goal is to create a legacy-value  \textit{Herschel}
    dataset that can be used by the scientific community in the future to    
    deepen our knowledge and understanding of these latest stages of the     
    evolution of low-to-intermediate mass stars.}


   \maketitle
%
%________________________________________________________________

\section{Introduction} \label{Introduction}
\textit{Herschel} \footnote{Herschel is an ESA space observatory with science
  instruments provided by European-led Principal Investigator consortia and 
  with important participation from NASA.} \citep{Pilbratt2010}, launched in May
  2009, has provided a new vision of the whole Universe in the far infrared
  (FIR) thanks to the capabilities of the three instruments on board: HIFI 
  (The Heterodyne Instrument for the Far-Infrared) \citep{deGraauw2010}, 
  SPIRE (Spectral and Photometric Imaging Receiver) \citep{Griffin2010} and 
  PACS \citep{Poglitsch2010}. In particular, \textit{Herschel} has been extremely
  useful in the study of the complex physical and chemical processes that take
 place in the final stages of stellar evolution. For example, observations
 from the MESS (Mass-loss of Evolved Stars) observing programme \citep{Groenewegen2011} have demonstrated the
 capability of \textit{Herschel} to image the circumstellar envelopes (CSEs) of
 evolved stars and the interaction regions between the stellar winds and the
 interstellar medium with unprecedented detail. In the spectroscopic area,
 studies have addressed the analysis of the circumstellar forsterite (e.g.
 \cite{deVries2011}) and the CO line emission \citep[e.g.][]{DeBeck2010,Danilovich2015,Maercker2016}.
 The detection of  rotational emission lines of OH$^{+}$ for the
 first time in three planetary nebulae (PNe) \citep{Aleman2014} with
 observations taken as part of the HerPlans observing programme \citep{Ueta2014},
 or the detection of warm water vapour around IRC +10216 \citep{Decin2010}, the
 closest C-rich star to our solar system, are also good examples of the contribution 
 of \textit{Herschel} to our understanding of the chemical complexity 
 of these CSEs and can also be considered as important
 highlights of the mission in the field.

 Towards the end of their lives, low-to-intermediate mass stars (1 $\msun$
 $\leq$ M $\leq$ 8 $\msun$) have burnt up their central hydrogen and helium,
 leaving a quiescent C-O core with H and He fusion reactions taking place in
 thin shells surrounding the inner core. These objects are ascending the
 asymptotic giant branch (AGB), a phase of stellar evolution during which their
 atmospheres expand and cool down, characterized by an intense mass loss
 (from 10$^{-7}$ to 10$^{-4}$ $\msun$ yr$^{-1}$) that results in the formation
 of a CSE of gas and dust around the central star, which emits
 very strongly in the mid-to far-infrared wavelengths \citep{Habing1996}.
  Once the star terminates the AGB phase, the mass-loss rate suddenly decreases
 and the temperature of the central object becomes, progressively, high enough to induce the onset of
 the ionization of the gas in the surrounding CSE. If the 
 temperature increases on a timescale shorter than the dispersion time of the
 matter previously ejected, we will observe an ionized planetary nebula (PN) \citep{Kwok2005}.

   The intermediate stage between the AGB and PN phases is called the 
 post-AGB phase. This phase is also known as pre-PN phase, although it is not
 clear whether all post-AGB stars will develop a PN. During the post-AGB phase, 
 the shell, formed in the AGB phase, detaches from the central star, the spherical symmetry
 is broken, and fast bipolar or multi-polar winds appear \citep[see e.g.][for a review]{Balick2002}.
 This is a very short-lived phase ($\sim$ 1000 years) and, as a consequence of it, a not 
 very well understood stage of stellar evolution.

  The infrared and sub-millimetre regions offer a rich variety of diagnostic,
 atomic, ionic, molecular, and solid-state spectral features, and are
 particularly well suited to study the complex physical and chemical 
 properties found in AGBs, post-AGBs, and PNe, which may be very different from
 source to source, depending on critical parameters like the initial mass or the
 metallicity. A large sample of low- and intermediate-mass evolved stars in our Galaxy were
 observed by the PACS instrument on-board \textit{Herschel} under many different
 observing programmes, and the associated automatically pipeline-processed data
 are now publicly available from the Herschel Science
 Archive\footnote{http://archives.esac.esa.int/hsa/whsa.}(HSA)
 after the end of the proprietary period of one year.
 In the vast majority of cases, unfortunately, the PACS spectroscopy pipeline
 products in the HSA cannot be considered science-ready, and they would
 strongly benefit from dedicated interactive data reduction to help remove
 the residual instrument artefacts and to improve the absolute flux calibration
 as a necessary further step in order to become publication-quality products. 

 With this aim, we have interactively processed in a systematic and homogeneous
 way all PACS range spectroscopic observations contained in the HSA corresponding to 
 stars that can be identified as low- or intermediate-mass evolved stars, with
 the exception of a small subset of nearby sources that show very extended
 emission and a few cases where the unchopped observing mode was used, as they
 require a special case-by-case treatment, which is beyond the scope of this
 project. The result of this interactive data reduction effort has been used to 
 compile the first version of THROES: "a caTalogue of HeRschel Observations of 
 Evolved Stars", through which our final data products are made available to
 the community for scientific exploitation. We are currently working on a 
 second version of the THROES catalogue that will also incorporate spectroscopic data 
 from the Herschel/SPIRE instrument (Ramos Medina et al., in preparation).

  This paper is organized as follows. In Section \ref{Observations} we describe
  how the observations were performed, the building of the THROES sample, and the main 
  characteristics of the sources included in it. In Section \ref{DataReduction} the main
  data reduction steps applied are described.  In Section
  \ref{DescriptionCatalogue}, we introduce the contents of the THROES 
  catalogue and its web interface. In Section \ref{MultiMission}, we try to characterize
 the quality of our science data products through comparisons with the
 standard pipeline products contained in the HSA and with observations taken by
 other space-based facilities in the past, like IRAS, AKARI, and ISO. The final
 summary is given in Section \ref{Summary}.

%__________________________________________________________________

\section{Observations} \label{Observations} 

\subsection{PACS spectroscopy} \label{PacsMode}
 
\subsubsection{The PACS spectrometer}

 The PACS spectrometer covers nominally the wavelength range from 51 to 210 $\mu$m in two different
channels that  operate simultaneously in the so-called blue (51 to 105 $\mu$m) and red (102 to 220$\mu$m) bands. 
The field of view (FoV) covers a 47''$\times$47'' region in the sky using an array composed of 
5$\times$5 square spatial pixels  (hereinafter \textit "spaxels"), each one 9.4''$\times$9.4'' 
in size, with 16 pixels along the spectral dimension that are shifted to sample the whole 
wavelength range to be covered.  PACS provides a resolving power between 940 and 5500 (i.e. a
spectral resolution of 75 to 300 km/s) depending on the wavelength range. As shown in the
PACS Observer's Manual \footnote{herschel.esac.esa.int/Docs/PACS/pdf/pacs\_om.pdf.}, the PSF 
(point spread function) of the PACS spectrometer ranges from $\sim$ 9'' in the blue band to $\sim$ 14'' 
 in the red band.

\subsubsection{Astronomical observing templates (AOTs)}

 Two different observation schemes or AOTs (astronomical observing
 templates) were offered to the users of the PACS spectrometer. The LineScan Mode 
 was intended for the observation of one or a limited number of
 narrow spectral line features, while the RangeScan Mode was 
 optimized for the observation of broad spectral lines or features, including the
 possibility to observe the full spectral range of the selected orders in 
 `SED (Spectral Energy Distribution)
 mode'. Both configurations generate data in a 3D cube format (flux versus wavelength and 
 spatial position).

%double check above statement with Jesus ``both configurations generate'' ...

\subsubsection{Observing modes}

%_____________________________________________________________
%                 A figure as large as the width of the column
%-------------------------------------------------------------
%   \begin{figure}
%   \centering
%   \includegraphics[width=\hsize]{PACSFoV.png}
%      \caption{Schematic view of the configuration of the Field of View in PACS. The 3D data cubes consist of two spatial dimensions (5x5 spaxels) and one spectral dimension (16 pixels).
%              }
%         \label{PACSFoV}
%   \end{figure}

For each of the AOTs above described, PACS also offered two different observing
modes: "Standard chopping-nodding mode" (hereafter `Chop/Nod') and "Unchopped
grating scan mode" (hereafter `Unchopped'); the difference between these two
modes lies in the observing technique used to allow the telescope and astronomical 
background to be subtracted from the signal coming from the source. `Chop/Nod' 
observations point, alternatively, to the On- and Off- positions and collect 
the On- and Off- data in a common `ObsID'. The `Unchopped' ones take one observation for the 
On-source position and another different observation for
the Off-source pointing; they are then processed independently. After that, the
user has to carry out the required On-Off subtraction.

%_____________________________________________________________
%                 A figure as large as the width of the column
%-------------------------------------------------------------
%   \begin{figure*}
%   \centering
%   \includegraphics[width=\textwidth]{ChopAB.png}
%      \caption{Schematic view of the \textit{Chop-Nod} technique. The chopper throw is chosen by the user; options are small (1'), medium (3') and large (6') 
%              }
%         \label{ChopNod}
%   \end{figure*}

\subsubsection{Pointing modes}

Every observation was also defined by the pointing mode, which could be `pointed' using a single pointing on the source;
or `mapping', a composition of different pointings that, combined, were used to generate maps with improved 
beam sampling and a larger FoV. More information about the different instrument AOTs, observing, and pointing modes 
can be found in the PACS Observer's Manual.

\subsection{Building the THROES sample}

 \textit{Herschel} successfully performed more than 37,000 science observations
 during its operational lifetime. The full list can be found in the Herschel
 Observing Log, available at {\tt http://herschel.esac.esa.int/obslog/}. The 
 Herschel Observing Log contains a total of 530 PACS spectroscopy observations 
 executed successfully, associated to 44 science proposals that were submitted  under the 
 `Evolved Stars/Planetary Nebulae/Supernova Remnants' category
 by the original proposers. Out of these original 530 observations, 347 were identified as corresponding to
 low- or intermediate-mass stars, according to the information available in the
 bibliography, of which a subset of 258 were taken in PACS RangeScan mode. 
 
% how many different proposals? 

  All these observations were originally included in the THROES sample. However, 
  in the final version of the catalogue, we discarded 7 observations of a 
  small group of extended, nearby sources taken in PACS `mapping' mode, as they
  would require a dedicated reprocessing adapted to the
  specific characteristics of each source, an effort that in most cases has been or is 
  being carried out by the research groups that requested the original
  observations. Similarly, we excluded 20 additional observations taken in
  `Unchopped' mode from our reprocessing as these are in general more complex
  observations affected by technical problems that would deserve a
  special case-by-case treatment, beyond the scope of this project. Finally, 11 observations 
  failed during the reprocessing for different reasons and were not included in this first 
  version of the THROES catalogue. In particular, one observation of 
 the planetary nebula NGC 6153 (ObsID 1342249998) and another one of NGC 7662
 (ObsID 1342246642) both failed because of the too narrow spectral range covered, centred
 at the forbidden [O III] emission line at 52 $\mu$m, at the edge of the spectral
 coverage of the PACS blue detector where the spectral response function is not well 
 characterized; six observations (ObsIDs: 1342230895 and 1342230905 to 1342230909) associated to 
 the Red Rectangle failed because they cover very narrow spectral regions affected by leakage; an 
 observation of the post-AGB star HR 4049 (ObsID 1342247550),
 covering a very short wavelength region between 103 and 116 $\mu$m, was extremely noisy 
 in the red channel and failed reprocessing in the blue channel; and finally, we could not reprocess two
 very long exposure SED mode observations of the proto-planetary nebula IRAS 01005+7910 
 (ObsIDs 1342247005 and 1342247006) as they demanded too 
 much memory, exceeding the capacity of our local hardware environment. In summary, a total of 220 Herschel 
  observations (ObsIDs) were finally considered for interactive data reduction, 
  corresponding to 114 individual targets (see full list in
  Table \ref{THROESSampleInfo}), comprising a total of 440 individual spectral
  ranges.

\subsection{Characteristics of the THROES sample}

\subsubsection{Evolutionary stage and IRAS colours}

\begin{figure}
 \centering
 \includegraphics[width=0.5\textwidth]{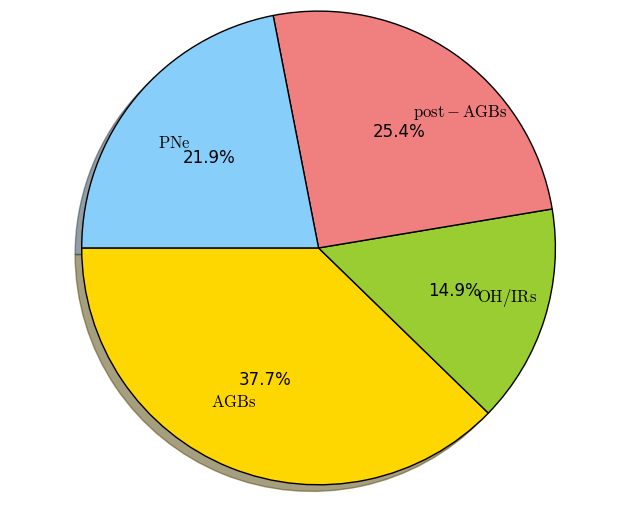}
   \caption{Pie chart illustrating the distribution of stars in the THROES sample according to their 
                evolutionary stage.}
 \label{THROESEv}
\end{figure}

 Information on the evolutionary stage of each object in the THROES sample was extracted 
 from the SIMBAD (Set of Indications, Measurements, and Bibliography for Astronomical Data) database and 
 from the literature. Accordingly, we have classified our 
 objects into four main groups: AGB stars, OH/IR stars (extreme 
 O-rich AGB stars with high mass-loss rates and long variability periods), post-AGB stars 
 (or pre-PNe), and PNe.  In its current version, the catalogue contains 
 PACS range spectra for 43 (38 $\%$) AGB stars, 17 (15 $\%$) OH/IR stars, 29 (25 $\%$) 
 post-AGB stars, and 25 (22 $\%$) PNe (see Fig. \ref{THROESEv}).

% The pie charts with the relative distribution of sources according to their
% evolutionary stage should go here - i.e. before any other plot
% They should include [ absolute number (relative number) and class of star] class of star below, centered

\begin{figure}
 \centering
 \includegraphics[width=\hsize]{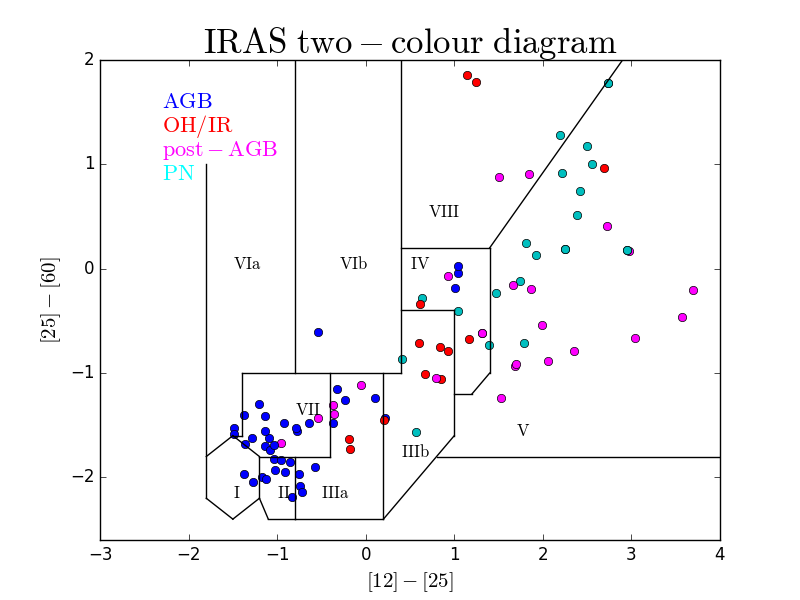}
  \caption{IRAS colour-colour diagram of the THROES sample with good quality
      IRAS data (Quality Flag=3) in the 12, 25, and 60 $\mu$m bands. The diagram is divided into different boxes
      where sources with different characteristics and evolutionary stage are contained
      (see van der Veen \& Habing 1988, for a detailed description).}
 \label{IRASDiag}
\end{figure}

Figure \ref{IRASDiag} shows the distribution of our targets in the IRAS
two-colour diagram, where the location of various types of sources is
indicated, following the original description presented by \cite{vanDerVeen1988}. 
This diagram illustrates the variety of evolutionary stages covered by the
THROES sample. The AGB stars are distributed along a sequence of 
increasing infrared excess, which represents the evolution expected during the
AGB as a result of the formation of thick and dense shells of dust and gas around
these mass-losing stars, with the reddest IRAS
colours corresponding to the most extreme OH/IR stars. Once the mass-loss phase
ends, objects evolve towards the right in the two-colour diagram (region IV, V, and VIII in the plot),
which are the areas populated by most of the sources in our sample 
classified as post-AGB stars and PNe, surrounded by detached cool dust shells.

%% La referencia de van der Veen et al. 1988 debe aparecer como van der Veen &
%% Habing 1988, si solo son dos los autores.

%                 A figure as large as the width of the column
%-------------------------------------------------------------

\subsubsection{Galactic distribution}

\begin{figure*}
   %\centering
 \begin{subfigure}{.33\textwidth}
    %\centering
 \includegraphics[width=62mm]{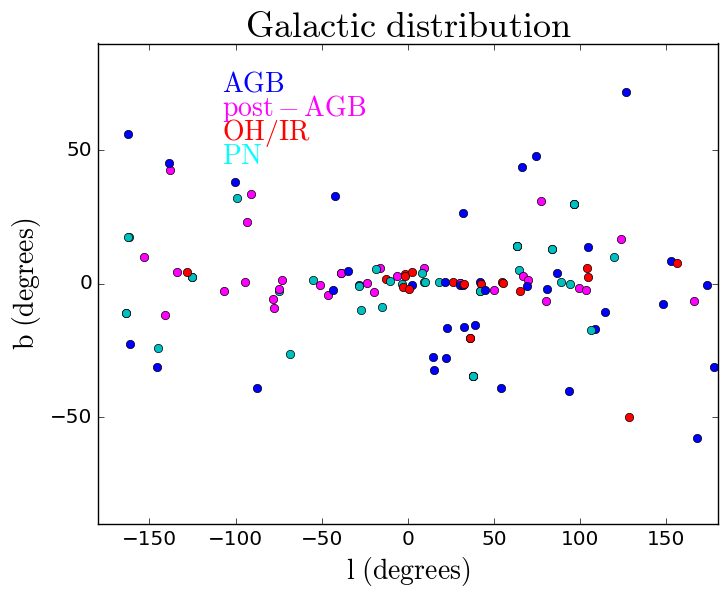}
  %\caption{1d}
 \label{fig:sfig1d}
 \end{subfigure}
 \begin{subfigure}{.33\textwidth}
    %\centering
 \includegraphics[width=62mm]{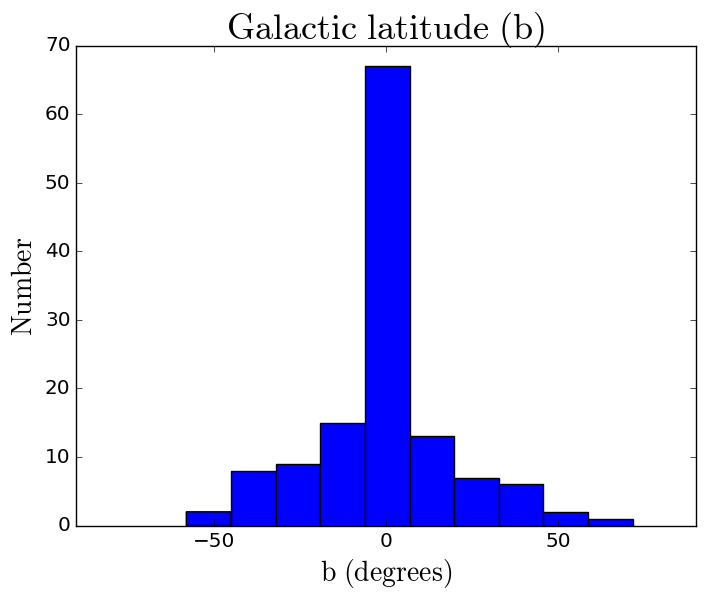}
  %\caption{1d}
 \label{fig:sfig1d}
 \end{subfigure}
 \begin{subfigure}{.33\textwidth}
    %\centering
 \includegraphics[width=62mm]{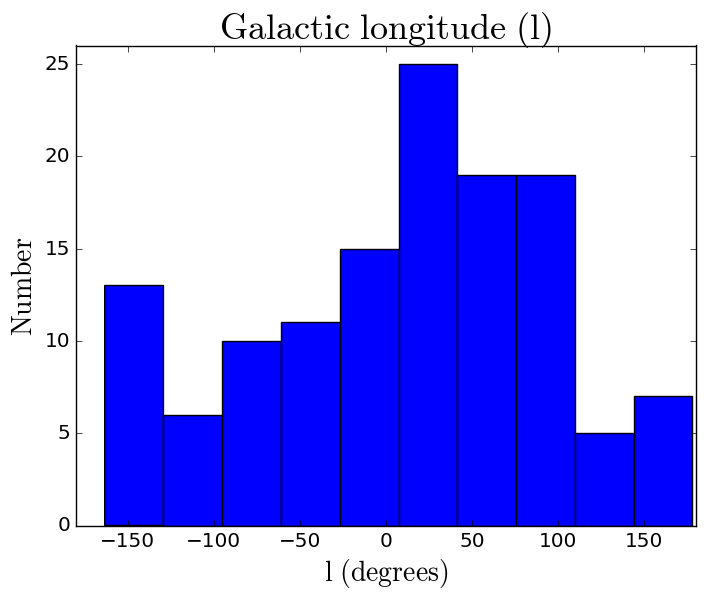}
  %\caption{1d}
 \label{fig:sfig1d}
 \end{subfigure}
 \caption{Galactic distribution of the THROES targets. Colour code as in Fig. \ref{IRASDiag}.}
 \label{THROESGal}
\end{figure*}

The galactic distribution of the THROES sample is shown in
Fig. \ref{THROESGal}. Our sources are strongly concentrated at relatively low 
galactic latitudes, as expected from a distant population of luminous
sources concentrated in the galactic disk, although a significant number of
them are also observed at high galactic latitudes, corresponding to the
small fraction of bright, nearby sources. Interestingly, most of the OH/IR
stars are found at very low galactic latitudes, as they correspond to a
population of stars that are proposed to represent the most massive precursors 
of PNe \citep{GarciaHernandez2007}.

%% Add reference Garcia-Hernandez et al. 2007 ApJ...666L..33G

%_____________________________________________________________
%                 A figure as large as the width of the column
%-------------------------------------------------------------

%_____________________________________________________________
%                                     Two column figure (place early!)
%______________________________________________ Gamma_1 (lg rho, lg e)

\subsubsection{Dominant chemistry}

\begin{figure}
 \centering
 \includegraphics[width=0.5\textwidth]{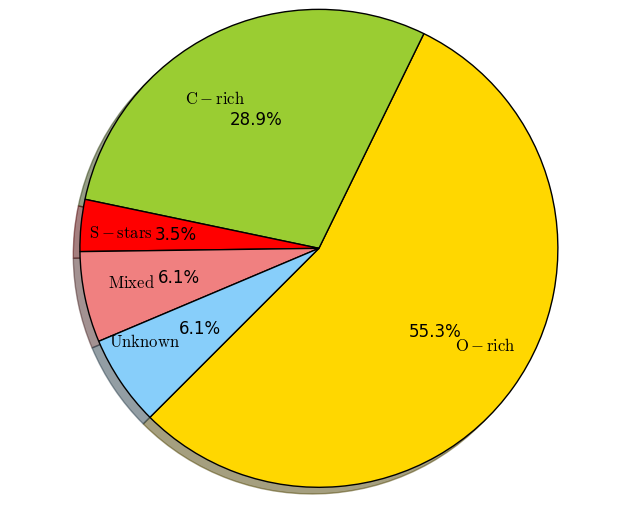}
 \caption{Pie chart illustrating the distribution of stars in the THROES sample according to their
                dominant chemistry.}
 \label{THROESChem}
\end{figure}

 Asymptotic giant branch stars are generally classified as O-rich (M-type) or C-rich (C-type) based on the C/O 
 ratio found in their outer envelope. Objects with C/O > 1 are considered C-rich stars under this
 criterion while objects with C/O < 1 are considered O-rich stars. Those sources showing C/O ratios
 of approximately one. are denoted as S-stars. In addition, there are also objects that present 
 chemical indicators from both kinds of chemistries, such as the simultaneous presence of 
 crystalline silicates, typical of O-rich chemistry, and policyclic aromatic hidrocarbons (PAHs),
 expected in C-rich targets, in their mid-infrared spectra.
 The C/O ratio in this kind of source may be different depending on the region of
 the source considered. Some of them are known to be objects in transition 
 between O-rich and C-rich objects \citep{Herwig2005}; others display a strong
 bipolar morphology, and may be surrounded by disks that could explain the mixed
 chemistry observed. In our sample, 29$\%$ of sources are C-rich stars, 55$\%$ are O-rich 
 stars, 3$\%$ are S-stars, 7$\%$ are sources with mixed chemistry, and 6$\%$ are sources with
 unknown chemistry (see Fig. \ref{THROESChem}).

% The pie charts with the relative distribution of sources according to their                               % evolutionary stage should go here - i.e. before any other plot                                            
% They should include [ absolute number (relative number) and class of star] class of star below, centered   

\section{Data reduction} \label{DataReduction}

 The data available in the HSA have been generated using the
 latest version of the Standard Product Generator (SPG), an automated pipeline that takes
 the data from level 0 (raw) to level 2 (spectral cubes). At the
 time of starting this project, the archive was populated with products generated with
 HIPE 13.0.0. A complete explanation of the pipeline processing steps applied in this version
 of the pipeline can be found in the PACS Data Reduction Guide: Spectroscopy. 
 \footnote{\tiny{herschel.esac.esa.int/twiki/pub/Public/PacsCalibrationWeb/pacs\_spec\_Hipe13.pdf.}}
%%The strategy followed to reduce the observations of the sample was determined by the way the observations
%%were taken, so it is crucial to know all the parameters that defined each observation.
 
 Interactive data reduction can help improve the quality of 
 the final products by applying certain data reduction tasks not available in the
 automated pipeline because they require a direct interaction of the user. Therefore, we used HIPE 13.0 
 and version 65 of the PACS calibration files to interactively reprocess all the observations in our sample, 
 introducing those tasks that at the moment of starting this project were not
 applied by the standard data reduction pipeline (version 13 at that time), with the aim of 
 obtaining better quality products than those offered by the pipeline.

 For PACS range spectroscopy observations, the interactive data reduction tasks
 that we have applied to improve the quality of the data can be split into two main
 groups: 

\begin{itemize}
 \item i) tasks applied to the spectral cubes before the level 2 data are generated, so-called, 
 interactive re-processing.
 \item ii) tasks applied to the level 2 data products to extract the best 1D spectrum, so-called, post-processing.
\end{itemize}

 The interactive re-processing tasks comprise:

\begin{itemize}

\item{FlatField correction:} This correction fits an n-grade polynomium to the PACS spectra 
  to obtain a higher signal-to-noise ratio (S/N) and a better shape of the continuum.
\item{Telescope background correction:} This correction uses the telescope background emission to flux 
calibrate the spectra instead of using the relative spectral response function
(RSRF) applied in the SPG. This correction was later implemented in version 14 of the SPG pipeline. 

\end{itemize}

To extract the final 1D spectrum from the level 2 data cubes, the following post-processing
corrections were applied:

\begin{itemize}

\item{Point source flux loss correction (PSFLC):} This is needed to extract the
  correctly-calibrated spectrum of a point source to account for PSF losses.
\item{Correction for semi-extended sources (which we refer to as "semi-extended 3x3" correction):} To recover the 
      whole flux of semi-extended sources,  defined as those sources with a FWHM (Full Width Half Maximum) larger
      than a single spaxel (9.4") but small enough to fill the field of view of a single pointing (about 47"x47"). 
      This correction is also applied to try to minimize the effects 
      of pointing jitter. The so-called semi-extended 3x3 correction is always applied after the 
      PSFLC and the final product is a 1D spectrum labelled as "PSFLC-3x3" corrected. 
\item{Correction for extended sources (which we refer to as "extended 5x5" correction):} This
       correction was also never made available in HIPE, but we had to create it to deal with 
       sources extended beyond the central 3x3 spaxels. It works in a similar way to the 
       semi-extended 3x3 correction, and is also  applied after PSFLC. In this case the
       final product is a 1D spectrum labelled as "PSFLC-5x5" corrected.

\end{itemize}

 In addition, several objects in our sample were significantly mispointed, that is, 
 the main target was clearly not located in the central spaxel of the 5$\times$5 spaxel array of PACS. 
 To extract the 1D spectra of these sources, we had to use a modified version of the semi-extended 3x3 
 correction, as we will explain in Section \ref{PostProcessing} in more in detail.

\subsection{Spectral flatfielding} \label{FlatField}

 As explained in the PACS Data Reduction Guide, spectral flatfielding can be crucial 
 for improving the continuum of the final spectra. For range scan observations, flatfielding will
 result in general in a better S/N, especially for the longer wavelength ranges. It will also remove the
 "fringing"-like  pattern that the SPG cubes often have and it will improve the 
 appearance of sharp rises or drops in the data.

 The HIPE task used for flatfield correction is \textit{specFlatFieldRange}, where the user can 
 select the order of the polynomial to fit. The default value is five but from our 
 experience, a lower value (three or four) usually  yields better results. For PacsRange observations
 that do not cover the whole SED of PACS, the best results are 
 obtained using an even lower value for the order of the polynomial (normally one). We have configured the
 \textit{specFlatFieldRange} task to exclude the regions affected 
 by leakage (70-73 $\mu$m, 98-105 $\mu$m, and 190-220 $\mu$m); by setting the 
 option \textit{excludeLeaks} to "True" we obtained "clean" spectra.
The exclusion of the regions affected by leaks is actually very important to obtain a correct shape of 
 the continuum and to improve the results obtained with the \textit{specFlatFieldRange} range. 
 We note that the  improvement in the S/N  ratio after flatfielding is most remarkable in relatively faint sources.

\subsection{Telescope background correction} 

 This correction is needed to remove the sky and telescope background contribution. Furthermore, it uses 
 the telescope background spectrum to flux calibrate the data, instead of using the standard calibration 
 blocks.
This process is applied using the HIPE task \textit{specDiffChop}. This task computes
 the pairwise difference ratio: 2*(On-Off)/(On+Off), rather than the pairwise differences (On-Off) as is the case for the standard pipeline. This computation eliminates in an optimal way the detector
 drifts that appear in both On- and Off- positions. After this step, the flux density in the cubes is 
 in units of "telescope background".

%_____________________________________________________________
%                 A figure as large as the width of the column
%-------------------------------------------------------------
%   \begin{figure*}
%   \centering
%   \includegraphics[width=\textwidth]{FFfigureCompleta2.png}
%      \caption{\textbf{Top)} Level 2 PACS spectra, without post-processing, in the 70-105 $\mu$m range 
%      for one of our targets (W Aql) with spectral flatfielding applied with \textit{PolyOrder=4} and 
%      \textit{PolyOrder=5} (left and right respectively). Note the artificial
%      bumps introduced in the shape of the continuum due to a \textit{PolyOrder} value too high.
%      \textbf{Bottom)} Level 2 PACS spectra, without post-processing, in the 70-105 $\mu$m range for W Aql with 
%      and without applying the FlatField correction (red and blue, respectively), see subsection 
%      \ref{FlatField}. The >=95 $\mu$m spectral range, which is adversely affected by the so-called red leak, 
%      is sistematically removed in our data reduction process for an improved flatfield correction.}
%         \label{Fig2}
%   \end{figure*}

\subsection{THROES post-processing} \label{PostProcessing}

\begin{figure*}
   \centering
   \includegraphics[width=\hsize]{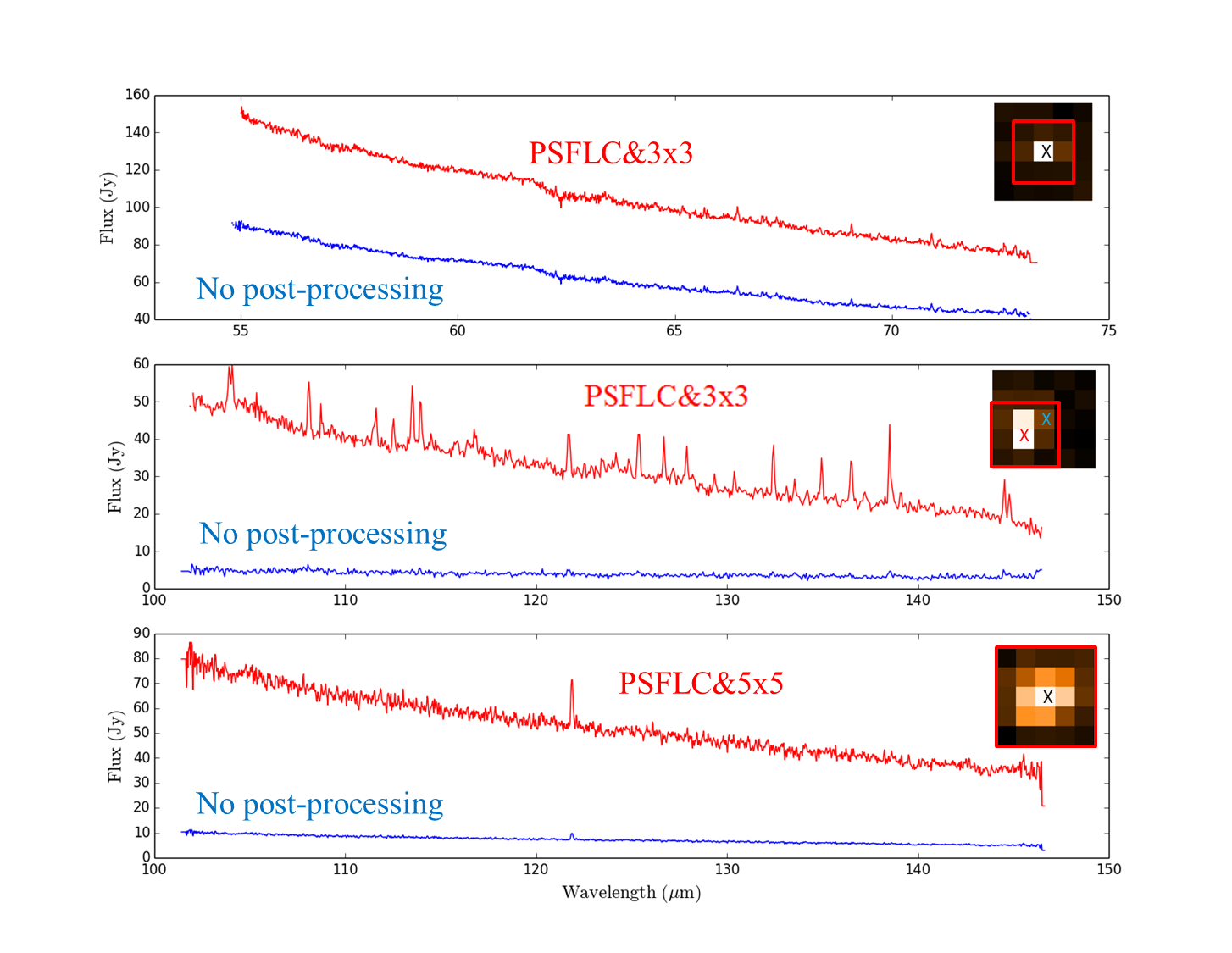}
      \caption{One-dimensional spectra extracted for three different targets in the THROES catalogue
      following the procedures described in Section \ref{DataReduction} (red). 
      The flux corrections and apertures have been chosen depending on the source extent and its location
      in the 5x5  spaxels FoV of PACS. For comparison, we show the corresponding spectrum obtained in each 
      case from the central spaxel before post-processing (blue). At the top right corner of each box, we show 
      a layer of the {\it FinalCube} with the 5x5 spaxels of the PACS spectrometer of the 
      observation. The crosses indicate the spaxel from which the spectra were taken; a black cross means that 
      both spectra were extracted from the central spaxel. The squares (red and blue) 
      enclose the spaxels used for the 3x3 correction or for the 5x5 correction. \textbf{Top)} AFGL 3116, blue 
      camera of PACS (ObsID: 1342212512): well pointed and semi-extended source; PSFLC-3x3 correction has been applied. 
      \textbf{Centre)} Same as above but for IRC$-$10529, red camera of PACS (ObsID: 1342208931): this 
      is a mispointed observation where the blue spectrum corresponds to that  extracted from the central 
      spaxel without post-processing, while the red one is the spectrum taken from the brightest spaxel 
      and with the post-processing applied (PSFLC-3x3). \textbf{Bottom)} Same as the first two panels but for
      NGC 6543, red camera (ObsID: 1342238389): this is an extended source; PSFLC-5x5 correction applied.}
         \label{PostAndMispointing}
   \end{figure*}

 All the observations have been post-processed in different ways taking into account 
 the nature of the object (extended or not) and the position of the source in the PACS FoV. 
 The main goal of the post-processing was to extract, from the spectral
 cubes previously reduced, a 1D spectrum recovering the whole emission of the source. Depending on the 
 post-processing tasks applied, the sources can be grouped into five main families:

\begin{itemize}

 \item{\textbf{Pointed observations, good pointing, point, or semi-extended sources (PSFLC-3x3)}}: To 
 recover the absolute flux of a point source from PACS data, it is necessary to apply first the point 
 source flux loss correction (PSFLC) to the spectrum extracted from the central spaxel of the final
 (level 2) spectral cube. This correction is needed to take into account the flux losses derived from
 the fraction of the PSF that falls out of the central spaxel. A theoretical PSF model is used to compute the 
 fraction of flux seen by the central spaxel and then recover the emission that has not been detected. After this, we 
 also need to apply the semi-extended 3x3 correction in order to recover the emission received by the spaxels around 
 the central one in case of semi-extended sources. Both corrections are performed by the HIPE task 
 \textit{extractCentralSpectrum}, which generates three different 1D spectra:
 
1) The first 1D spectrum returned by the task is simply the spectrum from the central spaxel corrected for
     point source flux losses.
     
2) The second 1D spectrum returned contains the integrated flux of the 3x3 central spaxels (also known 
    as "superspaxel") with the point source flux loss correction applied to the central one.

3) Finally, the third 1D spectrum is the same as the first one, corresponding to the central spaxel 
     with the point source flux loss correction applied, scaled to the flux level of the second spectrum,
     that of the 3x3 superspaxel. The result is the so-called semi-extended 3x3 corrected 
     spectrum.

Using the Semi-extended 3x3 corrected spectrum, the whole flux from sources that are slightly
extended is fully recovered. An example of the differences between the 1D spectrum generated after 
applying this correction and the spectrum taken directly from the central spaxel of the level 2 cube 
before post-processing is shown in Fig. \ref{PostAndMispointing}. The vast majority of the sources in 
our sample have been reprocessed in this standard way and the resulting 1D spectra are displayed in 
Fig. \ref{StandardSources}.
An exceptional case is OH 32.8$-$0.3, for which some spaxels around the central one show negative flux 
values and therefore these observations have only been corrected for PSFLC 
(i.e. the semi-extended 3x3 correction has not been applied).\\

%\end{itemize}

%_____________________________________________________________
%                 A figure as large as the width of the column
%-------------------------------------------------------------
%   \begin{figure*}
%   \centering
%   \includegraphics[width=\hsize]{PostProcessingFigureRectangularCompleta.png}
%      \caption{\textbf{Bottom)} One-dimensional spectra of AFGL 3116 obtained with the blue camera of PACS
%      (ObsID: 1342212512) extracted from the central spaxel of the FinalCubes before (blue) and after (red)
%      applying the post-processing tasks, PSC and 1-to-9 correction, described in section \ref{PostProcessing}.
%      Note the significant understimate of the continuum level if these corrections are not applied. At the top
%      right corner of the figure we show a layer of the FinalCube with the 5x5 spaxels. The crosses (red and 
%      blue) indicate that both spectra where extracted from the central spaxel.
%              }
%         \label{WAqlPSC3x3}
%   \end{figure*}

\item{\textbf{Pointed observations, extended sources (PSFLC-5x5):}} Based on PACS photometric data, 
 when this was available, the existing PACS spectroscopy, and the bibliography, we identified those 
 sources in our sample that appeared more extended than just the central 3x3 spaxels. These `extended'
 objects are: IRAS 16122$-$5122, NGC 3242, NGC 40, NGC 6445, NGC 6543, NGC 6781, NGC 6826, NGC 7009, 
 NGC 7026, and Mz3. Due to their extension, their emission can spread sometimes over the whole PACS FoV and, therefore, for
these objects, we developed a new task, the extended 5x5 correction, to scale the spectrum from 
the central spaxel, after applying the point source flux loss correction, to the continuum level of the
5x5 spectrum, as we did for the semi-extended 3x3 correction.\\

As we will see in Section \ref{MultiMission}, with this new task we were able to obtain a more realistic spectrum of 
the extended sources in our sample, better than the spectrum generated with the semi-extended 3x3 correction, as in.
this way we recovered the whole emission from the source. In Fig. \ref{Extended5x5}, we show the final 1D spectra of 
these extended sources after applying the PSFLC and the extended 5x5 correction. The effect of this
task in terms of continuum flux level recovery can be seen in Fig. \ref{PostAndMispointing}.\\

\item{\textbf{Pointed observations, mispointing (PSFLC-3x3):}} In our sample there were also seven cases of 
mispointed observations, namely, IRAS17347$-$3139, NGC 6302, IRC$-$10529, OH21.5+0.5, AFGL 5379, 
IRAS 16279$-$4757, and IRAS 13428$-$6232, for which the source is located on a spaxel different than the 
central one. In these cases, \textit{extractCentralSpectrum} cannot be 
applied because, in this task, the 3x3 superspaxel is always built around the central spaxel, instead 
of around the one where the source is located.\\
 
To deal with these mispointed cases, we developed a script that works in a similar way to the HIPE task 
\textit{extractCentralSpectrum} but which allows the user to select the spaxel (other than the central one) 
where the source is located. Our task was successfully applied to IRC$-$10529, OH 21.5+0.5, IRAS 13428-6232, NGC 6302, and
IRAS 17347-3139. In Fig. \ref{Mispointed3x3} we show the final 1D spectra obtained after applying this correction.

For AFGL 5379 and IRAS16279$-$4757 this correction did not work correctly as, due to the position of the 
sources being too far away from the central spaxel, it was impossible to generate the 3x3 superspaxel around
the off-centred spaxel where the source was located.

As was done for other observations in our sample, it is also necessary to correct these 
mispointed observations from PSF losses before applying the semi-extended 3x3 correction described above.
To do that, HIPE provides two different tasks: \textit{extractSpaxelSpectrum} and \textit{pointSourceLossCorrection}. The first one takes the spectrum from the spaxel where the source is located and, after that, the second one corrects for the point source flux losses. 

Again in Fig. \ref{PostAndMispointing} we show the significant improvement of the final 1D spectra after 
applying the semi-extended 3x3 correction to mispointed observations.\\
\item{\textbf{Sources corrected only for PSFLC:}} Three additional sources(AFGL 5379, IRAS 16279-4757, and OH 32.8-0.3)
were only corrected for PSFLC, due to different issues that prevented the application of the semi-extended 3x3 correction. 
These issues are related to the position of the source in the 5$\times$5 spaxels array and the presence of corrupted data
in some of the spaxels needed to create the 3$\times$3 superspaxel. Their spectra are displayed in Fig. \ref{OnlyPSC}. For these three sources the absolute flux level of the final 1D spectrum available in the THROES catalogue 
should be considered only as a lower limit. \\

\end{itemize}

%_____________________________________________________________
%                 A figure as large as the width of the column
%-------------------------------------------------------------

%\begin{itemize}

%\item{\textbf{Mapping observations}}: Mapping observations are those which present a raster of pointings at different 
%positions, in the same obsID. For mapping observations we can generate a new final cube named 
%\textit{projectedCubes} using the \textit{slicedDiffCubes} (for Unchopped observations) or the 
%textit{slicedFinalCubes} (for Chop/Nod ones). The Hipe task used in both cases is the same \textit{specProject}. The 
%output product is a new data cube with a higher spatial resolution due to the combination of the raster of individual 
%observations, the spatial resolution is given by the parameter \textit{outputPixelSize} that is, by default, set to 3.
%In Fig. \ref{ProjectedCube} we show an specific layer of a projected cube. 

%The 'Unchopped' observations associated to NGC 7009, NGC 6720, NGC 2022, NGC 2392 and NGC 2440 have been discarded after
%the reprocessing because the final products were not reliable.

%\begin{figure}
% \centering
% \includegraphics[width=0.5\textwidth]{NGC2440-177microns.png}
% \caption{Layer of the projected cube of NGC 2440 (ObsIDs: 1342231295 (On), 1342231294 (Off). Camera: blue) at 177 
% microns and a pixel size of 3''.}
% \label{ProjectedCube}
%\end{figure}

To summarize, after our interactive data reduction process, we generated for each object the following
final products:

\begin{itemize}

\item{1)} Final spectral cubes (level 2) with the improvements that result from flatfielding and telescope
background correction.
\item{2)} 1D spectra with the PSFLC applied, as well as the specific 
correction for slightly extended sources (semi-extended 3x3 correction) or extended sources 
(extended 5x5 correction). As a word of caution, we note that the scaling strategy followed assumes 
that the extension of the continuum and of the line-emitting region is roughly the same and that the spectral
lines emission does not vary within the nebula, which may not be the case in some of the more extended objects.
For those sources showing a more complex and extended morphology (particularly the PSFLC-5x5 corrected targets),
we recommend the users to create their own 1D spectra making use of the final cubes provided in the THROES 
catalogue if they want to perform a more detailed analysis.

\end{itemize}

%\subsection{Data Reduction Issues}

%Along the data reduction process some issues were reported to the PACS Helpdesk Team. They are:
%\begin{itemize}
% \item{\textbf{1)}} The observations associated to HR 4049 and IRAS 01005+7910, were not reprocessed due to errors
% in the reduction process that could not be solved.
% \item{\textbf{2)}} The spectra generated from the unchopped observations (On and Off) for the sources: NGC 7009, 
% NGC 6720, NGC 2022, NGC 2392 and NGC 2440, present, for some bands in PACS, flux values below zero. This issue 
% is explained in \textit{PACS: Data Reduction Guide (Spectroscopy)} and it is related to the way the observations were
% taken.
 
%\end{itemize}

%______________________________________________________________

\section{THROES catalogue}\label{DescriptionCatalogue}

\begin{figure*}
   \centering
   \includegraphics[width=\textwidth]{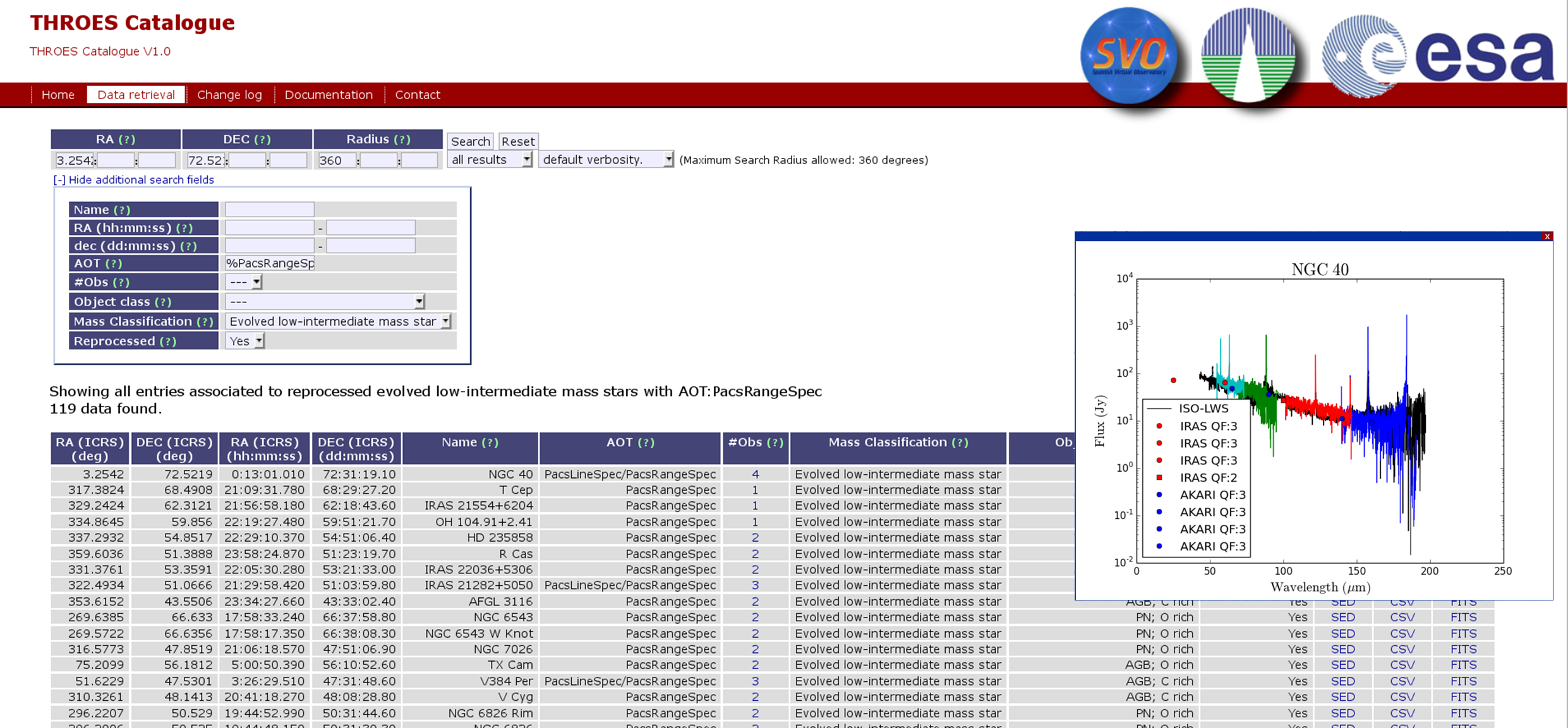}
      \caption{Screenshot of the THROES catalogue web interface. All the columns and search fields 
described in Section \ref{DescriptionCatalogue} are shown as well as an example of a complete SED 
with IRAS and AKARI data overplotted.
              }
         \label{THROESWeb}
   \end{figure*}

All the reprocessed PACS range spectroscopy observations have been compiled in a catalogue that is accessible via a web 
interface available at {\tt https://throes.cab.inta-csic.es/}. On this web page all the reprocessed
data are publicly available, including some general information about the objects and the observations made,
as well as plots with the reduced PACS spectroscopy data and complementary spectroscopic and photometric
observations from other observatories. A screenshot of the `Data Retrieval' interface of the catalogue
is shown in Fig. \ref{THROESWeb}.

The tool used to create the web-based catalogue is 
\textit{SVOCat}\footnote{http://svo2.cab.inta-csic.es/vocats/SVOCat-doc/.}. This application, developed by the 
Spanish Virtual Observatory (SVO\footnote{http://svo.cab.inta-csic.es.}) is designed to make the publication 
of an astronomical catalogue easier, both as a web page and as a Virtual Observatory \textit{Cone Search} service.
Our archive system has been designed following the IVOA (International Virtual Observatory
Alliance)\footnote{http://www.ivoa.net.} standards and requirements. In particular, it
implements the {\it Cone Search} protocol, a standard defined by the IVOA for retrieving
records from a catalogue of astronomical sources. Queries made through
the {\it Cone Search} service are based on the description of a sky position
and an angular distance, defining a cone on the sky. The response returns a 
list of astronomical sources from the catalogue whose positions in the sky lie
within the pre-defined cone, formatted as a VOTable (Virtual Observatory Table). The system as a whole is included in 
the IVOA Registry and can therefore be discovered by any VO-tool.

To structure our catalogue in a homogeneous and logical way, all the observations with the same pointing 
have been grouped into a unique entry in the archive. Therefore, there is a single line per region 
in the sky observed. As we can see in Fig. \ref{THROESWeb}, there are 13 columns for each entry, each providing 
information about the object or the observation as well as links to plots and to the reprocessed data. 
Each of the columns contains the following information:

\begin{itemize}
\item{\textbf{Columns 1 to 4}}: The equatorial coordinates (right ascension and declination) of the 
observation in different units: decimal degrees in columns 1 and 2, and hh:mm:ss and 
dd:mm:ss in columns 3 and 4.
\item{\textbf{Column 5}}: Name of the object in the THROES catalogue.
\item{\textbf{Column 6}}: The astronomical observation template (AOT). This could be PacsRange or PacsLine. 
For some objects there are spectra taken in both modes. These cases appear in the catalogue as PacsLine/PacsRange.
Only the PacsRange spectra have been reprocessed in the current version of the THROES catalogue.
\item{\textbf{Column 7}}: The number of observations taken in that position of the sky. 
By clicking on the number shown in this column, a new table is deployed with detailed information for each 
observation, such as: target name, equatorial coordinates (RA and Dec), proposal name, AOT, observation ID, 
observing date and time, and original AOR (Astronomical Observation Request) label.
\item{\textbf{Column 8}}: Mass classification (based on the bibliography and 
 SIMBAD\footnote{http://simbad.u-strasbg.fr/simbad/.}). The options are: evolved low-intermediate mass star,
 evolved massive star, or unknown.
\item{\textbf{Column 9}}: Object classification including the evolutionary stage and
 its dominant chemistry (based on the bibliography and SIMBAD), the options 
are: O-rich AGBs, C-rich AGBs, S stars, OH/IR stars, O-rich post-AGBs,  C-rich post-AGBs, mixed chemistry 
post-AGBs, O-rich PNe, C-rich PNe, mixed chemistry PNe, or Unknown.
\item{\textbf{Column 10}}: This column indicates if the observations have been reprocessed under the THROES
 project or not. This is because THROES may be expanded in the future to other observing modes of PACS and/or 
 massive evolved stars.
\item{\textbf{Column 11}}: By double clicking on `SED', a pop-up window is displayed showing the 1D PACS spectra 
generated after the  interactive data reduction and subsequent post-processing,  
together with complementary photometric (IRAS and AKARI) and spectroscopic 
(Infrared Space Observatory-Long-Wave Spectrometer (ISO-LWS)) data, when available.
\item{\textbf{Column 12}}: By double clicking on `CSV' (Comma-Separated Values), a compressed folder with the name 
of the target is downloaded containing a gzipped tar file with the 1D PACS spectra, in CSV format.
\item{\textbf{Column 13}}: Similarly, by double clicking on `FITS' (Flexible Image Transport System), a compressed 
gzipped tar file is downloaded containing the THROES reprocessed final spectral cubes (level2) in FITS format and the 
1D PACS spectra, also in FITS format, derived from these final spectral cubes after post-processing.\\

\end{itemize}

More information about how to query the catalogue through the different search fields is available in the
documentation available on the THROES catalogue web page.
% The inner structure of the folder is:

%\begin{itemize}

%\item{Pointed observations, well pointed:} 
%\\
%\\
%\verb+-TargetName/Final1D/hpacsObsID_TargetName_+
%\verb+Date_MinWavelength_MaxWavelengthmicrons-+
%\verb+cameraSlice_centralSpaxel_PSC_3x3correction+
%\verb+YES_1D.fits+
%\\
%\\
%Example:
%\\
%\verb+IC418/Final1D/hpacs1342265942_IC418_2013-+
%\verb+03-04_51_73microns-blue00_centralSpaxel_PSC_+
%\verb+3x3correctionYES_1D.fits+
%\\
%\\
%\verb+-TargetName/FinalCube/hpacsObsID_TargetName+
%\verb+_Date_MinWavelength_MaxWavelengthmicrons+
%\verb+-camera.fits+
%\\
%\\
%Example:
%\\
%\verb+IC418/FinalCube/hpacs1342265942_IC418_2013+
%\verb+-03-04_51_73microns-blue.fits+
%\\
%\\
%\item{Mispointed without SemiExtended3x3 correction:} 
%\\
%\\
%\verb+-TargetName/Final1D/hpacsObsID_TargetName+
%\verb+_Date_MinWavelength_MaxWavelengthmicrons-+
%\verb+cameraSlice_MISPOINTING_PSC_1D.fits+
%\\
%\\
%Example:
%\\
%\verb+AFGL5379/Final1D/hpacs1342228538_AFGL5379_+
%\verb+2011-09-13_69_105microns-blue00_+
%\verb+MISPOINTING_PSC_1D.fits+
%\\
%\\
%\verb+-TargetName/FinalCube/hpacsObsID_TargetName+
%\verb+_Date_MinWavelength_MaxWavelengthmicrons+
%\verb+-camera.fits+
%\\
%\\
%Example:
%\\
%\verb+AFGL5379/Final1D/hpacs1342228538_AFGL5379_+
%\verb+2011-09-13_69_105microns-blue00.fits+
%\\
%\\
%\item{Mispointed with SemiExtended3x3 correction:}
%\\
%\\
%\verb+-TargetName/Final1D/hpacsObsID_TargetName+
%\verb+_Date_MinWavelength_MaxWavelengthmicrons-+
%\verb+cameraSlice_MISPOINTING_PSC_3x3_1D.fits+
%\\
%\\
%Example:
%\\
%\verb+IRC-10529/Final1D/hpacs1342208931_IRC-10529_+
%\verb+2010-11-14_51_73microns-blue00_+
%\verb+MISPOINTING_PSC_3x3_1D.fits+
%\\
%\\
%\verb+-TargetName/FinalCube/hpacsObsID_TargetName+
%\verb+_Date_MinWavelength_MaxWavelengthmicrons+
%\verb+-camera.fits+
%\\
%\\
%Example:
%\\
%\verb+IRC-10529/Final1D/hpacs1342208931_IRC-10529_+
%\verb+2010-11-14_51_73microns-blue00.fits+
%\\
%\\
%\item{Pointed observations, extended:} 
%\\
%\\
%\verb+-TargetName/Final1D/hpacsObsID_TargetName_+
%\verb+Date_MinWavelength_MaxWavelengthmicrons-+
%\verb+cameraSlice_centralSpaxel_PSC_5x5correction+
%\verb+YES_1D.fits+
%\\
%\\
%Example:
%\\
%\verb+NGC3242/Final1D/hpacs1342232279_NGC3242_2011-+
%\verb+11-12_69_105microns-blue00_centralSpaxel_PSC_+
%\verb+5x5correctionYES_1D.fits+
%\\
%\\
%\verb+-TargetName/FinalCube/hpacsObsID_TargetName+
%\verb+_Date_MinWavelength_MaxWavelengthmicrons+
%\verb+-camera.fits+
%\\
%\\
%Example:
%\\
%\verb+NGC3242/FinalCube/hpacs1342232279_NGC3242_2011+
%\verb+-11-12_69_105microns-blue.fits+
%\\
%\\
%\end{itemize}

%\verb+-TargetName/Final1D/hpacsObsID_TargetName_Date_+
%\verb+MinWavelength_MaxWavelengthmicrons-cameraSlice_+
%verb+MISPOINTING_PSC_1D.csv+
%\\
%\\
%Example:
%\\
%\verb+IRC-10529/Final1D/hpacs1342208931_IRC-10529_+
%\verb+2010-11-14_51_73microns-blue00_+
%\verb+MISPOINTING_PSC_1D.csv+

%\end{itemize}

%_____________________________________________________________
%                 A figure as large as the width of the column
%-------------------------------------------------------------

%
%______________________________________________________________

\section{Characterization of the THROES sample} \label{MultiMission}

All the individual THROES spectra are shown from Figs. \ref{StandardSources} to \ref{OnlyPSC}. To illustrate 
the quality of the resulting spectra, every plot contains not only the final PACS 1D spectra, but also the 
photometric (IRAS and AKARI) and spectroscopic ISO-LWS data, whenever these are available.
In Table. \ref{THROESSampleInfo} we display information associated to the PACS, ISO (Infrared Space Observatory), 
IRAS, and AKARI data used for the generation of the SEDs.

\subsection{Comparison with IRAS and AKARI photometry}

All the objects in the THROES catalogue have complementary IRAS photometric information
while, for AKARI data, the percentage of common objects decrease to 89$\%$.
The Infrared Astronomical Satellite (IRAS) (http://irsa.ipac.caltech.edu/IRASdocs/iras.html) 
point source catalogue presents photometric data in four bands centred at 12, 25, 60, and 100 $\mu$m. 
The beam size of IRAS observations, which is much larger than that of PACS, varies 
from 2' at shorter wavelengths to 5' at longer ones \citep{Jeong2007}.
AKARI (http://www.ir.isas.jaxa.jp/AKARI/index.html) covers longer wavelengths, 65, 
90, 140, and 160 $\mu$m, and the beam size varies from 0.5' to 0.9' \citep{Jeong2007}.
The different beam sizes of the instruments is a key point that has to be kept in mind when comparing PACS 
spectroscopy to IRAS and AKARI photometric data.

In order to check the quality of the final processed PACS spectroscopy data and to obtain information about the effect 
of the post-processing tasks applied to generate the final 1D spectra, a comparison between the
synthetic photometry derived from the PACS spectroscopic data and the IRAS and AKARI photometric data at 100 and 160
$\mu$m, respectively, has been done.

Synthetic photometry is generated by convolving the transmission curve of the photometric filters,
 IRAS$_{100}$ and AKARI$_{160}$, with the 1D PACS spectra. The transmission curve of IRAS$_{100}$ 
has been obtained from:\\ \verb+http://irsa.ipac.caltech.edu/IRASdocs/exp.sup/ch2/+
\verb+tabC5.html+ and the transmission curve of AKARI$_{160}$ has been estimated from the relative response function
available in:\\ \verb+http://svo2.cab.inta-csic.es/theory/fps3/+
\verb+?id=AKARI/FIS.N160+.

As the synthetic photometry requires a convolution between the 1D PACS spectrum and the curve of the 
photometric filters, it is necessary that the PACS spectrum spectral coverage extends, at least, along the 
whole wavelength subrange covered by the photometric filters. For this reason, the synthetic photometry can be 
estimated for the subsample of 71 THROES targets with full spectral coverage of the PACS wavelength range.

As mentioned in Section \ref{DataReduction}, due to the instrument configuration and the exclusion of 
some sub-regions affected by leakage, all the 1D PACS spectra show a small gap (less than 10 $\mu$m) around 100 $\mu$m,
even for those sources that present a complete spectral coverage. As the photometric curve of IRAS 100 $\mu$m is 
centred on this region, it is important to find a solution. To cope  with that in the synthetic photometry estimation,
a linear interpolation was done to approximate the continuum flux level in this region. This assumption is good enough 
as intense emission lines are not expected in this region.

On the AKARI 160 $\mu$m side, due to the exclusion of the wavelength regions affected by leakage, 
there are no PACS data to cover the wavelength range from 190 to 220 $\mu$m. To solve that, a power law 
($\lambda ^{\alpha}$) was fitted to the red bands of the PACS SED and, after that, we extrapolated the flux values to
the 190-220 $\mu$m region. At these wavelengths, we are mainly tracing the Rayleigh-Jeans emission, so this 
extrapolation is reasonable.

\begin{figure*}
\centering
\includegraphics[width=\hsize]{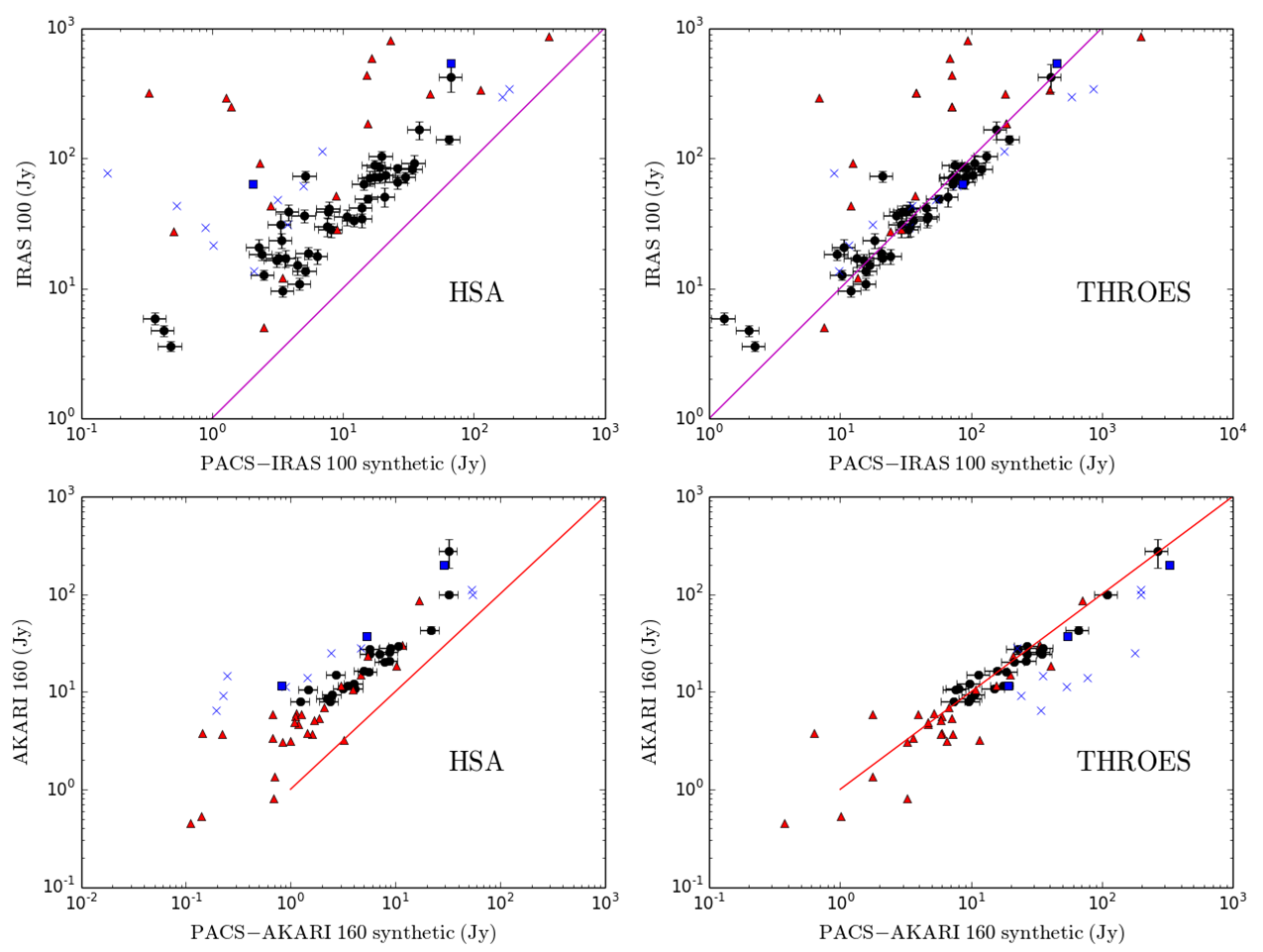}
\caption{Comparison of IRAS (100 $\mu$m) and AKARI (160 $\mu$m) photometry with synthetic PACS photometry at these 
wavelengths before (left) and after (right) the interactive reprocessing of the data carried out in this work and 
that is available through the THROES catalogue. The objects plotted here are those which present a complete coverage of 
the PACS spectral range. We have established a code to distinguish the extended objects 
(blue crosses), the mispointed ones (red squares), the objects with bad IRAS$_{100}$ or AKARI$_{160}$ data
(Quality Flag$\neq$3, red triangles), and objects with good IRAS$_{100}$ or AKARI$_{160}$ data (Quality Flag=3, black
circles with error bars). The solid line represents a perfect match (1:1 ratio) of the synthetic (PACS) and 
observational photometric (IRAS and AKARI) points.}
\label{PhotoSintetico}

\end{figure*}

To show the effect of the post-processing tasks applied in the THROES reduction process, the synthetic 
photometry has been estimated for 1D PACS spectra before and after applying the post-processing tasks. 
In Fig. \ref{PhotoSintetico} the comparison between IRAS$_{100}$ and AKARI$_{160}$ photometric data and
their synthetic counterparts using PACS data with and without post-processing is shown. We can see clearly that a 
disagreement between the photometric and the synthetic photometric data, before post-processing, is obtained.
This disagreement is significantly reduced after post-processing tasks are applied, so it confirms that the 
tasks introduced in the reduction process of THROES are needed for reliable absolute flux calibration. The few points
that fall out of a one-to-one observational-to-synthetic photometry relation are points that have bad 
IRAS and/or AKARI photometric data, (Quality Flag$\neq$3, red triangles) or are associated to extended objects 
(blue crosses).

\subsection{ISO-LWS spectroscopy} \label{ISOComp}

It is interesting to compare our PACS spectroscopic data to ISO-LWS spectra, since this
instrument covers a wavelength range from 43 to 197 $\mu$m that almost fully overlaps with PACS. However, it is important 
to keep in mind that ISO-LWS presents a worse spectral resolution, R=200 (medium resolution) and R=1000 (high resolution),
and a larger beam size (80''$\times$100'').

Whenever available, ISO spectra are shown in Figs \ref{StandardSources} to \ref{OnlyPSC} together with our 
1D PACS spectra and IRAS/AKARI photometry points. After a visual inspection of those objects which present both 
PACS and ISO data, objects can be grouped into five main families:

\textbf{1. Good agreement between PACS and ISO:} For most of the sources, a good agreement between PACS and ISO-LWS
spectroscopic data is found. For example, HD 161796, CIT 6, AFGL 618, or CPD -568032 (see Fig. \ref{StandardSources}).

\textbf{2. Mispointed PACS observations without 3x3 correction:} Due to its location close to the edge of the PACS 
5$\times$5 array, it was not possible to apply the semi-extended 3x3 correction to one source, AFGL 5379. This could 
explain why the continuum flux level of the ISO-LWS is higher than that of the PACS spectrum (see Fig. \ref{OnlyPSC}).

\textbf{3. Background contamination of the ISO spectra:} One of the most remarkable facts found with this comparison is 
that, for some sources, the ISO-LWS spectra show a continuum level higher than the one of PACS. This effect seems to 
be more evident at longer wavelengths (> 100 $\mu$m). Besides, for all these spectra we find an emission line 
at 158 $\mu$m associated to [C~{\sc ii]} due to interstellar emission, so the origin of this extra contribution could be 
the contamination of the interstellar medium. There are 11 sources that have been classified 
in this group: AFGL6815, IRAS 16342-3814, IRAS 16594-4656, IRAS 22036+5306, IRAS 21282+5050, MWC 922, NGC 6537, 
NML Cyg, OH 26.5+0.6, OH 32.8-0.3, and V Cyg. The reason why ISO-LWS spectra show this contamination is the larger 
FoV of ISO-LWS with respect to the PACS FoV (see Figs. \ref{StandardSources} to \ref{OnlyPSC}).

\textbf{4. Very extended sources:} Extended sources in THROES sample have been corrected using the extended 5x5 correction to recover the whole flux taken by PACS spectroscopy. However, there are some objects, such as NGC 6781, 
that are even larger than the PACS FoV and, therefore, PACS spectroscopic data do not measure all the flux 
emission from the source in contrast to ISO-LWS, which has a larger beam. For that reason, for these very extended 
sources, the continuum flux level of the ISO-LWS spectra is higher than that of PACS (see Fig. \ref{Extended5x5}).

\textbf{5. Bad ISO data:} Finally, for some sources like IRAS 07027-7934 or HD 56126, ISO-LWS data present evident
artefacts that make the ISO spectroscopic data unreliable (see Fig. \ref{StandardSources}).

\section{Summary} \label{Summary}

We made an inventory of all the observations of 
all evolved stars taken in standard mode by \textit{Herschel} with PACS spectroscopy. From all of them, we  selected pointed, Chop/Nod, and PacsRange observations of evolved low-to-intermediate mass
stars and we interactively processed the resulting 220 individuals \textit{Herschel}/PACS Obs IDs, corresponding 
to 114 different targets.

% Check numbers!! They are inconsistent 220 or 226?

Along the reduction process we introduced new tasks to improve the final PACS spectroscopy 
products. These tasks can be divided into two main groups:

\begin{itemize}
 \item 1) Tasks applied to the spectral cubes before the level 2 data are generated: flatfield correction
 and telescope background correction.
 \item 2) Tasks applied to the level 2 data products to extract the best 1D spectrum (post-processing). 
 These tasks  try to recover the whole flux of the sources within the PACS 5$\times$5 spaxels array, taking into account 
 the extension of the emission and the position of the object in the PACS FoV. They are: 
 point source flux loss correction (PSFLC), semi-extended 3x3 correction, and extended 5x5 correction.
\end{itemize}

After the interactive data reduction, we generated synthetic photometry to compare our final THROES 1D
spectra to photometric, IRAS$_{100}$ , and AKARI$_{160}$ photometric data. THROES 1D spectra were 
compared also with spectroscopic (ISO-LWS) data. From these comparisons we can conclude that our re-processing 
generates improved quality products compared with those routinely generated by the SPG in an automated mode, 
available in the HSA, and are in good agreement with IRAS and AKARI photometric data. Furthermore, the comparison 
of PACS with ISO-LWS spectra has highlighted the presence of interstellar contamination in some ISO-LWS data.

To ease the access to the final spectral cubes (level 2) and to the post-processed 1D spectra, we have
created a web-based interface using \textit{SVOCat}. Through  
{\tt https://throes.cab.inta-csic.es/}, all the products generated as a result of the THROES interactive data 
reduction process are available.
The THROES catalogue is expected to be updated in the near future by extending the analysis to new spectroscopic
data from other instruments such as SPIRE. Furthermore, there is the potential to extend the analysis to 
evolved massive stars.

%\end{document}

%-------------------------------------------------------------------
\begin{acknowledgements}
The \textit{Herschel} spacecraft was designed, built, tested, and
launched under a contract to ESA managed by the Herschel/Planck Project team
by an industrial consortium under the overall responsibility of the prime contractor
Thales Alenia Space (Cannes), and including Astrium (Friedrichshafen)
responsible for the payload module and for system testing at spacecraft level,
Thales Alenia Space (Turin) responsible for the service module, and Astrium
(Toulouse) responsible for the telescope, with in excess of a hundred subcontractors.
HCSS and HIPE are joint developments by the Herschel Science Ground
Segment Consortium, consisting of ESA, the NASA Herschel Science Center,
and the HIFI, PACS, and SPIRE consortia. We are grateful to the entire spectroscopy
group of PACS for their help and support, especially to E. Puga and K. Exter, as well as the anonymous referee
for their careful reading of the manuscript and very useful suggestions.

This research has been supported by the funding of the ESAC Faculty and the Herschel
Science Division. CSC is partially funded by the Spanish MINECO 
through grants AYA2012-32032 and AYA2016-750066-C2-1-P.

This research has made use of the Spanish Virtual Observatory
(svo.cab.inta-csic.es) supported from the
Spanish MINECO through grants AyA2014-55216.
\end{acknowledgements}

\bibliographystyle{aa}

\bibliography{References}

%\end{document}

%\section{Appendix figures}

%--- Standard sources --------------

\begin{figure*}
\begin{subfigure}{.55\textwidth}
\centering
  \includegraphics[width=0.9\linewidth,left]{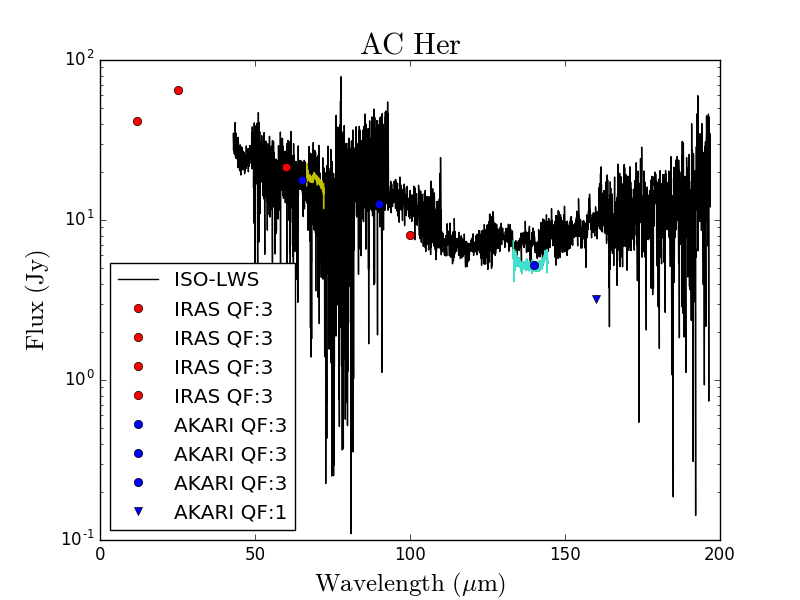}
  %\caption{1a}
  \label{fig:sfig1a}
\end{subfigure}
\begin{subfigure}{.55\textwidth}
\centering
  \includegraphics[width=0.9\linewidth,left]{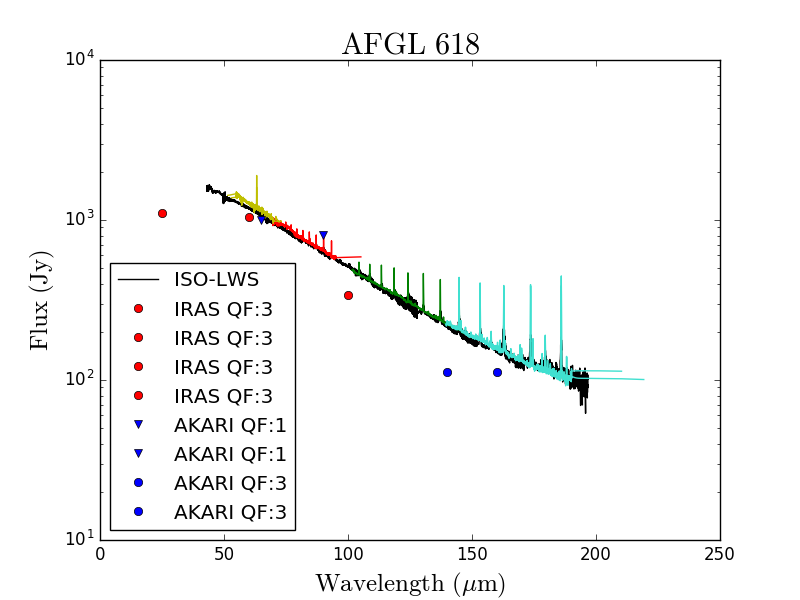}
  %\caption{1d}
  \label{fig:sfig1d}
\end{subfigure}
\begin{subfigure}{.55\textwidth}
\centering
  \includegraphics[width=0.9\linewidth,left]{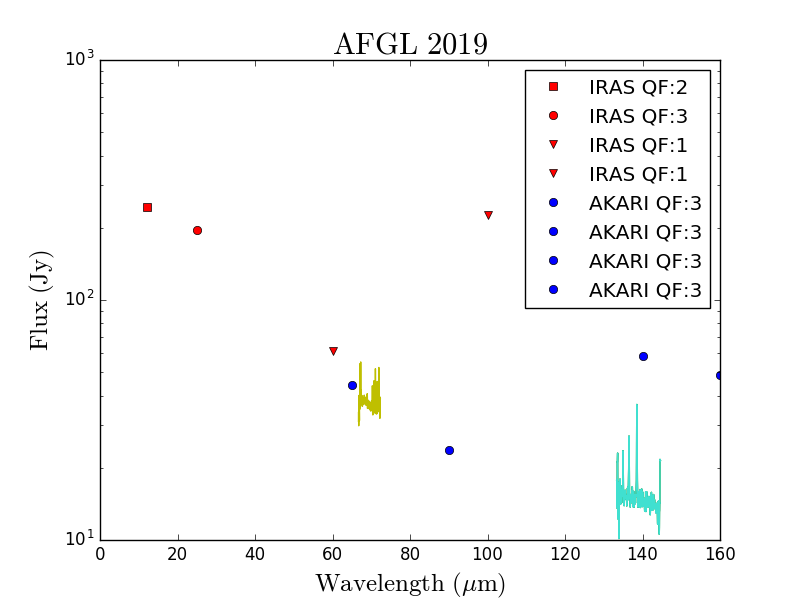}
  %\caption{1d}
  \label{fig:sfig1d}
\end{subfigure}
\begin{subfigure}{.55\textwidth}
\centering
  \includegraphics[width=0.9\linewidth,left]{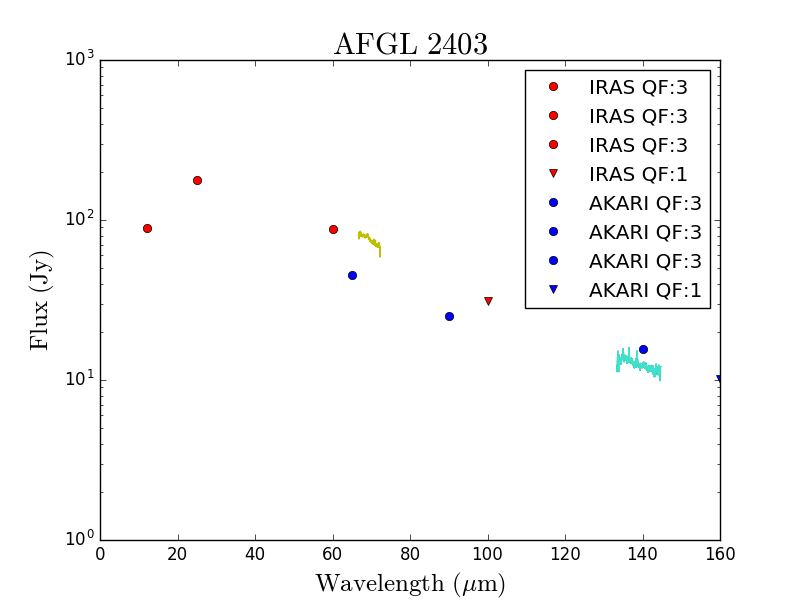}
  %\caption{1d}
  \label{fig:sfig1d}
\end{subfigure}
\begin{subfigure}{.55\textwidth}
\centering
  \includegraphics[width=0.9\linewidth,left]{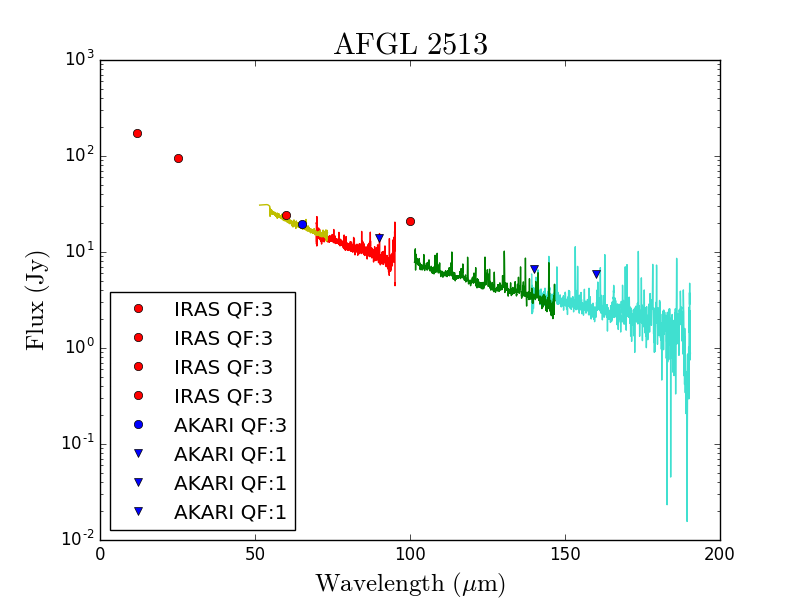}
  %\caption{1d}
  \label{fig:sfig1d}
\end{subfigure}
\begin{subfigure}{.55\textwidth}
\centering
  \includegraphics[width=0.9\linewidth,left]{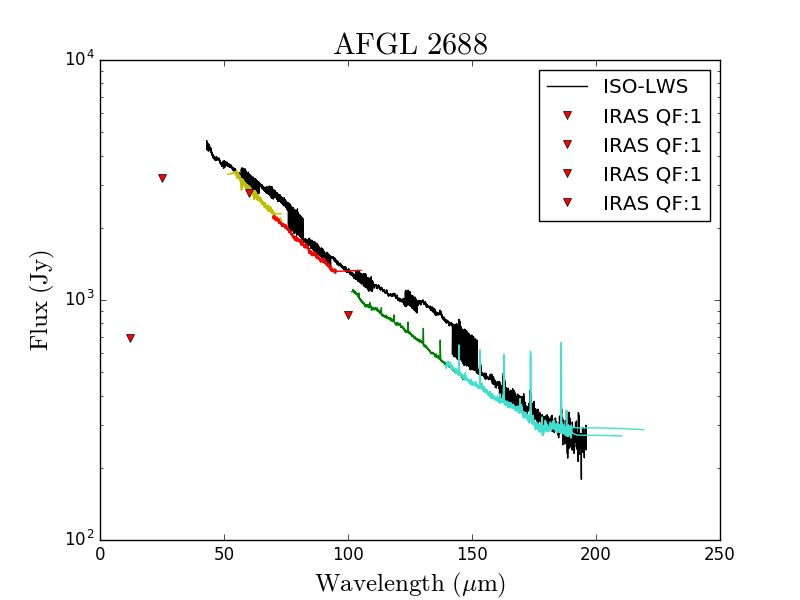}
  %\caption{1d}
  \label{fig:sfig1d}
\end{subfigure}
\caption{PACS spectroscopy SED of well-pointed, non-extended sources, after applying PSFLC and semi-extended 3x3 
correction (PSFLC-3x3). PACS data is colour-coded according to the spectral region covered by each
subrange, as follows: $\sim$50-70 $\mu$m (yellow), $\sim$70-100 $\mu$m (red), $\sim$100-145 $\mu$m (green), and 
$\sim$145-200 $\mu$m (turquoise). IRAS (red points) and AKARI (blue points) photometric 
data and ISO (black) spectroscopic data are also displayed when available. We do not show those observations
pointed to a specific region in the case of very extended sources such as rims or knots.}
\label{StandardSources}
\end{figure*}

\clearpage

\addtocounter{figure}{-1}
\begin{figure*}
\begin{subfigure}{.55\textwidth}
\centering
  \includegraphics[width=0.9\linewidth,left]{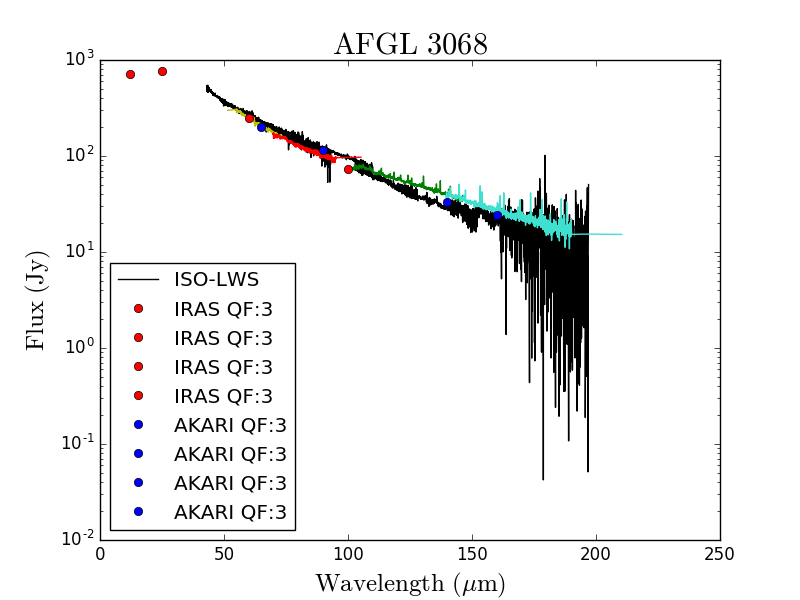}
  %\caption{1b}
  \label{fig:sfig1b}
\end{subfigure}
\begin{subfigure}{.55\textwidth}
\centering
  \includegraphics[width=0.9\linewidth,left]{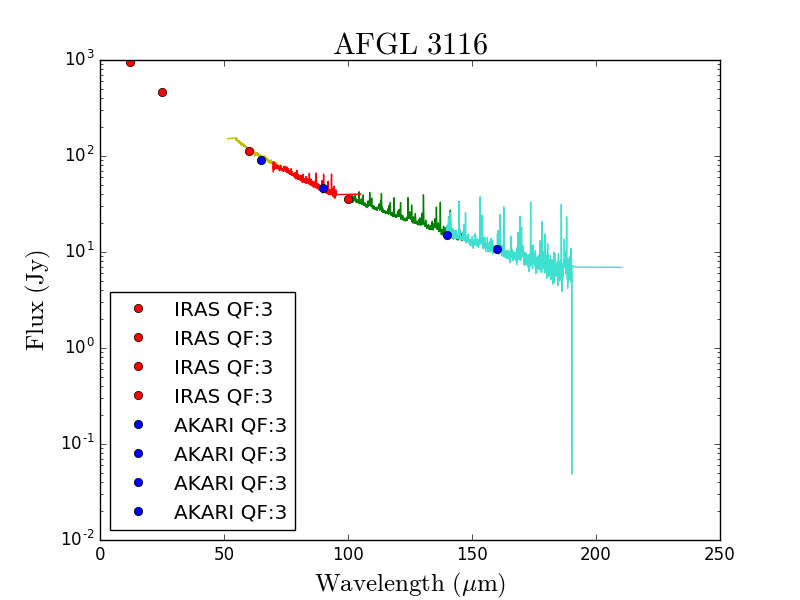}
  %\caption{1d}
  \label{fig:sfig1d}
\end{subfigure}
\begin{subfigure}{.55\textwidth}
\centering
  \includegraphics[width=0.9\linewidth,left]{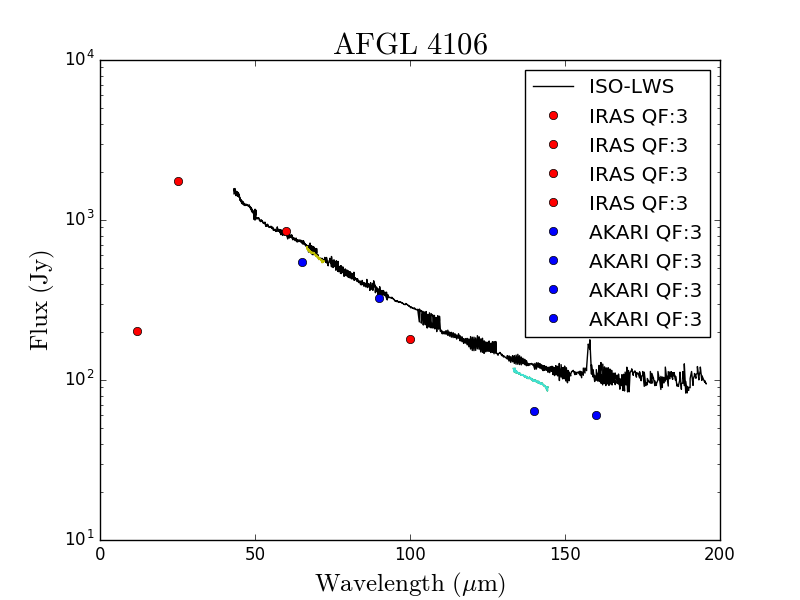}
  %\caption{1a}
  \label{fig:sfig1a}
\end{subfigure}
\begin{subfigure}{.55\textwidth}
\centering
  \includegraphics[width=0.9\linewidth,left]{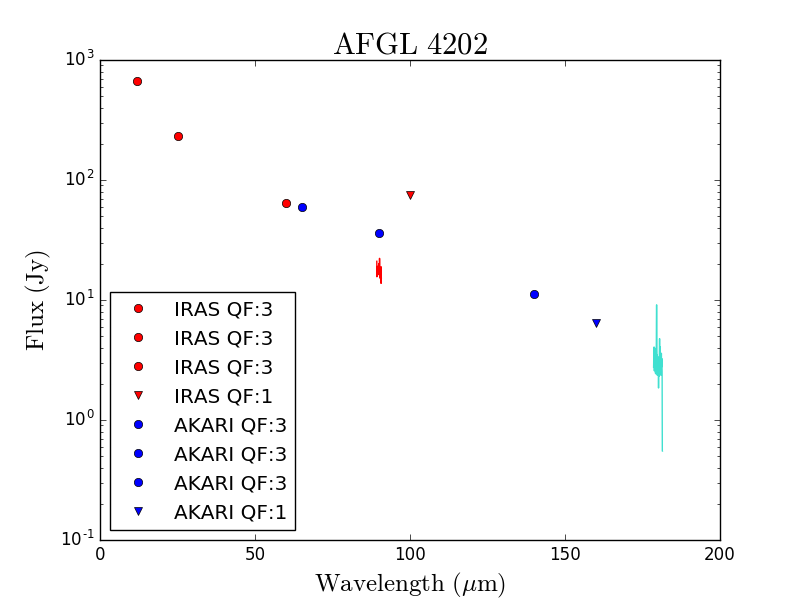}
  %\caption{1c}
  \label{fig:sfig1c}
\end{subfigure}
\begin{subfigure}{.55\textwidth}
\centering
  \includegraphics[width=0.9\linewidth,left]{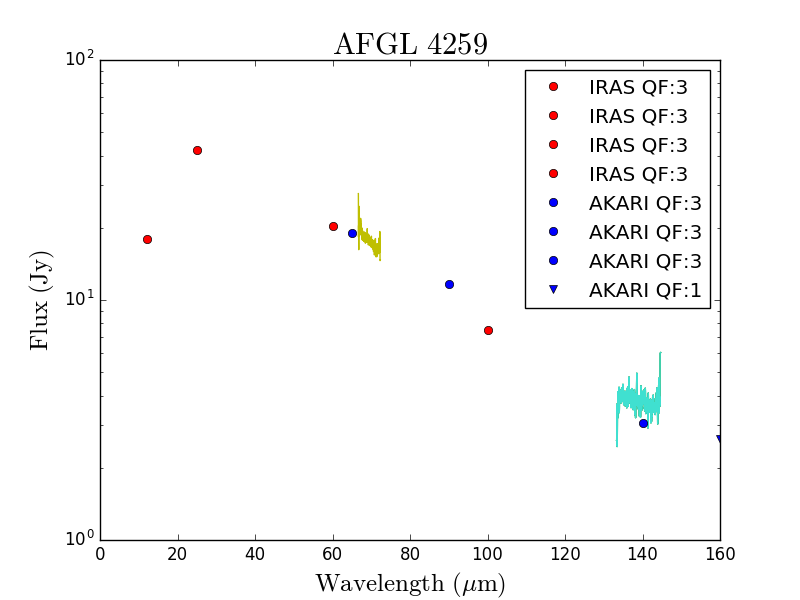}
  %\caption{1b}
  \label{fig:sfig1b}
\end{subfigure}
\begin{subfigure}{.55\textwidth}
\centering
  \includegraphics[width=0.9\linewidth,left]{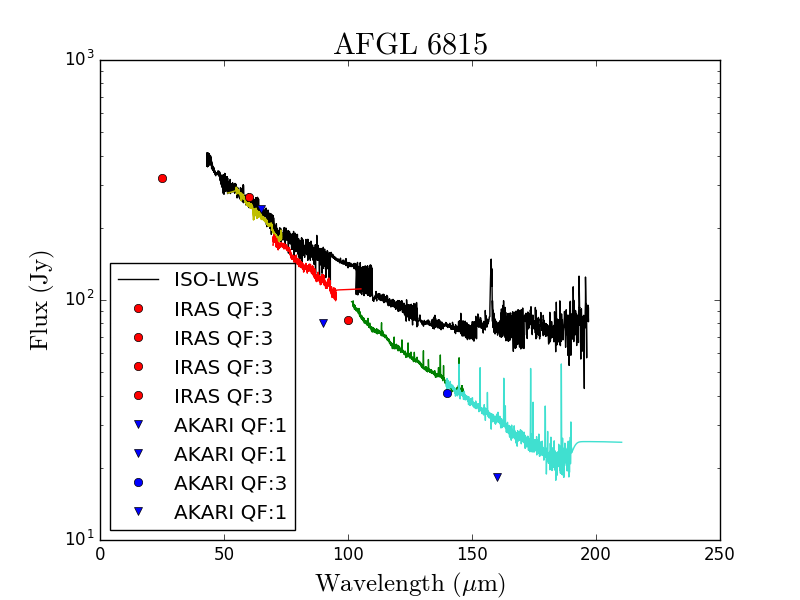}
  %\caption{1d}
  \label{fig:sfig1d}
\end{subfigure}
\caption{Continued.}
\end{figure*}

\clearpage

\addtocounter{figure}{-1}
\begin{figure*}
\begin{subfigure}{.55\textwidth}
\centering
  \includegraphics[width=0.9\linewidth,left]{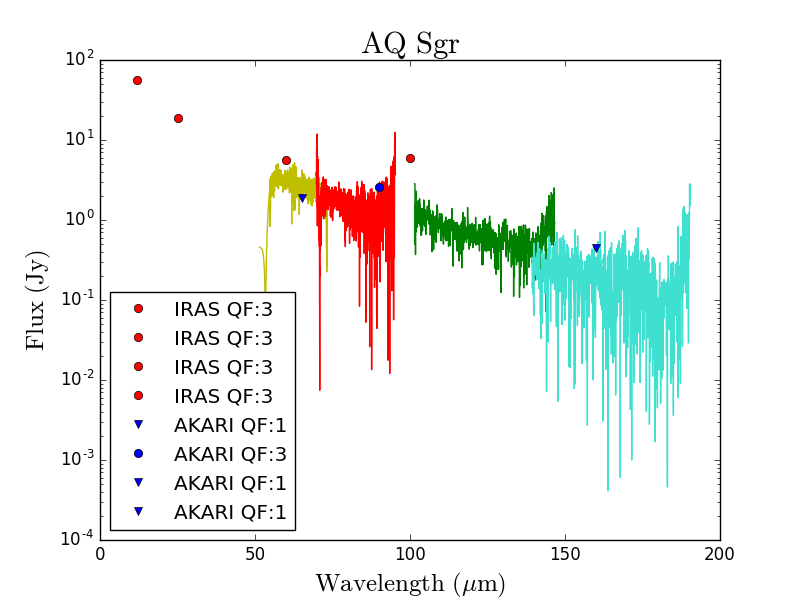}
  %\caption{1d}
  \label{fig:sfig1d}
\end{subfigure}
\begin{subfigure}{.55\textwidth}
\centering
  \includegraphics[width=0.9\linewidth,left]{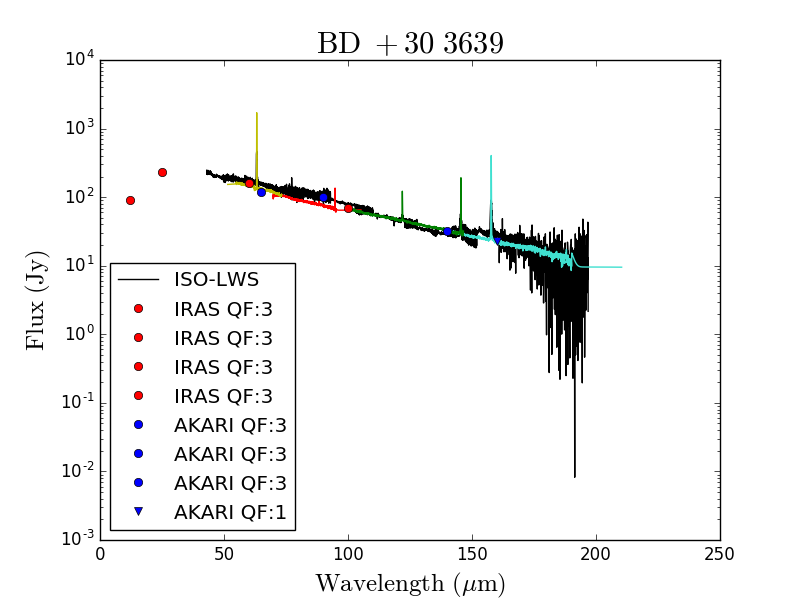}
  %\caption{1d}
  \label{fig:sfig1d}
\end{subfigure}
\begin{subfigure}{.55\textwidth}
\centering
  \includegraphics[width=0.9\linewidth,left]{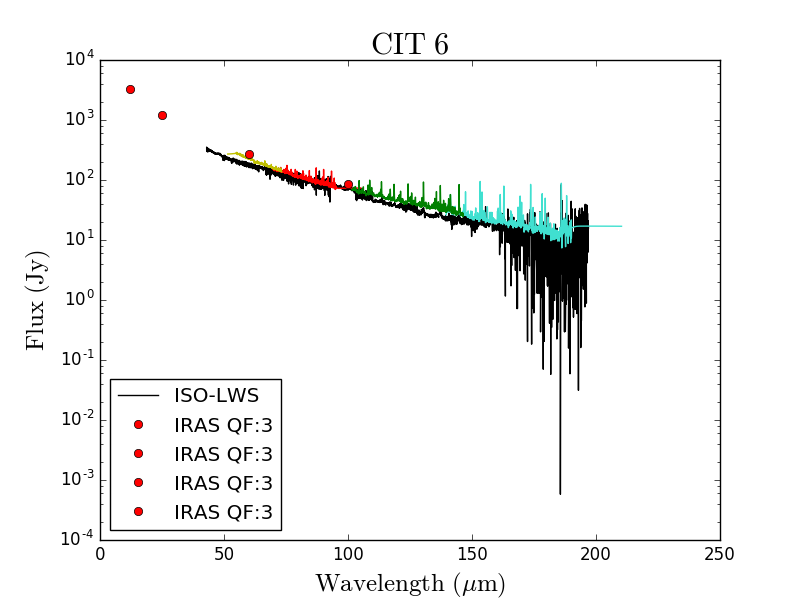}
  %\caption{1d}
  \label{fig:sfig1d}
\end{subfigure}
\begin{subfigure}{.55\textwidth}
\centering
  \includegraphics[width=0.9\linewidth,left]{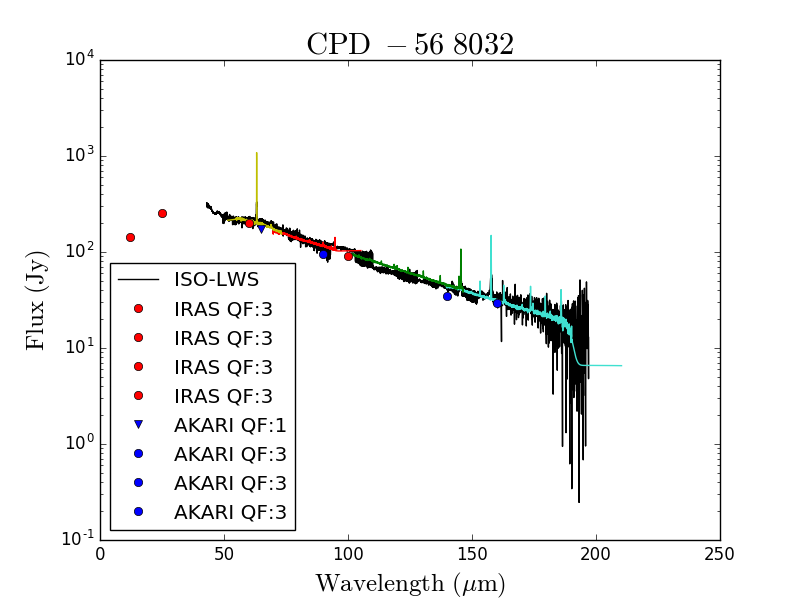}
  %\caption{1a}
  \label{fig:sfig1a}
\end{subfigure}
\begin{subfigure}{.55\textwidth}
\centering
  \includegraphics[width=0.9\linewidth,left]{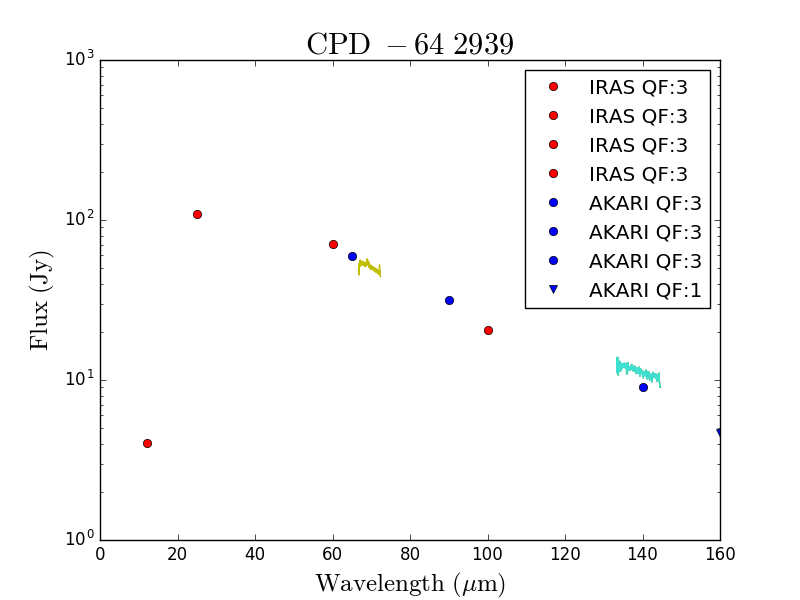}
  %\caption{1b}
  \label{fig:sfig1b}
\end{subfigure}
\begin{subfigure}{.55\textwidth}
\centering
  \includegraphics[width=0.9\linewidth,left]{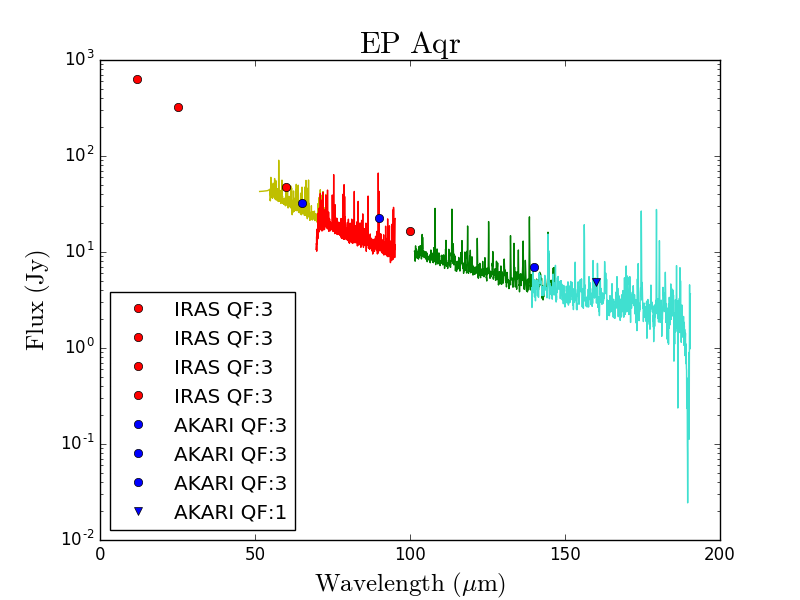}
  %\caption{1c}
  \label{fig:sfig1c}
\end{subfigure}
\caption{Continued.}
\end{figure*}

\clearpage
\addtocounter{figure}{-1}
\begin{figure*}
%\label{StandardSources}
\begin{subfigure}{.55\textwidth}
\centering
  \includegraphics[width=0.9\linewidth,left]{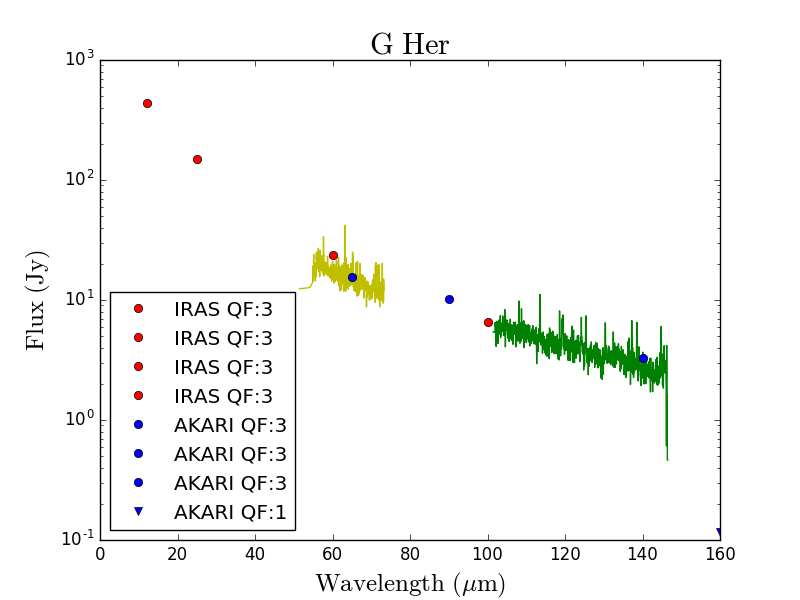}
  %\caption{1d}
  \label{fig:sfig1d}
\end{subfigure}
\begin{subfigure}{.55\textwidth}
\centering
  \includegraphics[width=0.9\linewidth,left]{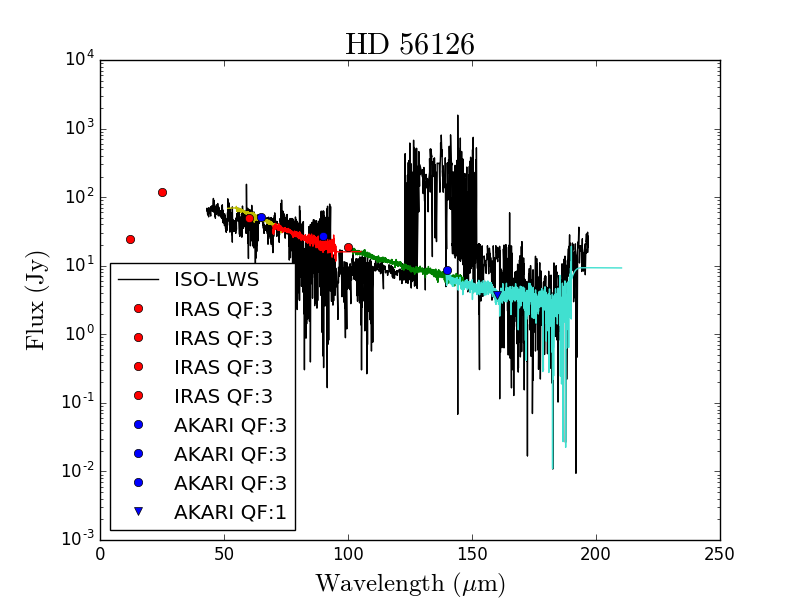}
  %\caption{1a}
  \label{fig:sfig1a}
\end{subfigure}
\begin{subfigure}{.55\textwidth}
\centering
  \includegraphics[width=0.9\linewidth,left]{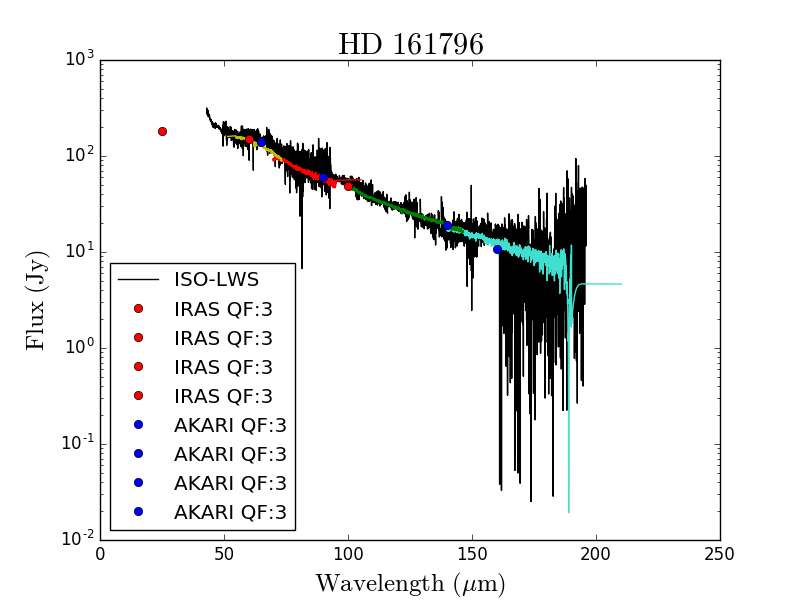}
  %\caption{1d}
  \label{fig:sfig1d}
\end{subfigure}
\begin{subfigure}{.55\textwidth}
\centering
  \includegraphics[width=0.9\linewidth,left]{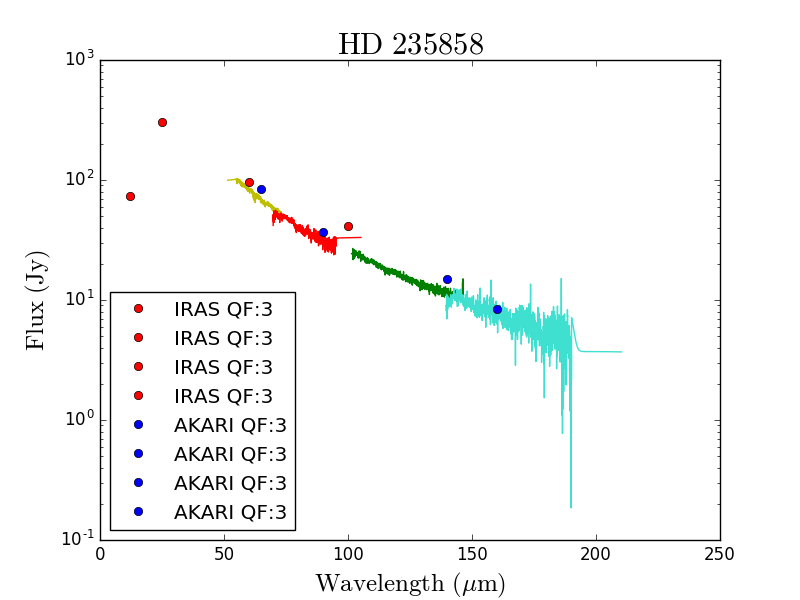}
  %\caption{1d}
  \label{fig:sfig1d}
\end{subfigure}
\begin{subfigure}{.55\textwidth}
\centering
  \includegraphics[width=0.9\linewidth,left]{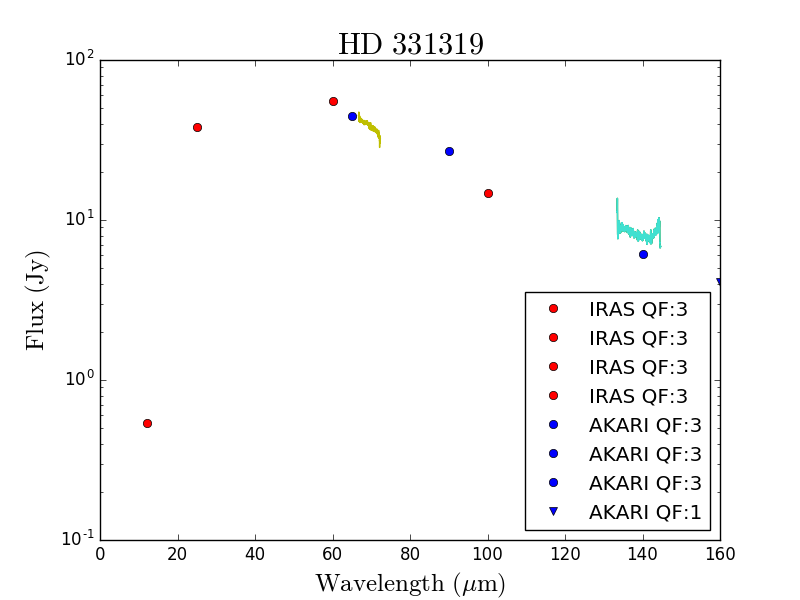}
  %\caption{1d}
  \label{fig:sfig1d}
\end{subfigure}
\begin{subfigure}{.55\textwidth}
\centering
  \includegraphics[width=0.9\linewidth,left]{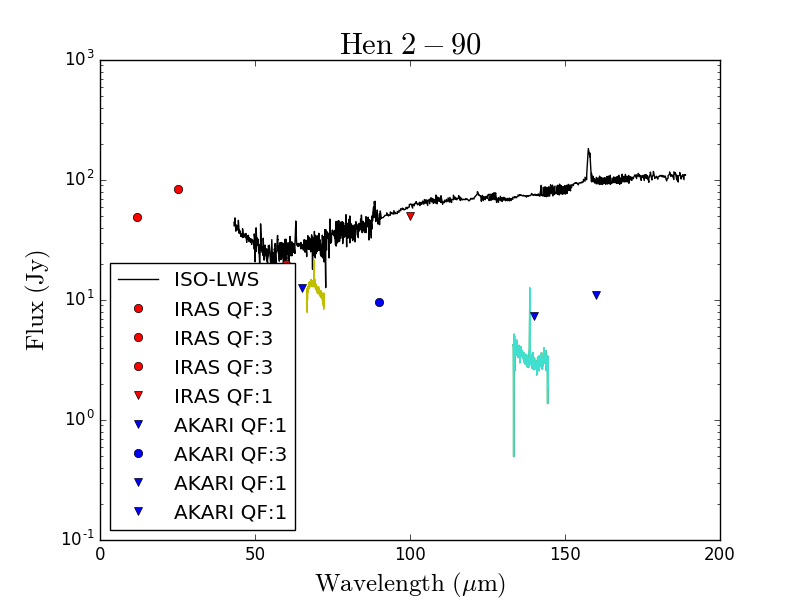}
  %\caption{1c}
  \label{fig:sfig1c}
\end{subfigure}
\caption{Continued.}
\end{figure*}

\clearpage
\addtocounter{figure}{-1}
\begin{figure*}
\begin{subfigure}{.55\textwidth}
\centering
  \includegraphics[width=0.9\linewidth,left]{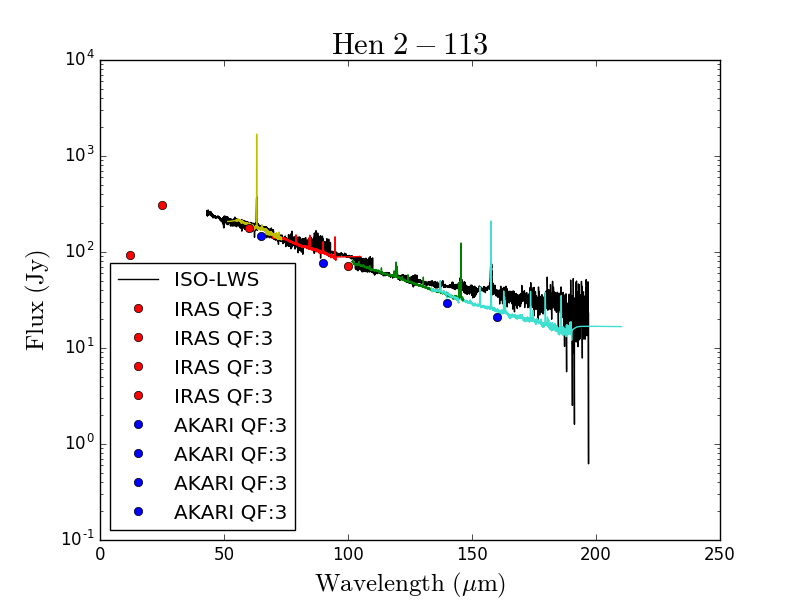}
  %\caption{1d}
  \label{fig:sfig1d}
\end{subfigure}
\begin{subfigure}{.55\textwidth}
\centering
  \includegraphics[width=0.9\linewidth,left]{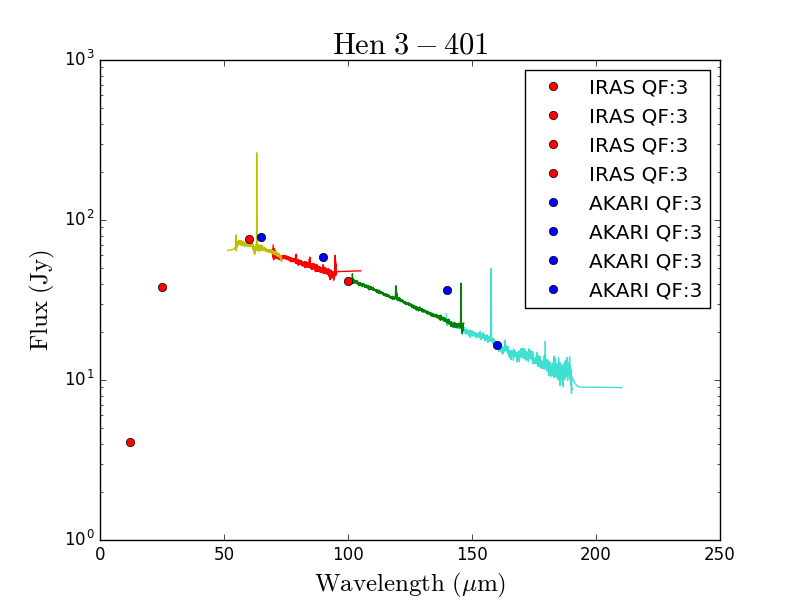}
  %\caption{1d}
  \label{fig:sfig1d}
\end{subfigure}
\begin{subfigure}{.55\textwidth}
\centering
  \includegraphics[width=0.9\linewidth,left]{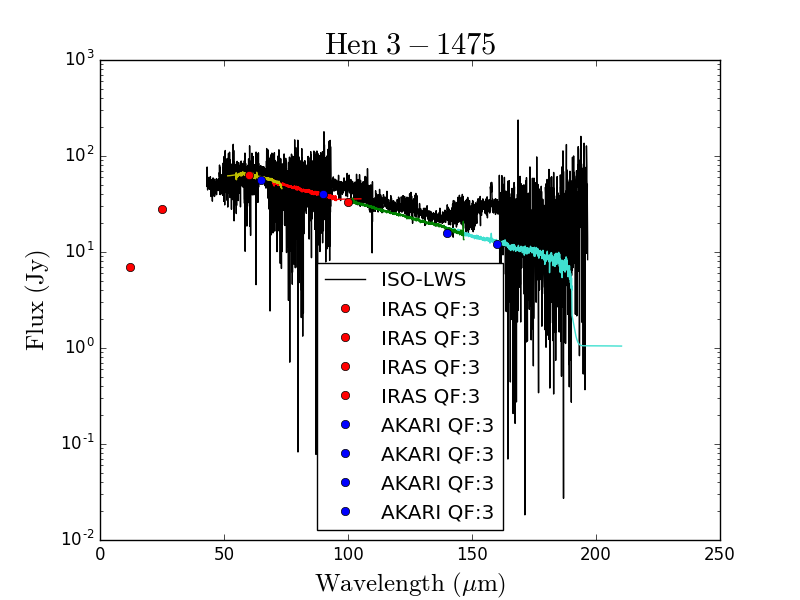}
  %\caption{1d}
  \label{fig:sfig1d}
\end{subfigure}
\begin{subfigure}{.55\textwidth}
\centering
  \includegraphics[width=0.9\linewidth,left]{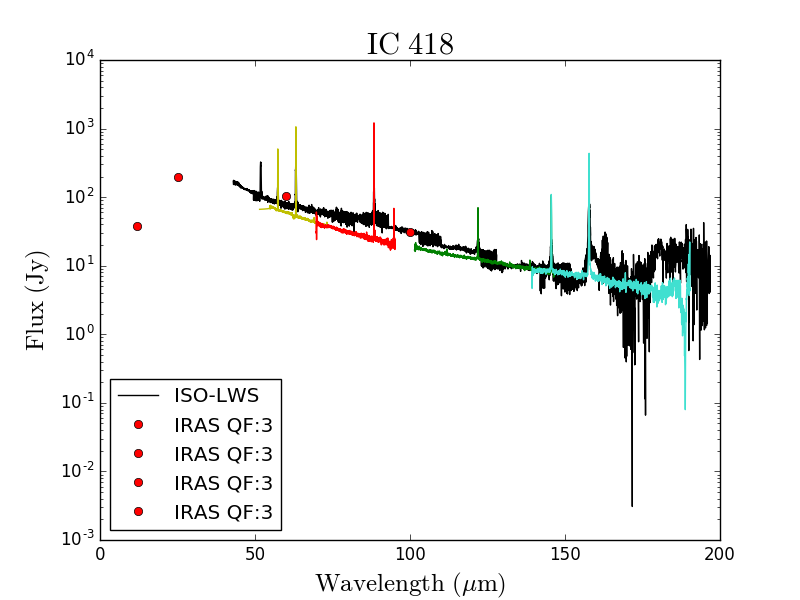}
  %\caption{1d}
  \label{fig:sfig1d}
\end{subfigure}
\begin{subfigure}{.55\textwidth}
\centering
  \includegraphics[width=0.9\linewidth,left]{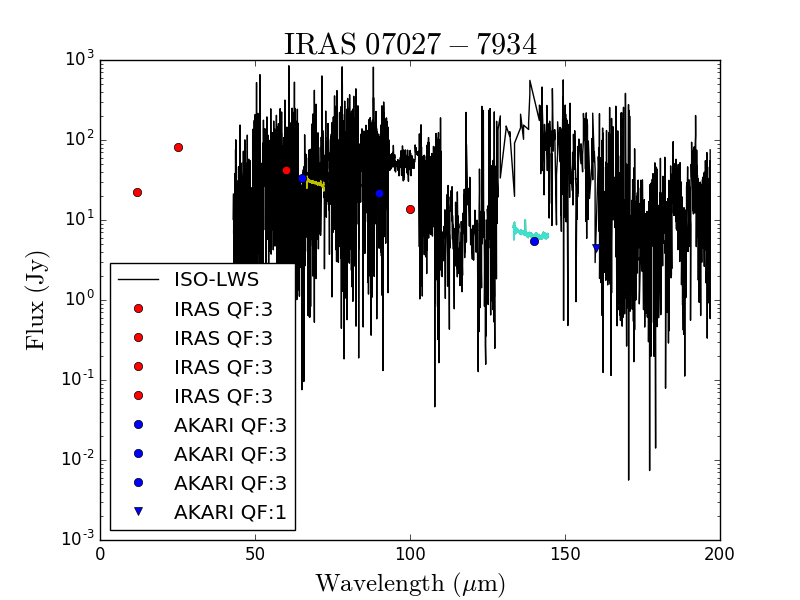}
  %\caption{1b}
  \label{fig:sfig1b}
\end{subfigure}
\begin{subfigure}{.55\textwidth}
\centering
  \includegraphics[width=0.9\linewidth,left]{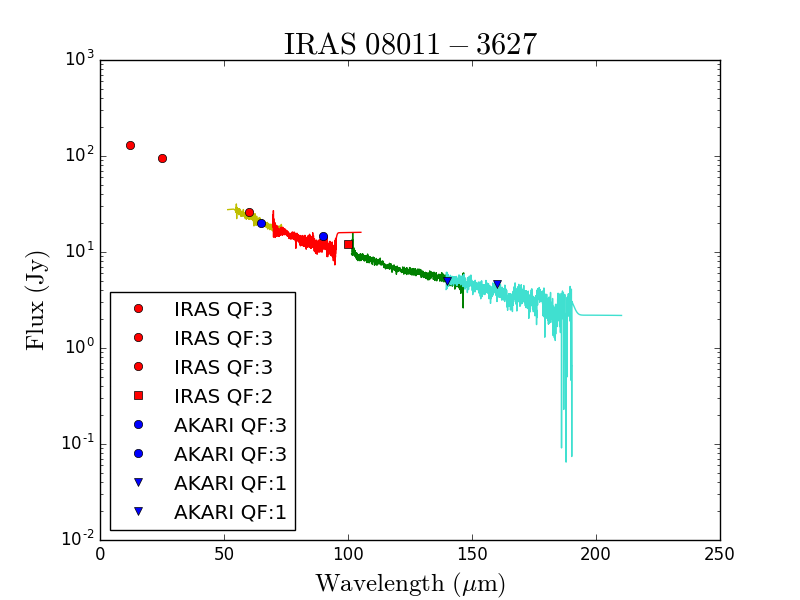}
  %\caption{1c}
  \label{fig:sfig1c}
\end{subfigure}
\caption{Continued.}
\end{figure*}

\clearpage
\addtocounter{figure}{-1}
\begin{figure*}
\begin{subfigure}{.55\textwidth}
\centering
  \includegraphics[width=0.9\linewidth,left]{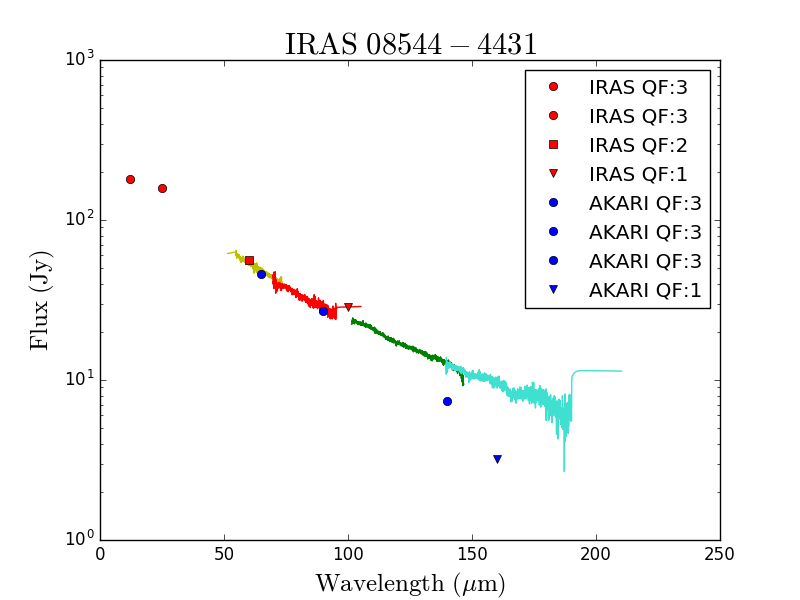}
  %\caption{1d}
  \label{fig:sfig1d}
\end{subfigure}
\begin{subfigure}{.55\textwidth}
\centering
  \includegraphics[width=0.9\linewidth,left]{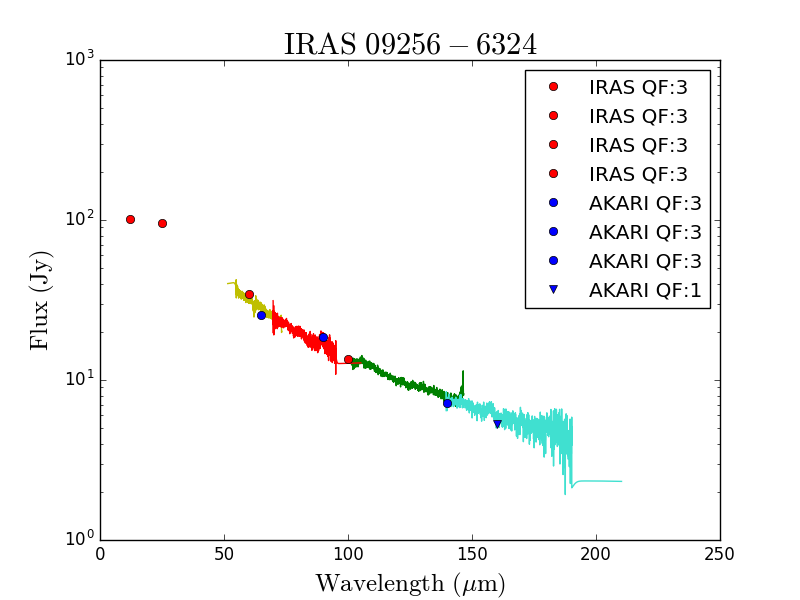}
  %\caption{1d}
  \label{fig:sfig1d}
\end{subfigure}
\begin{subfigure}{.55\textwidth}
\centering
  \includegraphics[width=0.9\linewidth,left]{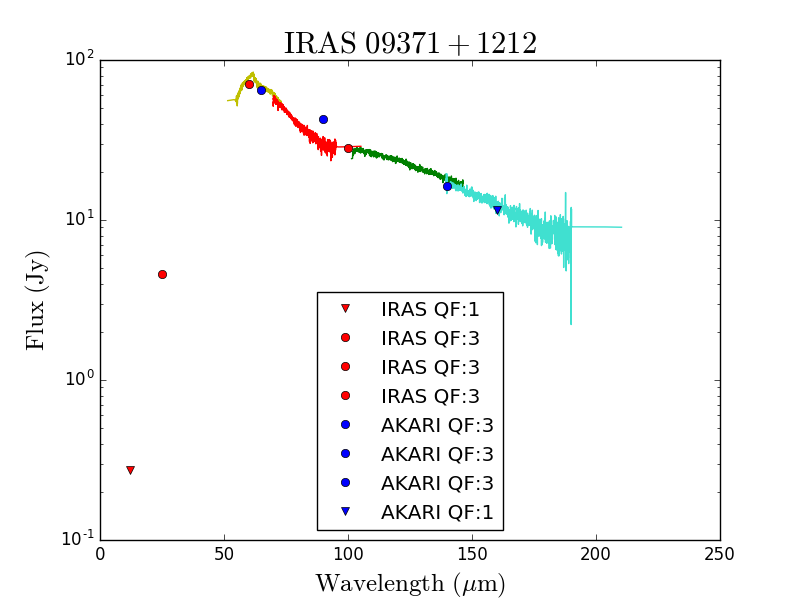}
  %\caption{1d}
  \label{fig:sfig1d}
\end{subfigure}
\begin{subfigure}{.55\textwidth}
\centering
  \includegraphics[width=0.9\linewidth,left]{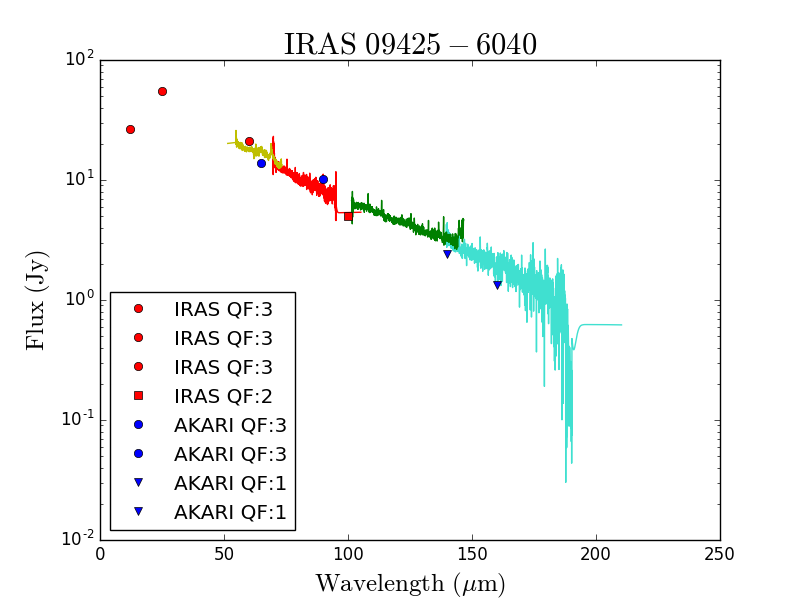}
  %\caption{1d}
  \label{fig:sfig1d}
\end{subfigure}
\begin{subfigure}{.55\textwidth}
\centering
  \includegraphics[width=0.9\linewidth,left]{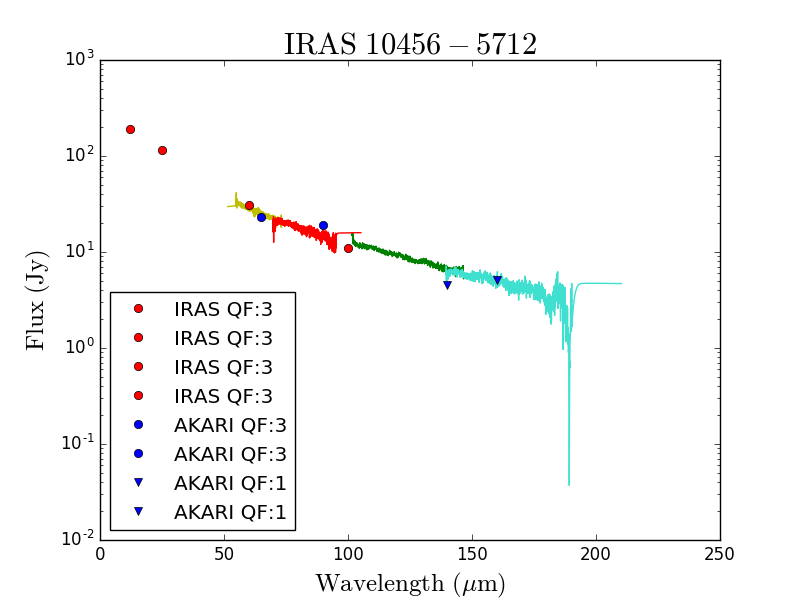}
  %\caption{1d}
  \label{fig:sfig1d}
\end{subfigure}
\begin{subfigure}{.55\textwidth}
\centering
  \includegraphics[width=0.9\linewidth,left]{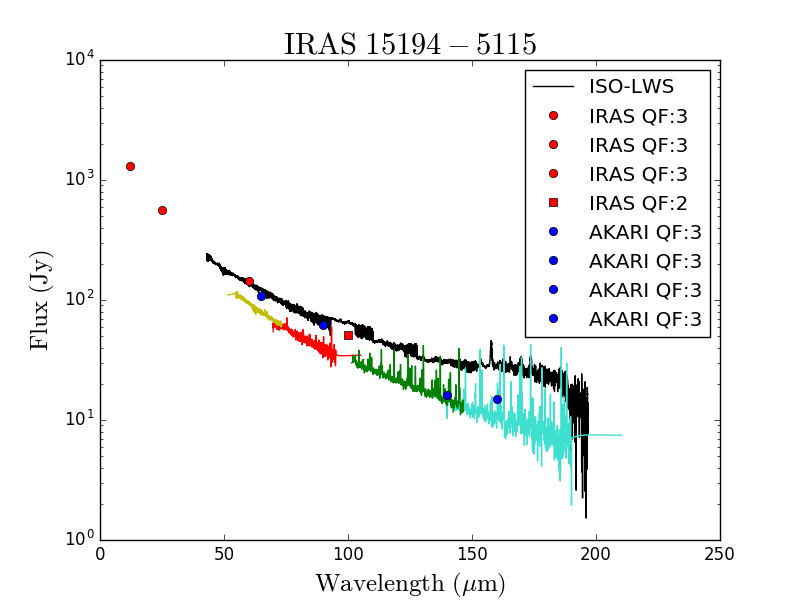}
  %\caption{1b}
  \label{fig:sfig1b}
\end{subfigure}
\caption{Continued.}
\end{figure*}

\clearpage
\addtocounter{figure}{-1}
\begin{figure*}
\begin{subfigure}{.55\textwidth}
\centering
  \includegraphics[width=0.9\linewidth,left]{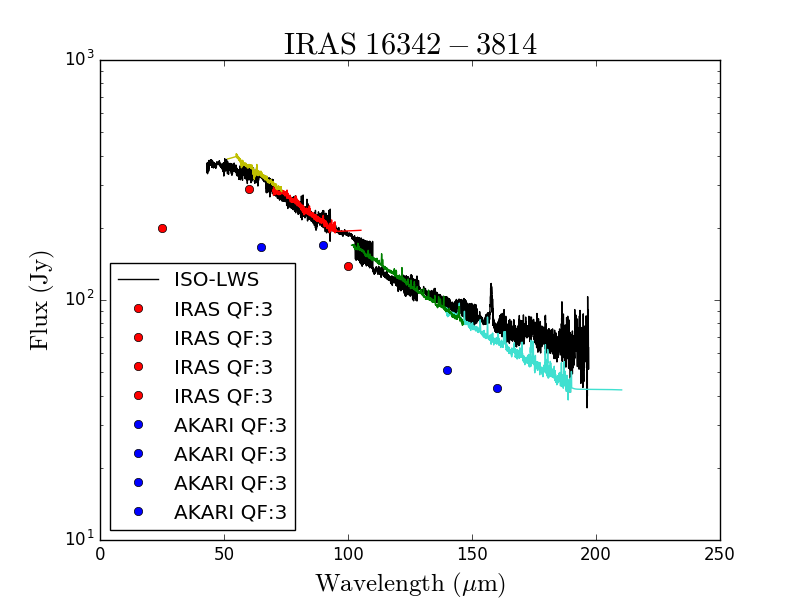}
  %\caption{1d}
  \label{fig:sfig1d}
\end{subfigure}
\begin{subfigure}{.55\textwidth}
\centering
  \includegraphics[width=0.9\linewidth,left]{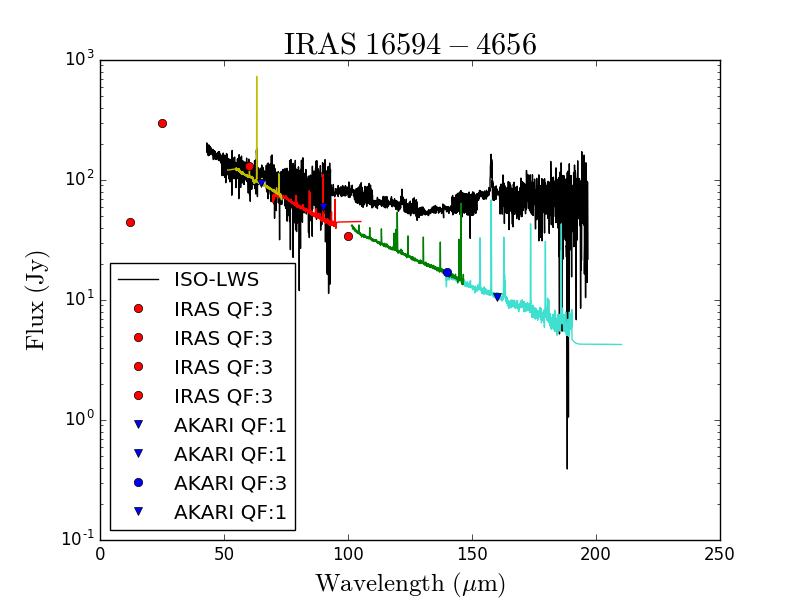}
  %\caption{1d}
  \label{fig:sfig1d}
\end{subfigure}
\begin{subfigure}{.55\textwidth}
\centering
  \includegraphics[width=0.9\linewidth,left]{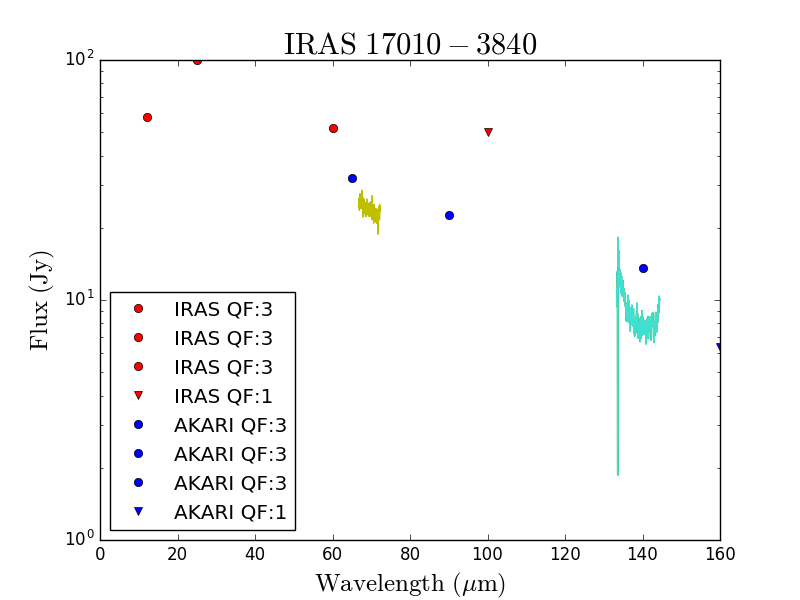}
  %\caption{1d}
  \label{fig:sfig1d}
\end{subfigure}
\begin{subfigure}{.55\textwidth}
\centering
  \includegraphics[width=0.9\linewidth,left]{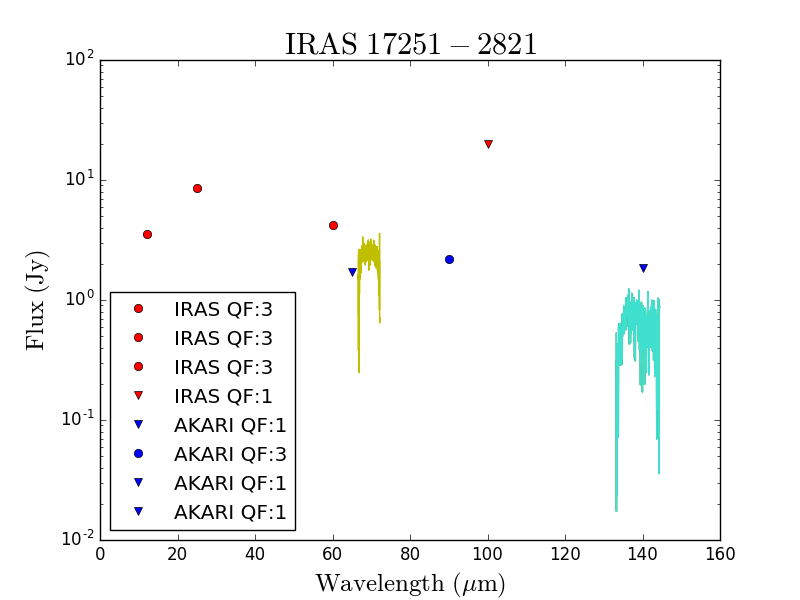}
  %\caption{1d}
  \label{fig:sfig1d}
\end{subfigure}
\begin{subfigure}{.55\textwidth}
\centering
  \includegraphics[width=0.9\linewidth,left]{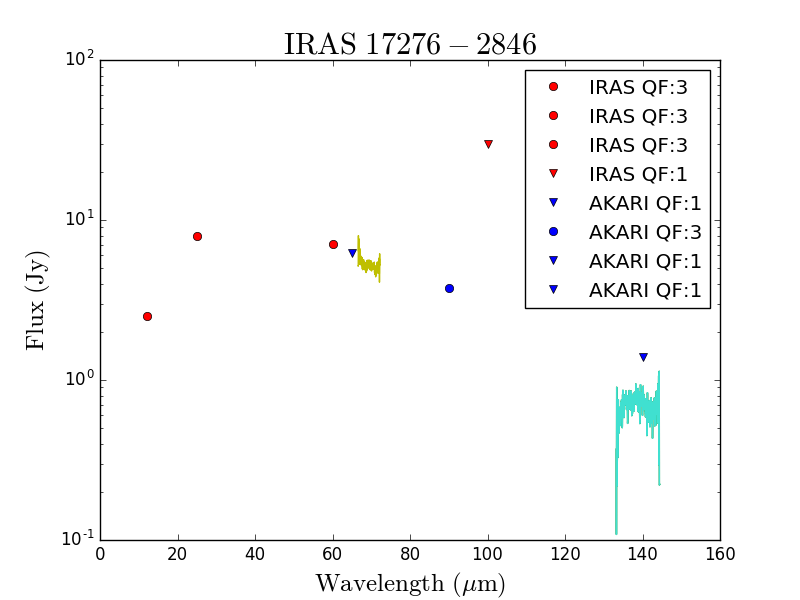}
  %\caption{1d}
  \label{fig:sfig1d}
\end{subfigure}
\begin{subfigure}{.55\textwidth}
\centering
  \includegraphics[width=0.9\linewidth,left]{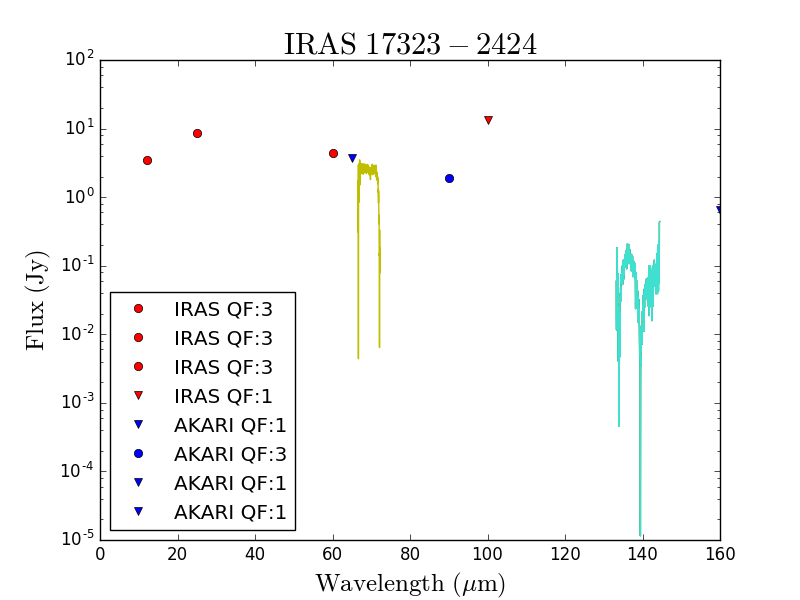}
  %\caption{1a}
  \label{fig:sfig1a}
\end{subfigure}
\caption{Continued.}
\end{figure*}

\clearpage
\addtocounter{figure}{-1}
\begin{figure*}
\begin{subfigure}{.55\textwidth}
\centering
  \includegraphics[width=0.9\linewidth,left]{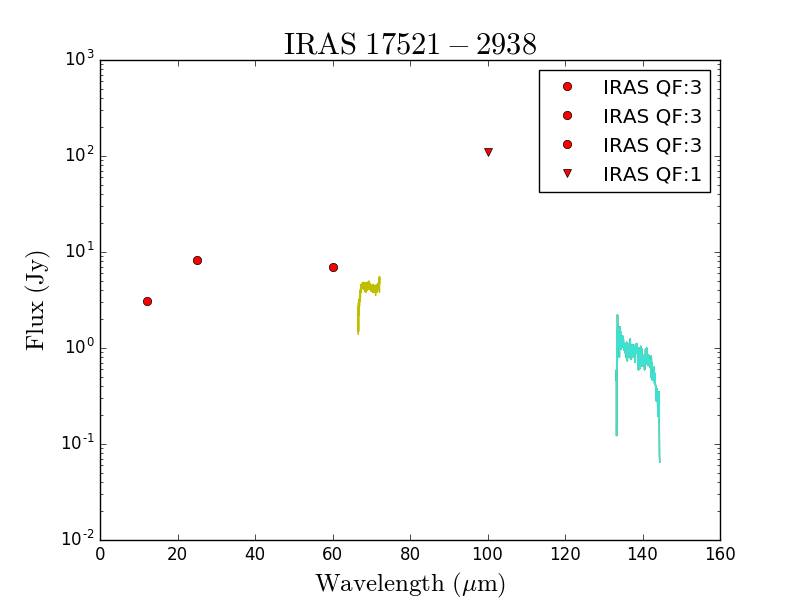}
  %\caption{1d}
  \label{fig:sfig1d}
\end{subfigure}
\begin{subfigure}{.55\textwidth}
\centering
  \includegraphics[width=0.9\linewidth,left]{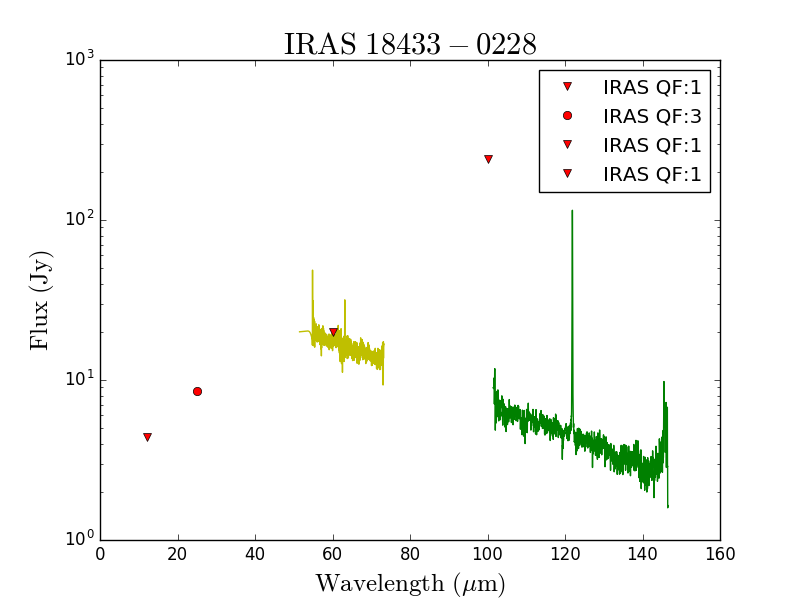}
  %\caption{1d}
  \label{fig:sfig1d}
\end{subfigure}
\begin{subfigure}{.55\textwidth}
\centering
  \includegraphics[width=0.9\linewidth,left]{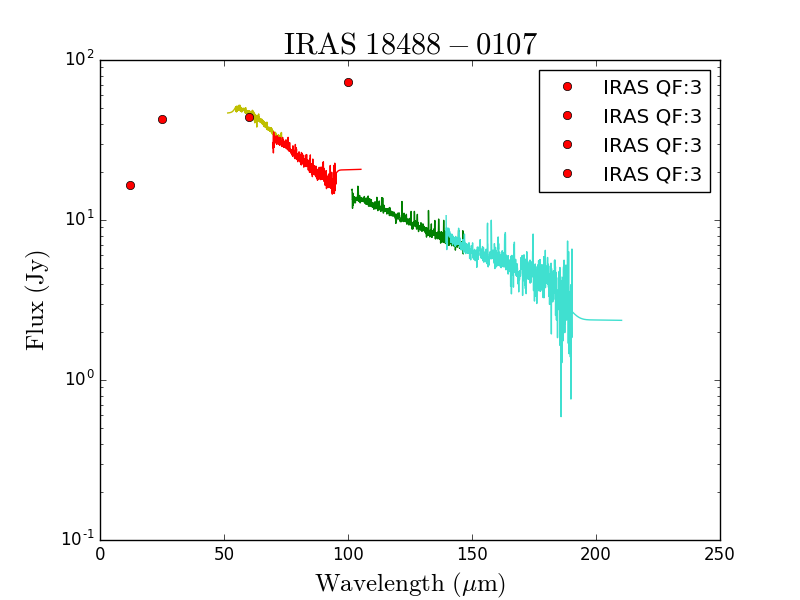}
  %\caption{1c}
  \label{fig:sfig1c}
\end{subfigure}
\begin{subfigure}{.55\textwidth}
\centering
  \includegraphics[width=0.9\linewidth,left]{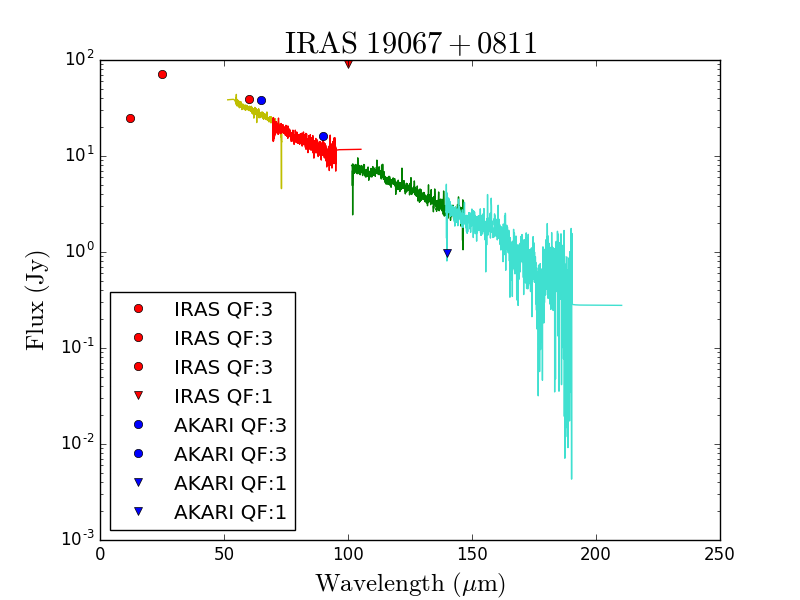}
  %\caption{1d}
  \label{fig:sfig1d}
\end{subfigure}
\begin{subfigure}{.55\textwidth}
\centering
  \includegraphics[width=0.9\linewidth,left]{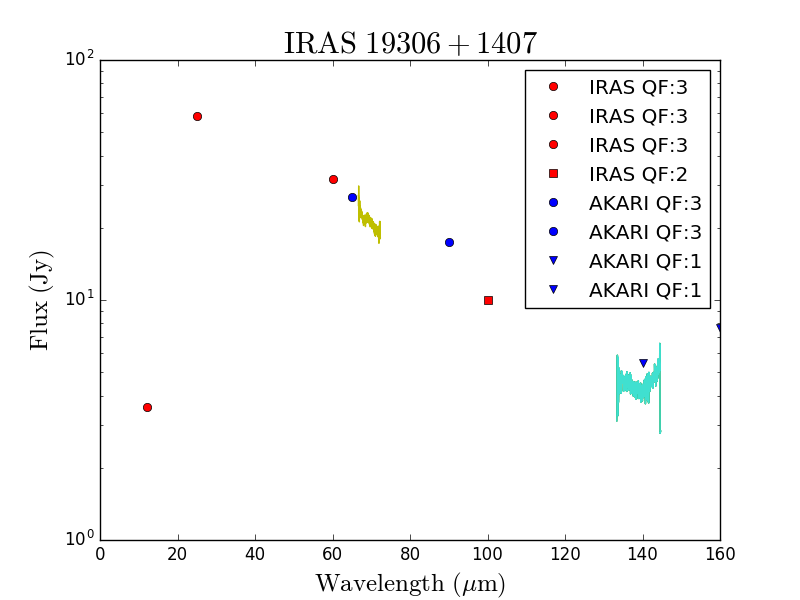}
  %\caption{1d}
  \label{fig:sfig1d}
\end{subfigure}
\begin{subfigure}{.55\textwidth}
\centering
  \includegraphics[width=0.9\linewidth,left]{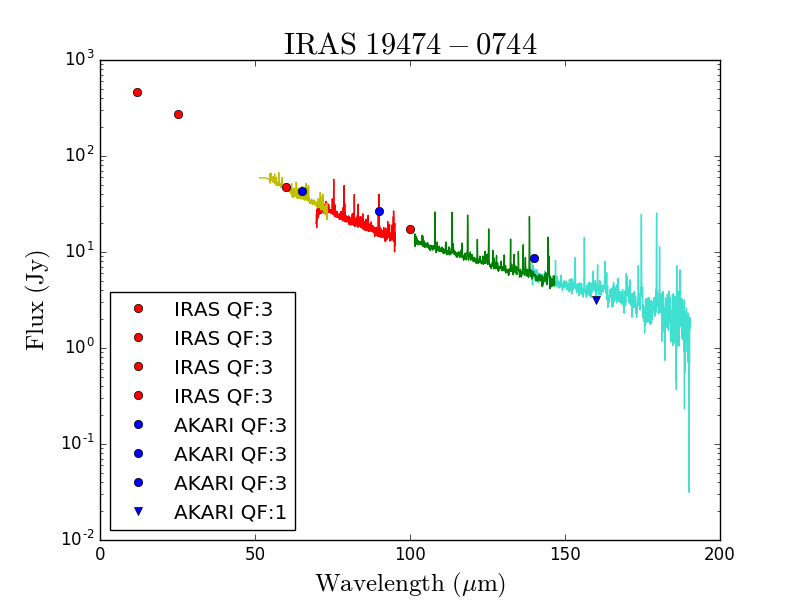}
  %\caption{1d}
  \label{fig:sfig1d}
\end{subfigure}
\caption{Continued.}
\end{figure*}

\clearpage
\addtocounter{figure}{-1}
\begin{figure*}
\begin{subfigure}{.55\textwidth}
\centering
  \includegraphics[width=0.9\linewidth,left]{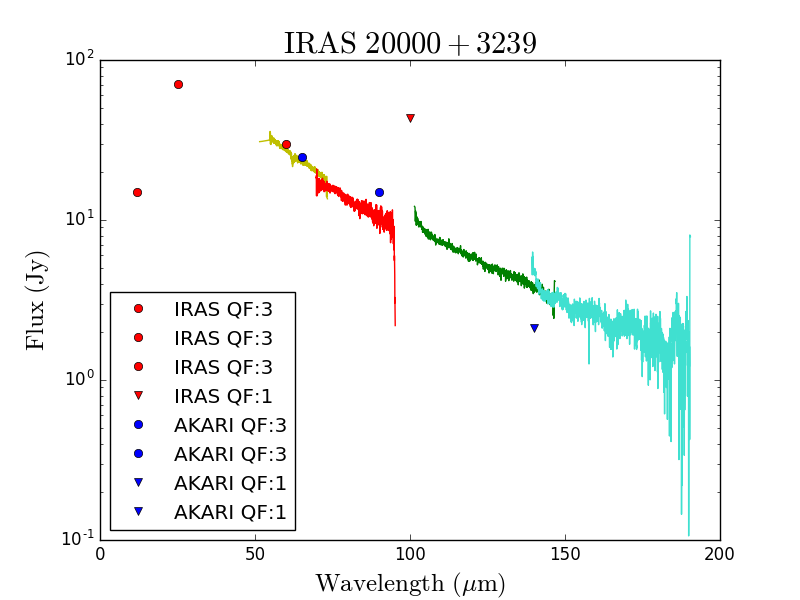}
  %\caption{1d}
  \label{fig:sfig1d}
\end{subfigure}
\begin{subfigure}{.55\textwidth}
\centering
  \includegraphics[width=0.9\linewidth,left]{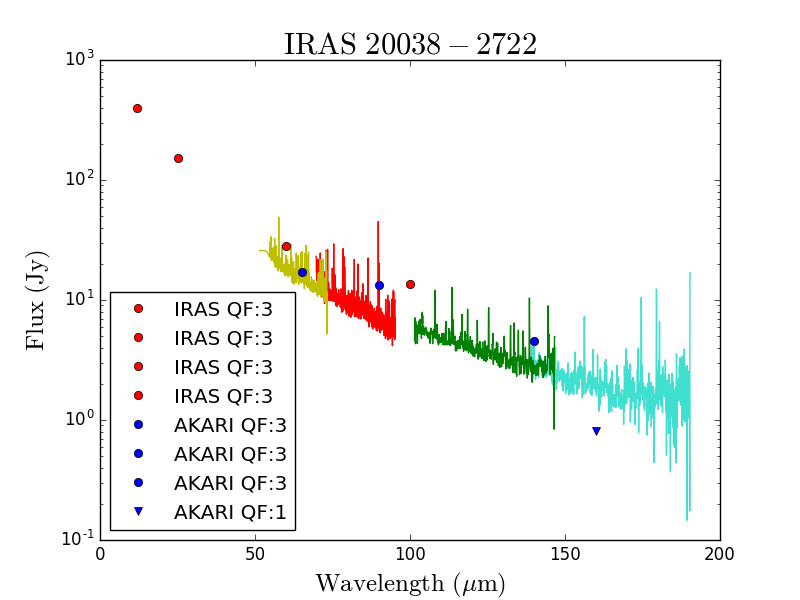}
  %\caption{1a}
  \label{fig:sfig1a}
\end{subfigure}
\begin{subfigure}{.55\textwidth}
\centering
  \includegraphics[width=0.9\linewidth,left]{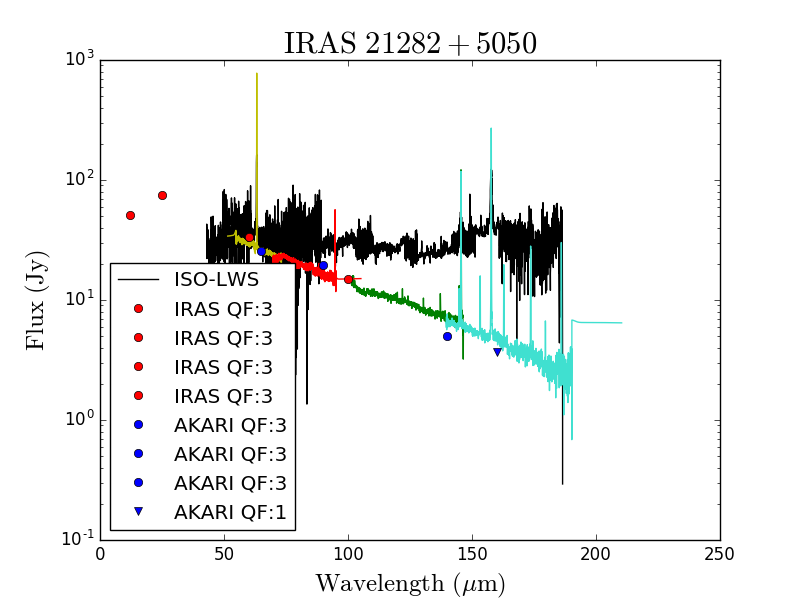}
  %\caption{1d}
  \label{fig:sfig1d}
\end{subfigure}
\begin{subfigure}{.55\textwidth}
\centering
  \includegraphics[width=0.9\linewidth,left]{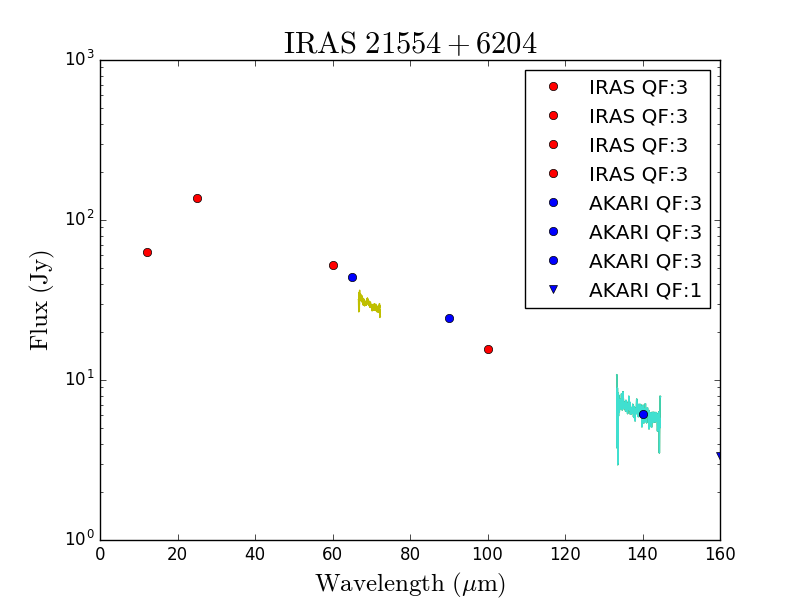}
  %\caption{1a}
  \label{fig:sfig1a}
\end{subfigure}
\begin{subfigure}{.55\textwidth}
\centering
  \includegraphics[width=0.9\linewidth,left]{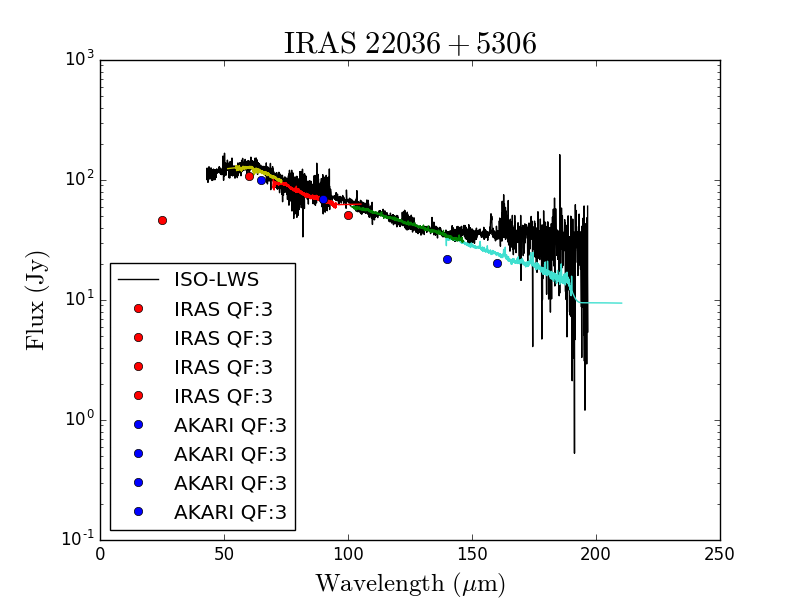}
  %\caption{1b}
  \label{fig:sfig1b}
\end{subfigure}
\begin{subfigure}{.55\textwidth}
\centering
  \includegraphics[width=0.9\linewidth,left]{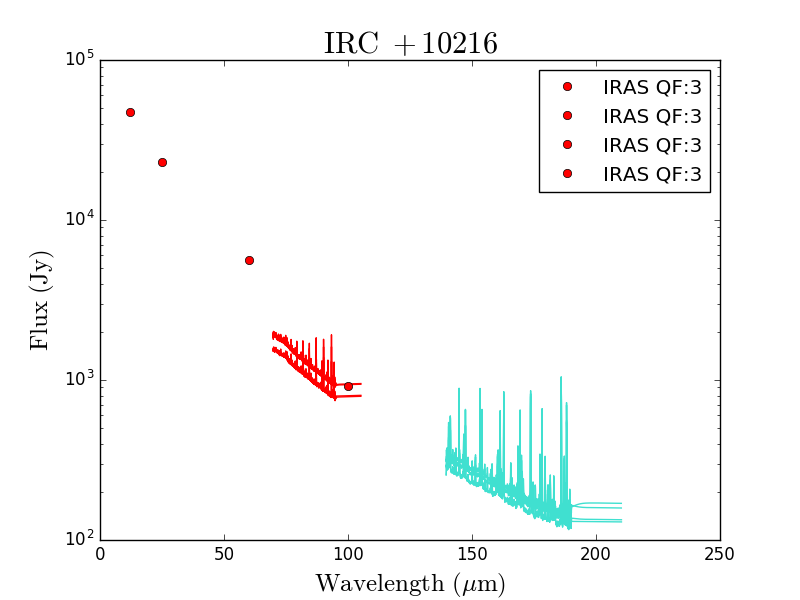}
  %\caption{1c}
  \label{fig:sfig1c}
\end{subfigure}
\caption{Continued.}
\end{figure*}

\clearpage
\addtocounter{figure}{-1}
\begin{figure*}
\begin{subfigure}{.55\textwidth}
\centering
  \includegraphics[width=0.9\linewidth,left]{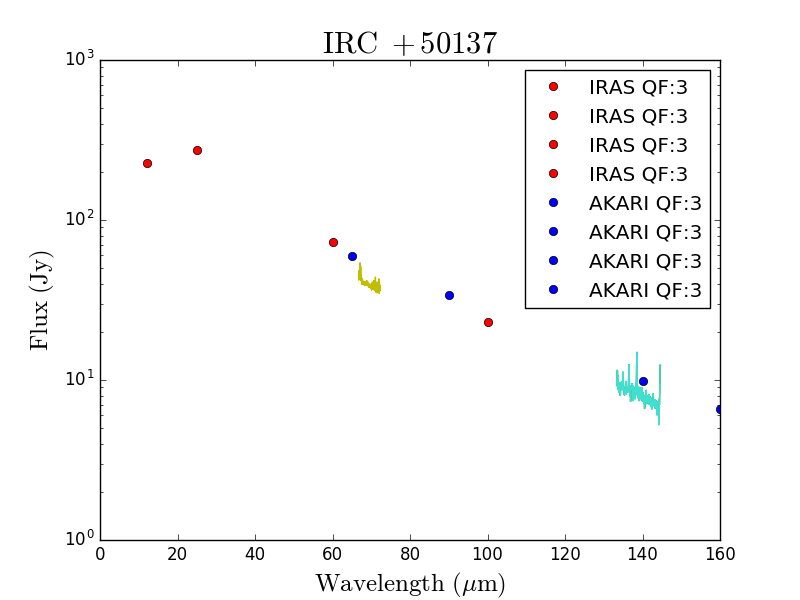}
  %\caption{1d}
  \label{fig:sfig1d}
\end{subfigure}
\begin{subfigure}{.55\textwidth}
\centering
  \includegraphics[width=0.9\linewidth,left]{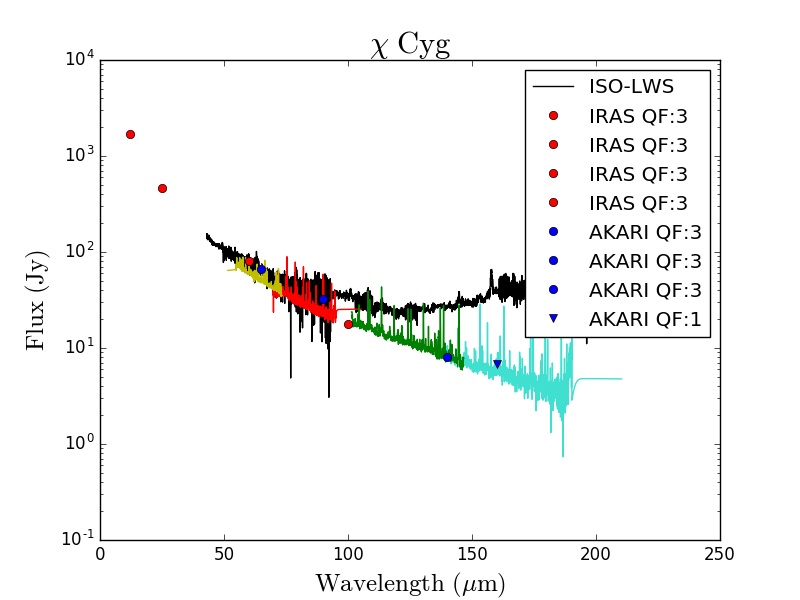}
  %\caption{1d}
  \label{fig:sfig1d}
\end{subfigure}
\begin{subfigure}{.55\textwidth}
\centering
  \includegraphics[width=0.9\linewidth,left]{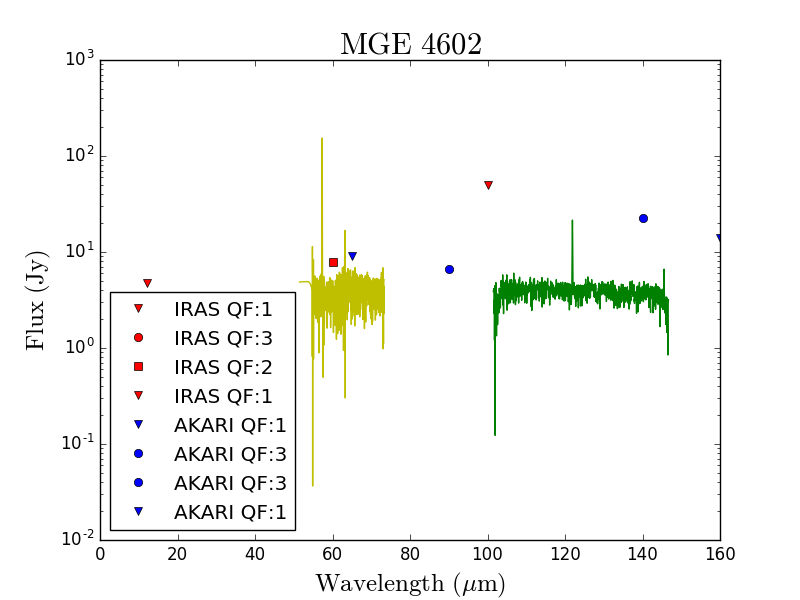}
  %\caption{1b}
  \label{fig:sfig1b}
\end{subfigure}
\begin{subfigure}{.55\textwidth}
\centering
  \includegraphics[width=0.9\linewidth,left]{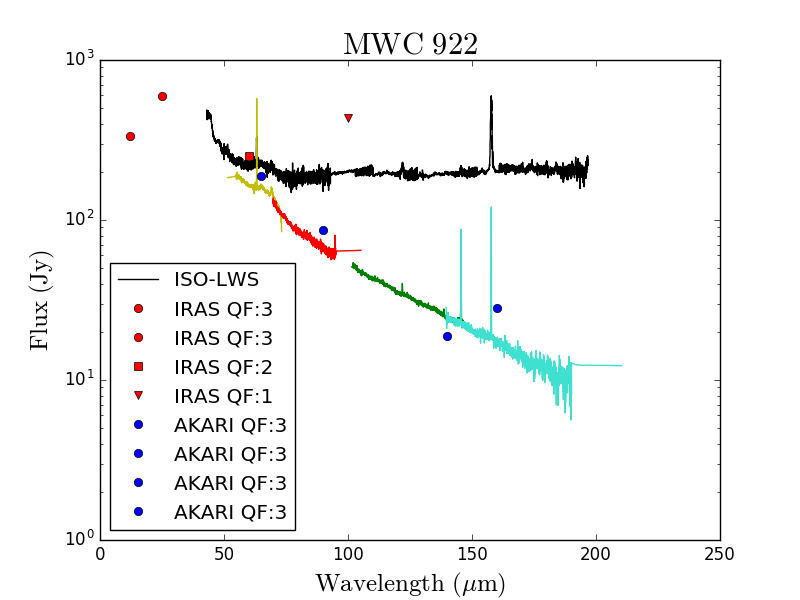}
  %\caption{1c}
  \label{fig:sfig1c}
\end{subfigure}
\begin{subfigure}{.55\textwidth}
\centering
  \includegraphics[width=0.9\linewidth,left]{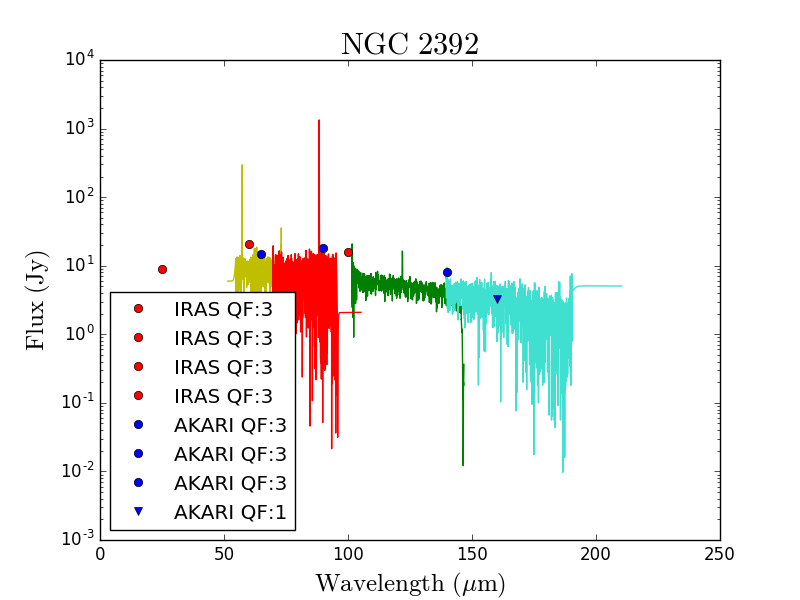}
  %\caption{1b}
  \label{fig:sfig1b}
\end{subfigure}
\begin{subfigure}{.55\textwidth}
\centering
  \includegraphics[width=0.9\linewidth,left]{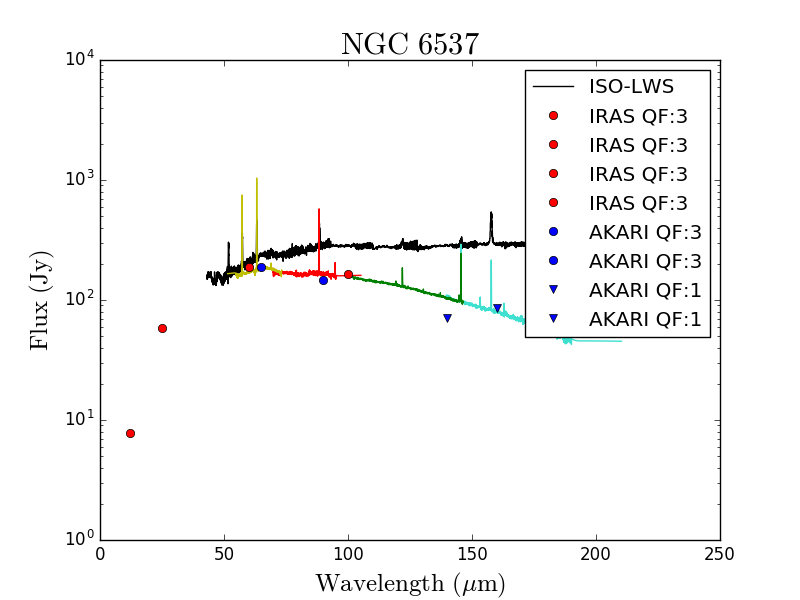}
  %\caption{1d}
  \label{fig:sfig1d}
\end{subfigure}
\caption{Continued.}
\end{figure*}

\clearpage
\addtocounter{figure}{-1}
\begin{figure*}
\begin{subfigure}{.55\textwidth}
\centering
  \includegraphics[width=0.9\linewidth,left]{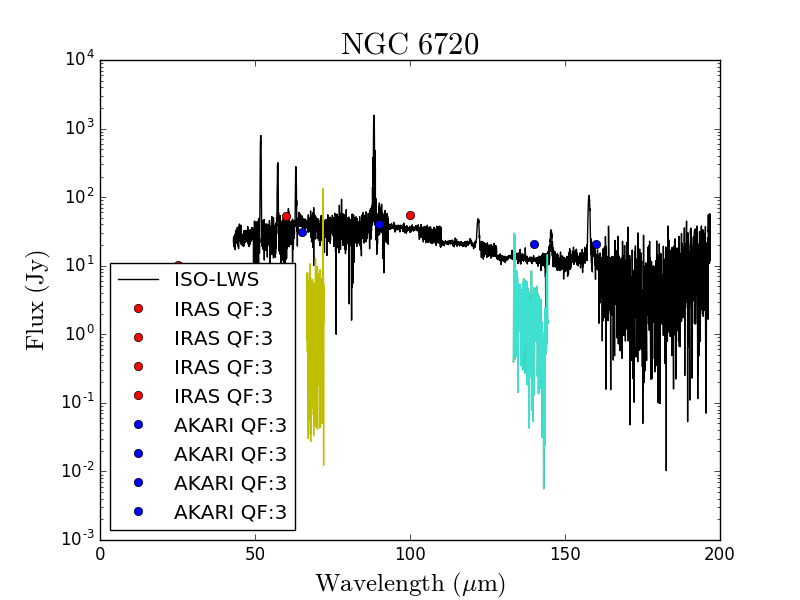}
  %\caption{1d}
  \label{fig:sfig1d}
\end{subfigure}
\begin{subfigure}{.55\textwidth}
\centering
  \includegraphics[width=0.9\linewidth,left]{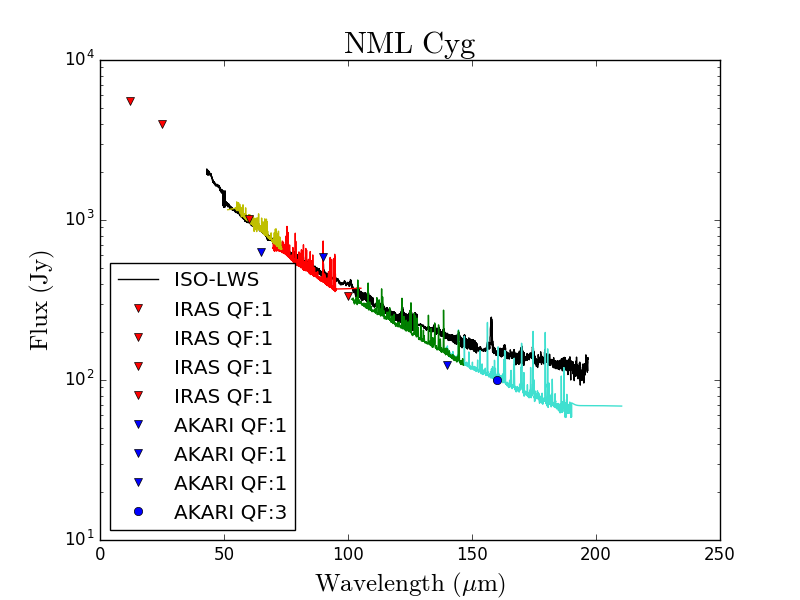}
  %\caption{1d}
  \label{fig:sfig1d}
\end{subfigure}
\begin{subfigure}{.55\textwidth}
\centering
  \includegraphics[width=0.9\linewidth,left]{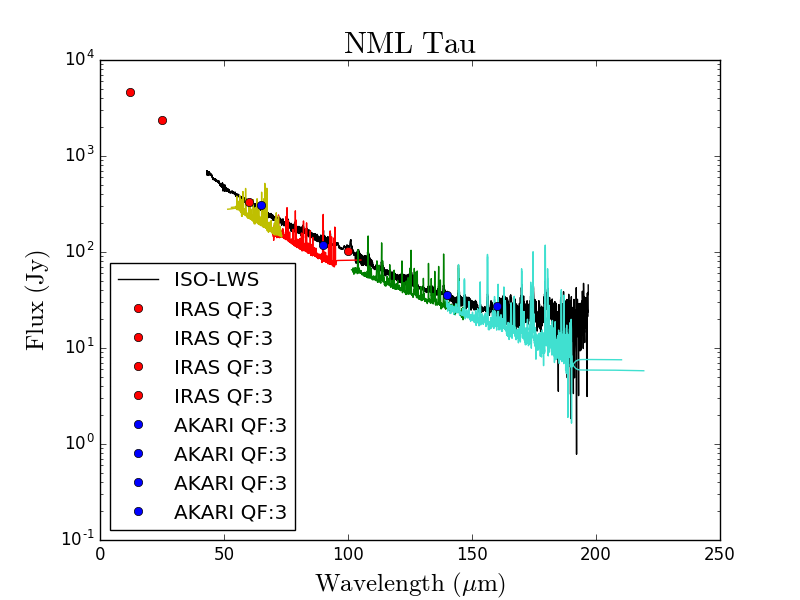}
  %\caption{1d}
  \label{fig:sfig1d}
\end{subfigure}
\begin{subfigure}{.55\textwidth}
\centering
  \includegraphics[width=0.9\linewidth,left]{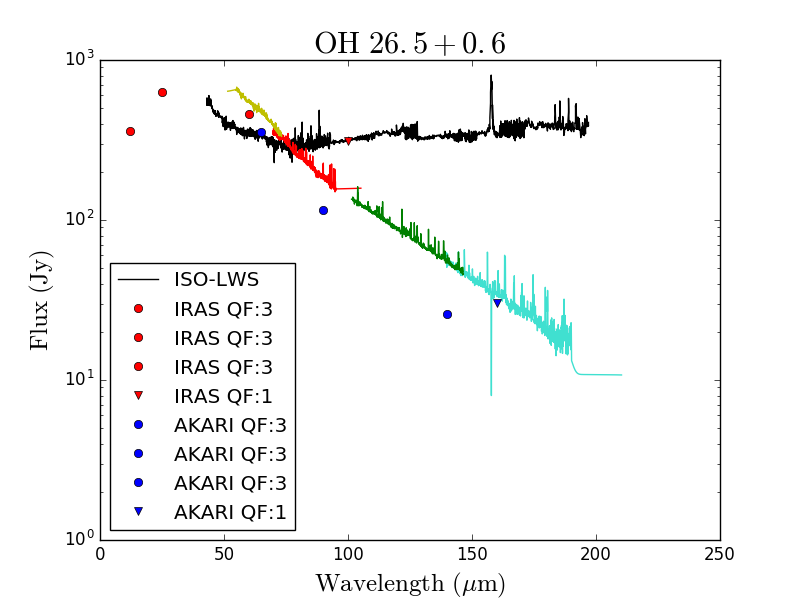}
  %\caption{1d}
  \label{fig:sfig1d}
\end{subfigure}
\begin{subfigure}{.55\textwidth}
\centering
  \includegraphics[width=0.9\linewidth,left]{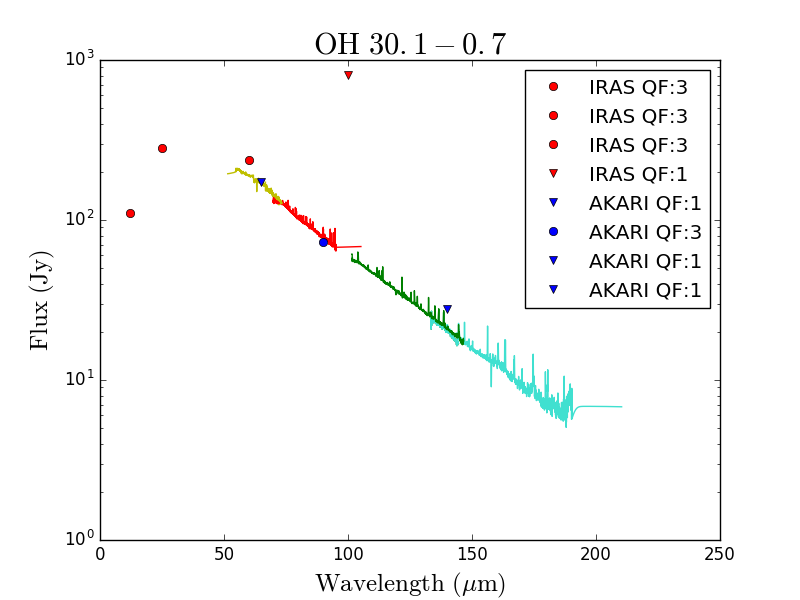}
  %\caption{1d}
  \label{fig:sfig1d}
\end{subfigure}
\begin{subfigure}{.55\textwidth}
\centering
  \includegraphics[width=0.9\linewidth,left]{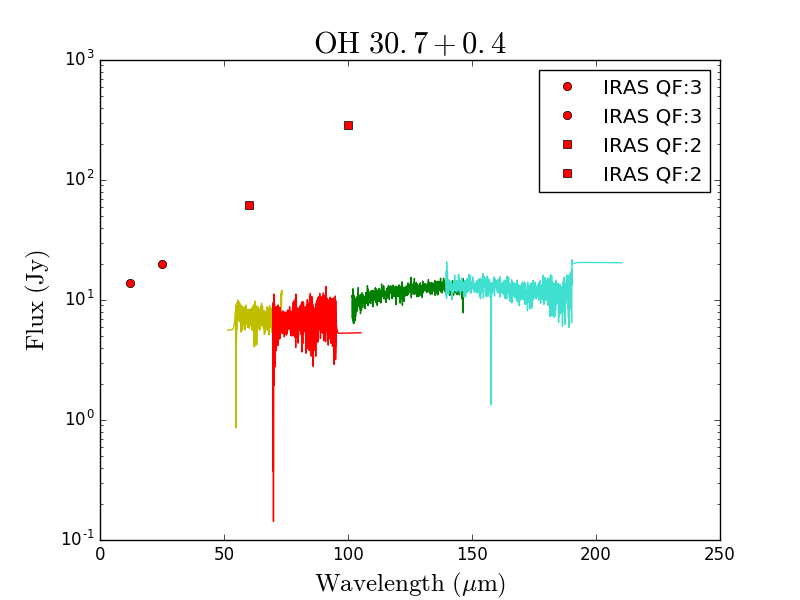}
  %\caption{1b}
  \label{fig:sfig1b}
\end{subfigure}
\caption{Continued.}
\end{figure*}

\clearpage

\addtocounter{figure}{-1}
\begin{figure*}
\begin{subfigure}{.55\textwidth}
\centering
  \includegraphics[width=0.9\linewidth,left]{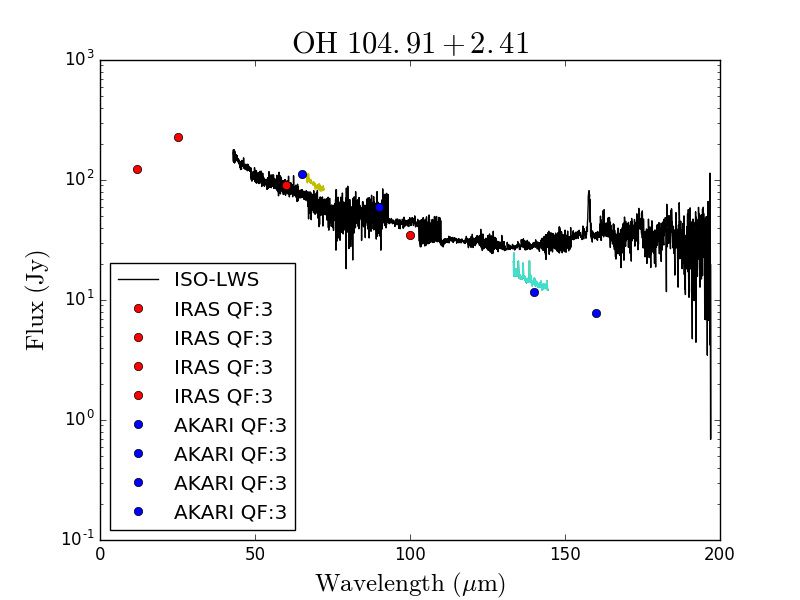}
  %\caption{1d}
  \label{fig:sfig1d}
\end{subfigure}
\begin{subfigure}{.55\textwidth}
\centering
  \includegraphics[width=0.9\linewidth,left]{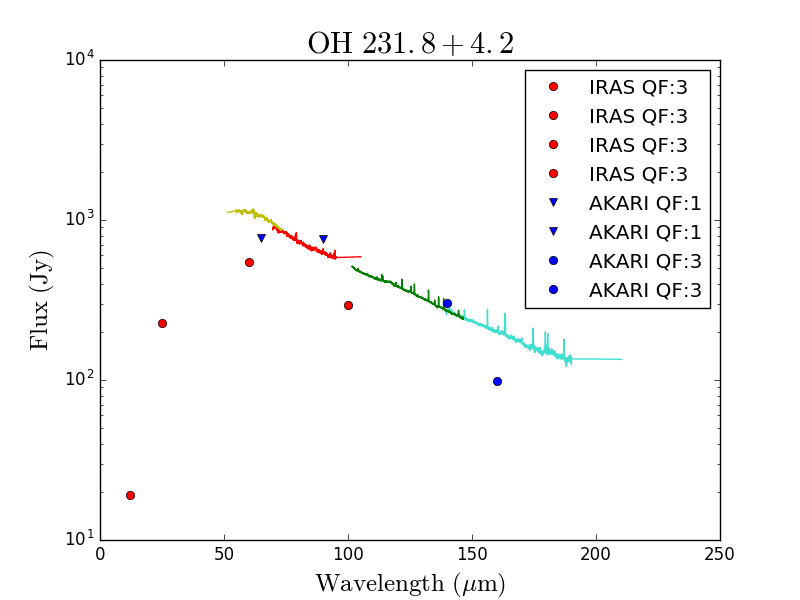}
  %\caption{1a}
  \label{fig:sfig1a}
\end{subfigure}
\begin{subfigure}{.55\textwidth}
\centering
  \includegraphics[width=0.9\linewidth,left]{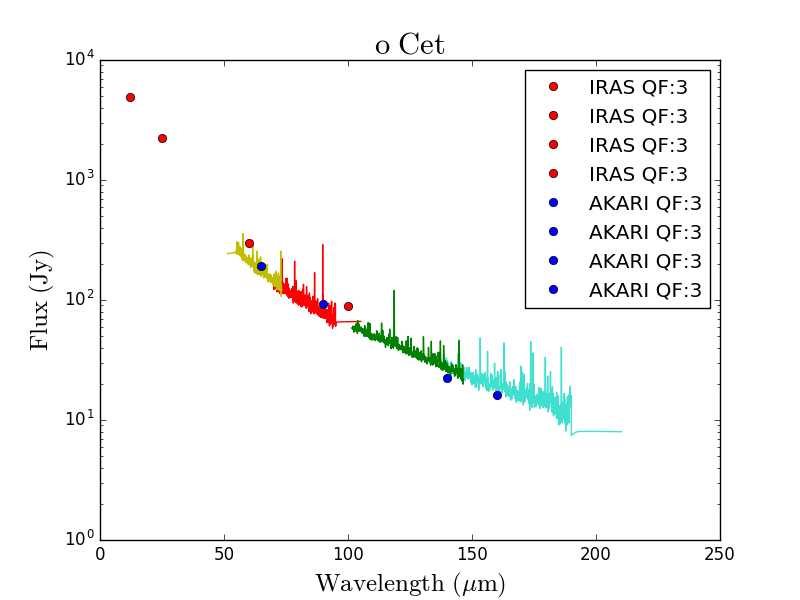}
  %\caption{1d}
  \label{fig:sfig1d}
\end{subfigure}
\begin{subfigure}{.55\textwidth}
\centering
  \includegraphics[width=0.9\linewidth,left]{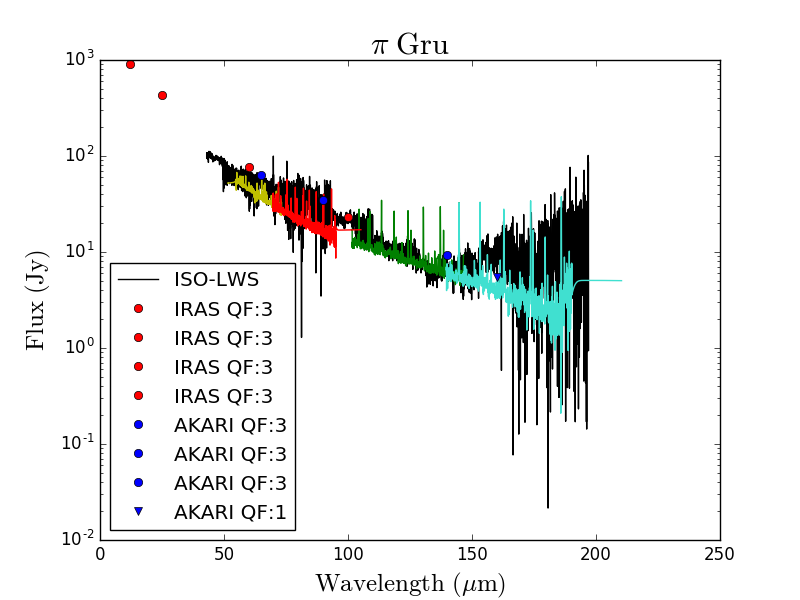}
  %\caption{1d}
  \label{fig:sfig1d}
\end{subfigure}
\begin{subfigure}{.55\textwidth}
\centering
  \includegraphics[width=0.9\linewidth,left]{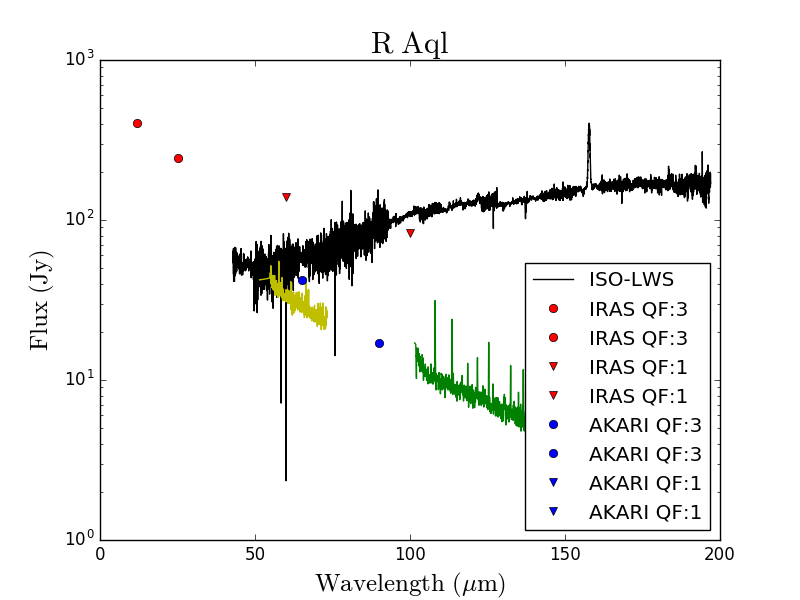}
  %\caption{1c}
  \label{fig:sfig1c}
\end{subfigure}
\begin{subfigure}{.55\textwidth}
\centering
  \includegraphics[width=0.9\linewidth,left]{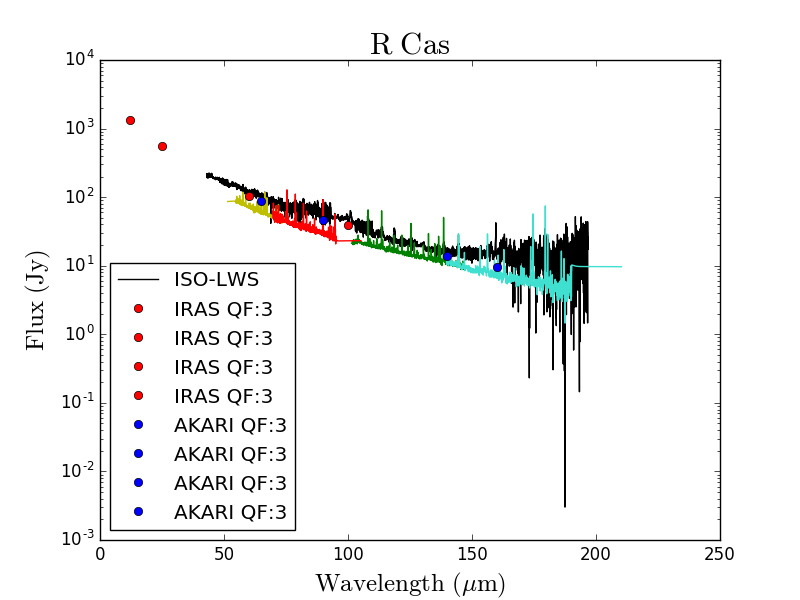}
  %\caption{1d}
  \label{fig:sfig1d}
\end{subfigure}

\caption{Continued.}
\end{figure*}

\clearpage

\addtocounter{figure}{-1}
\begin{figure*}
\begin{subfigure}{.55\textwidth}
\centering
  \includegraphics[width=0.9\linewidth,left]{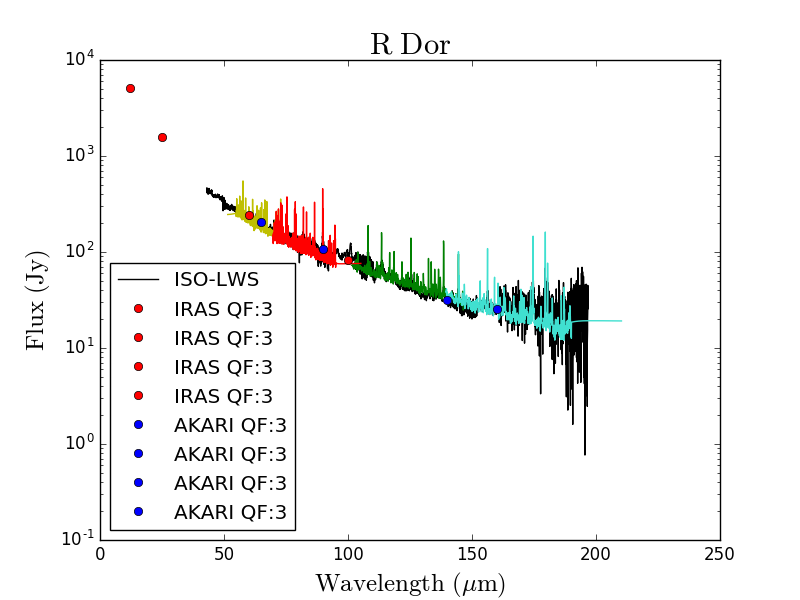}
  %\caption{1d}
  \label{fig:sfig1d}
\end{subfigure}
\begin{subfigure}{.55\textwidth}
\centering
  \includegraphics[width=0.9\linewidth,left]{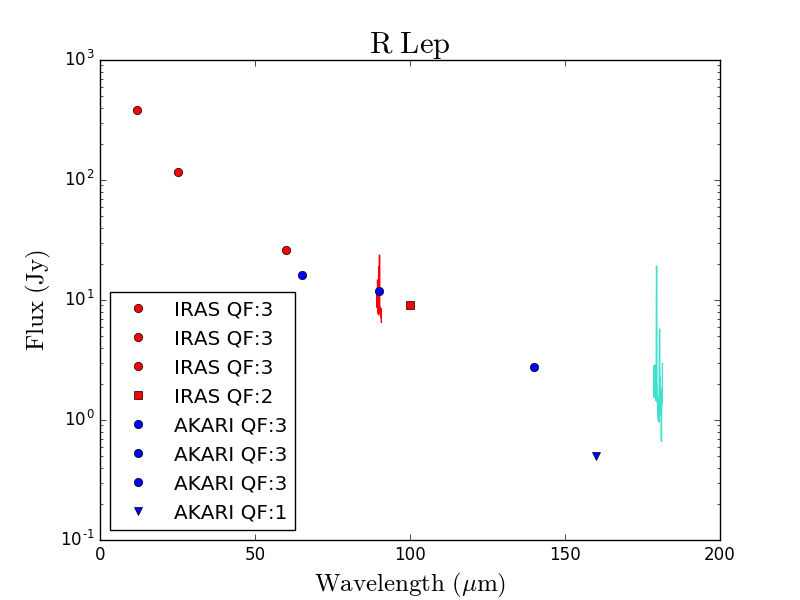}
  %\caption{1d}
  \label{fig:sfig1d}
\end{subfigure}
\begin{subfigure}{.55\textwidth}
\centering
  \includegraphics[width=0.9\linewidth,left]{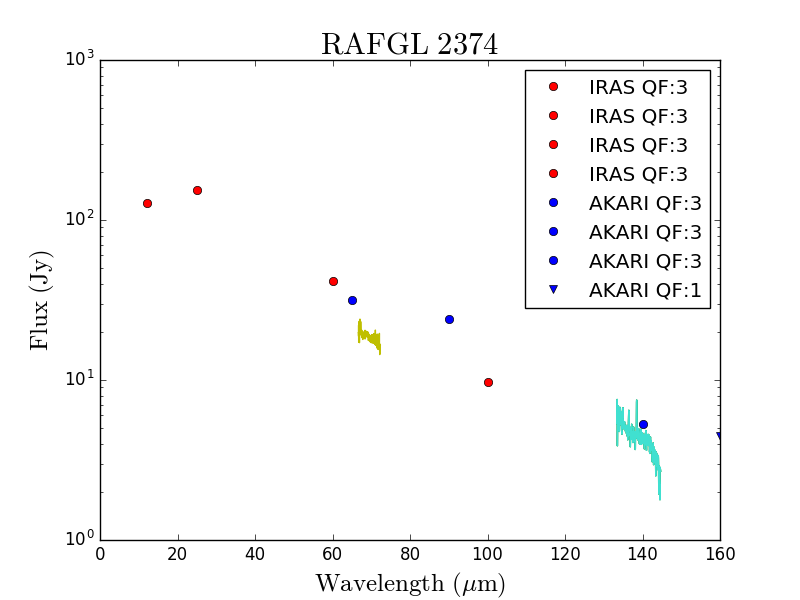}
  %\caption{1a}
  \label{fig:sfig1a}
\end{subfigure}
\begin{subfigure}{.55\textwidth}
\centering
  \includegraphics[width=0.9\linewidth,left]{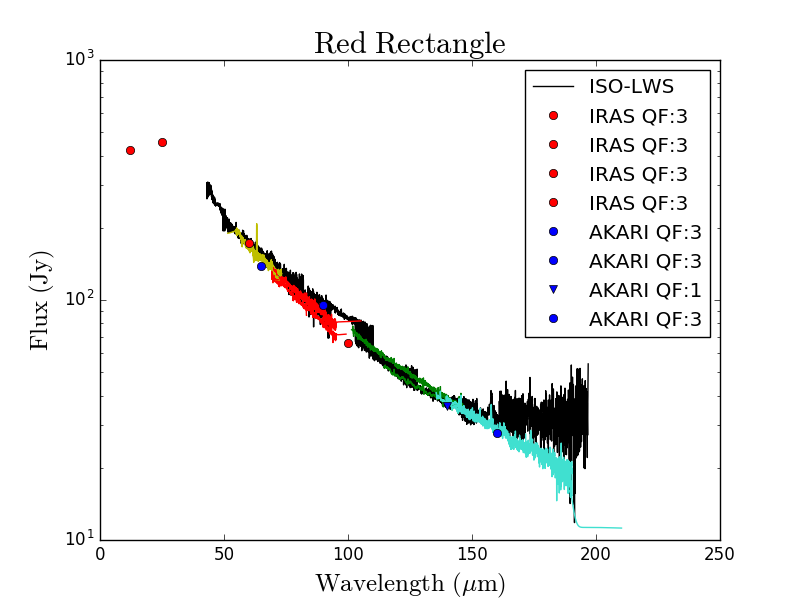}
  %\caption{1b}
  \label{fig:sfig1b}
\end{subfigure}
\begin{subfigure}{.55\textwidth}
\centering
  \includegraphics[width=0.9\linewidth,left]{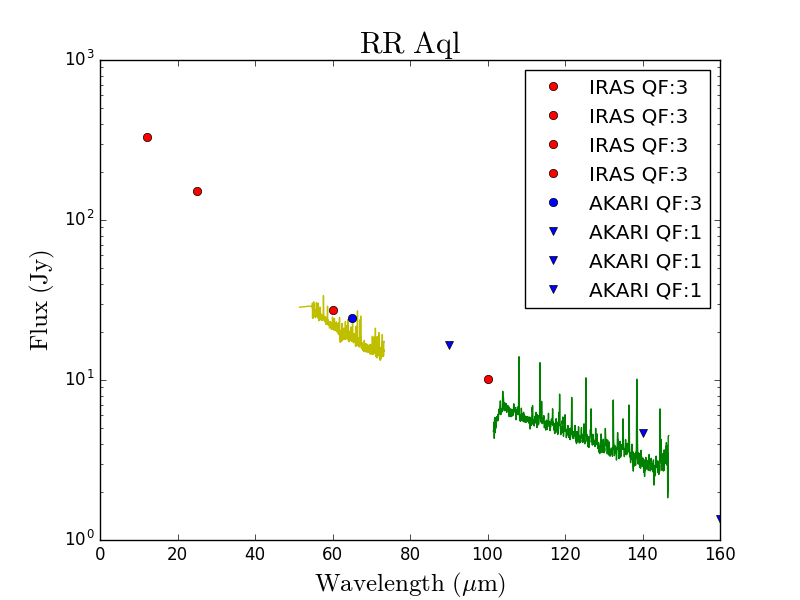}
  %\caption{1d}
  \label{fig:sfig1d}
\end{subfigure}
\begin{subfigure}{.55\textwidth}
\centering
  \includegraphics[width=0.9\linewidth,left]{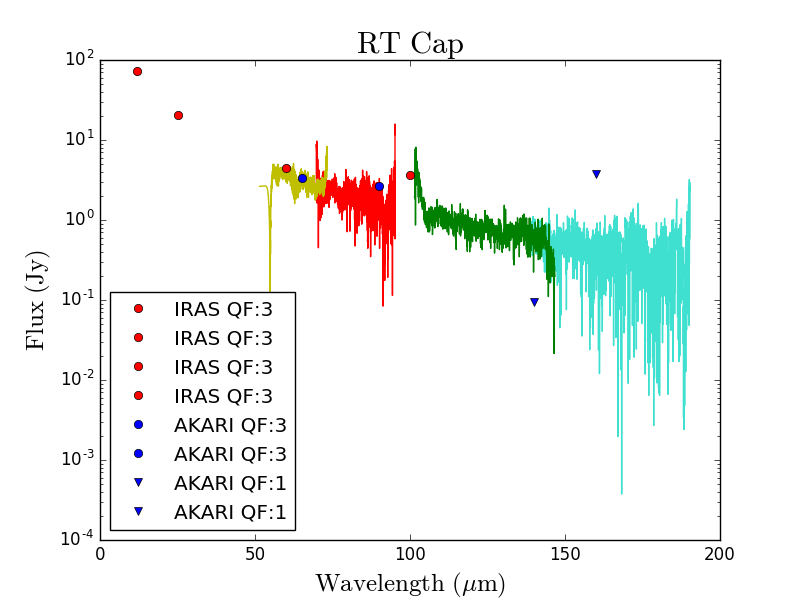}
  %\caption{1d}
  \label{fig:sfig1d}
\end{subfigure}
\caption{Continued.}
\end{figure*}

\clearpage
\addtocounter{figure}{-1}
\begin{figure*}
\begin{subfigure}{.55\textwidth}
\centering
  \includegraphics[width=0.9\linewidth,left]{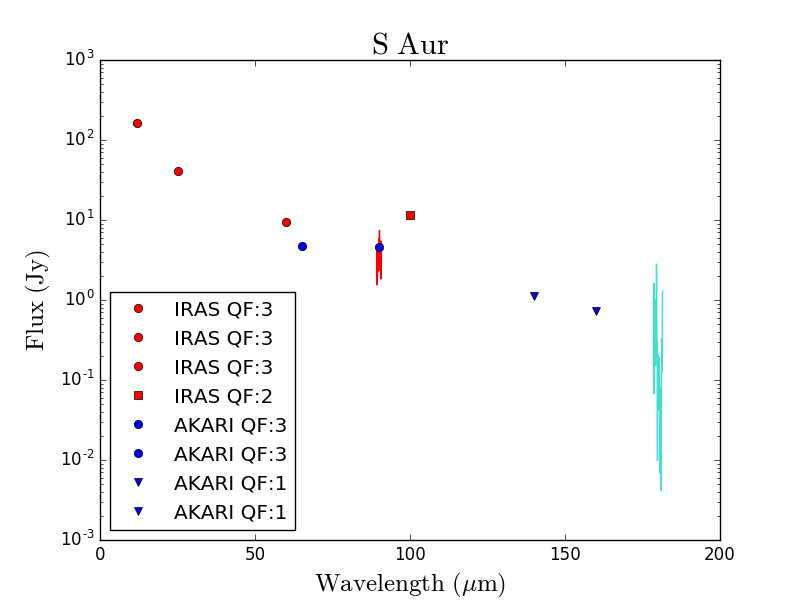}
  %\caption{1a}
  \label{fig:sfig1a}
\end{subfigure}
\begin{subfigure}{.55\textwidth}
\centering
  \includegraphics[width=0.9\linewidth,left]{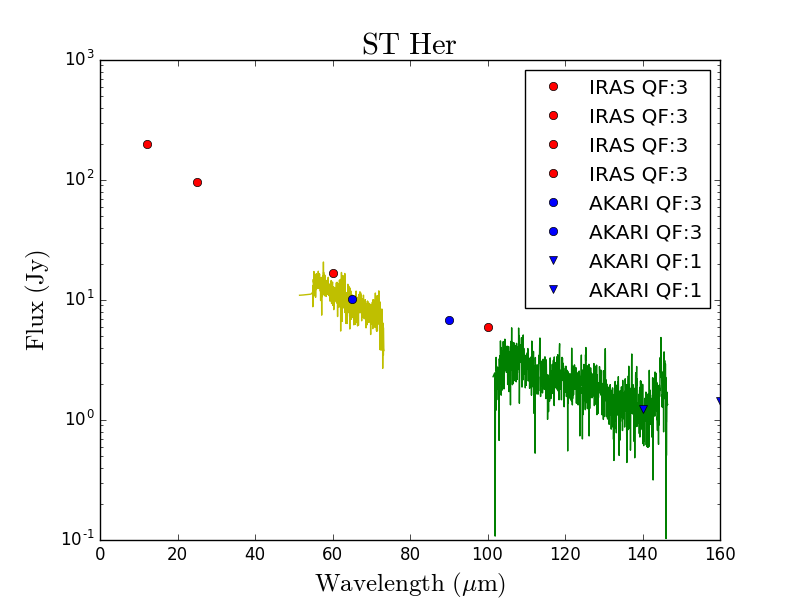}
  %\caption{1b}
  \label{fig:sfig1b}
\end{subfigure}
\begin{subfigure}{.55\textwidth}
\centering
  \includegraphics[width=0.9\linewidth,left]{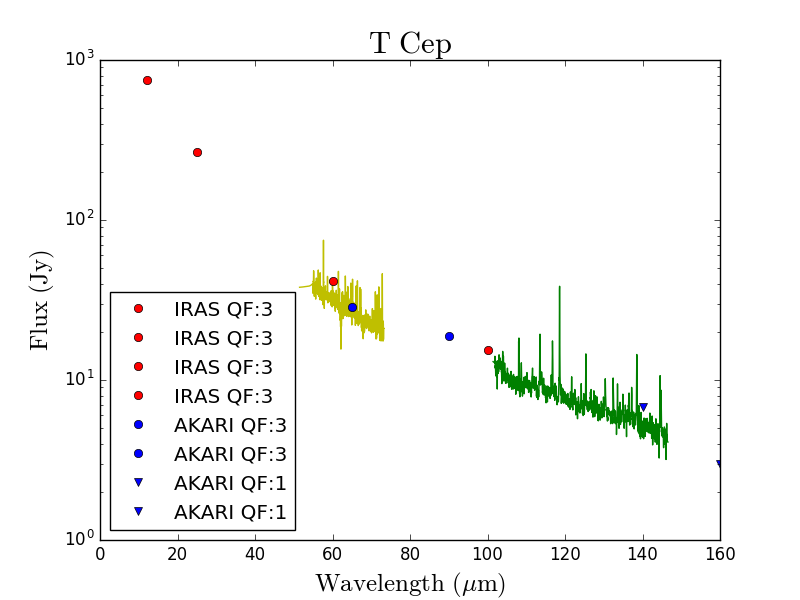}
  %\caption{1d}
  \label{fig:sfig1d}
\end{subfigure}
\begin{subfigure}{.55\textwidth}
\centering
  \includegraphics[width=0.9\linewidth,left]{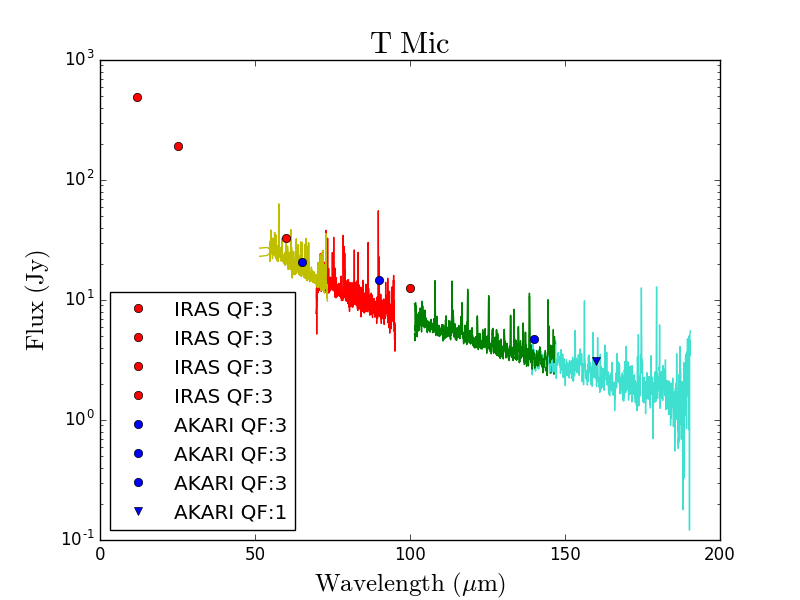}
  %\caption{1d}
  \label{fig:sfig1d}
\end{subfigure}
\begin{subfigure}{.55\textwidth}
\centering
  \includegraphics[width=0.9\linewidth,left]{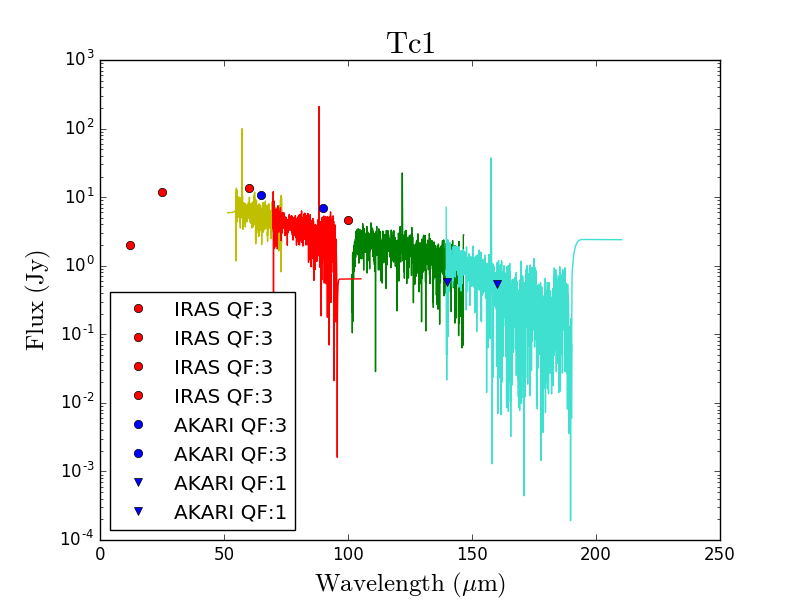}
  %\caption{1d}
  \label{fig:sfig1d}
\end{subfigure}
\begin{subfigure}{.55\textwidth}
\centering
  \includegraphics[width=0.9\linewidth,left]{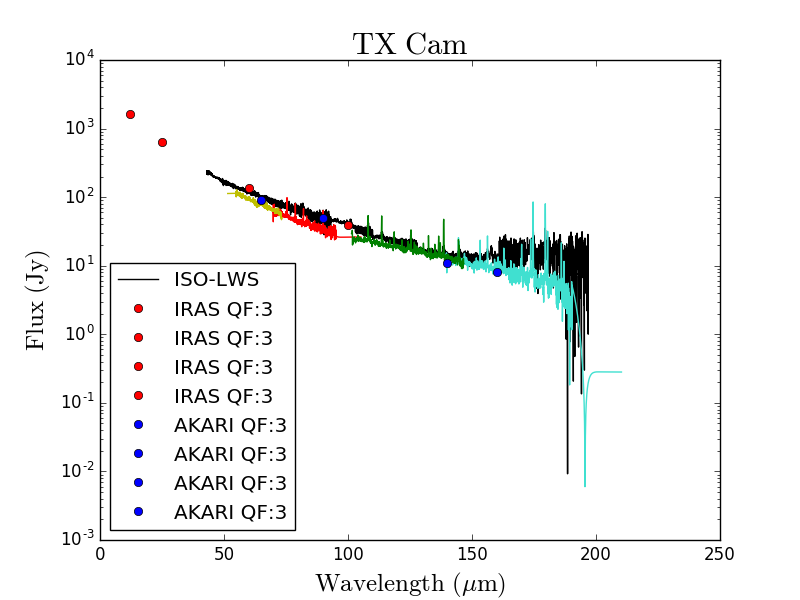}
  %\caption{1c}
  \label{fig:sfig1c}
\end{subfigure}
\caption{Continued.}
\end{figure*}

\clearpage
\addtocounter{figure}{-1}
\begin{figure*}
\begin{subfigure}{.55\textwidth}
\centering
  \includegraphics[width=0.9\linewidth,left]{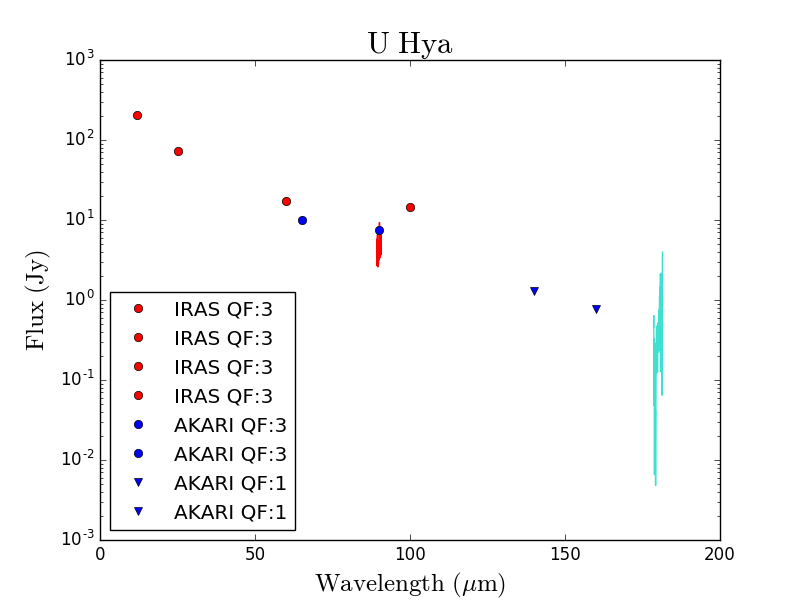}
  %\caption{1d}
  \label{fig:sfig1d}
\end{subfigure}
\begin{subfigure}{.55\textwidth}
\centering
  \includegraphics[width=0.9\linewidth,left]{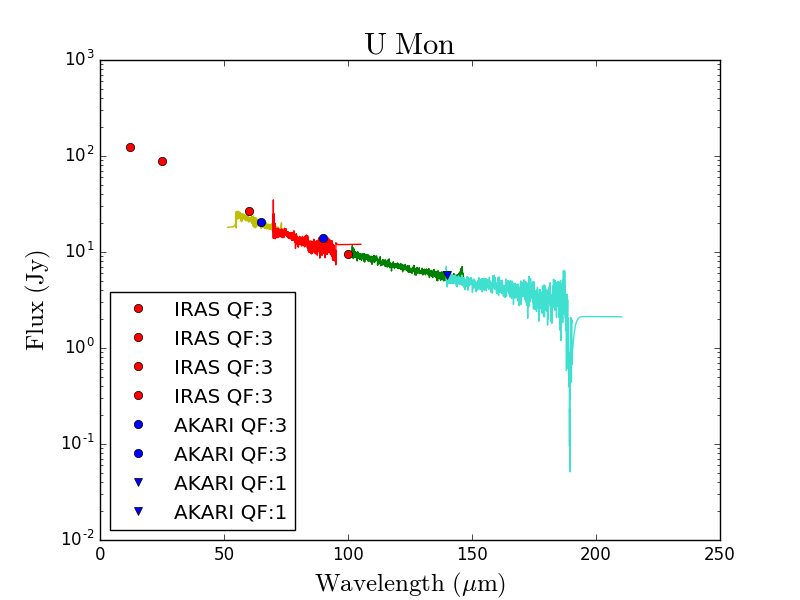}
  %\caption{1d}
  \label{fig:sfig1d}
\end{subfigure}
\begin{subfigure}{.55\textwidth}
\centering
  \includegraphics[width=0.9\linewidth,left]{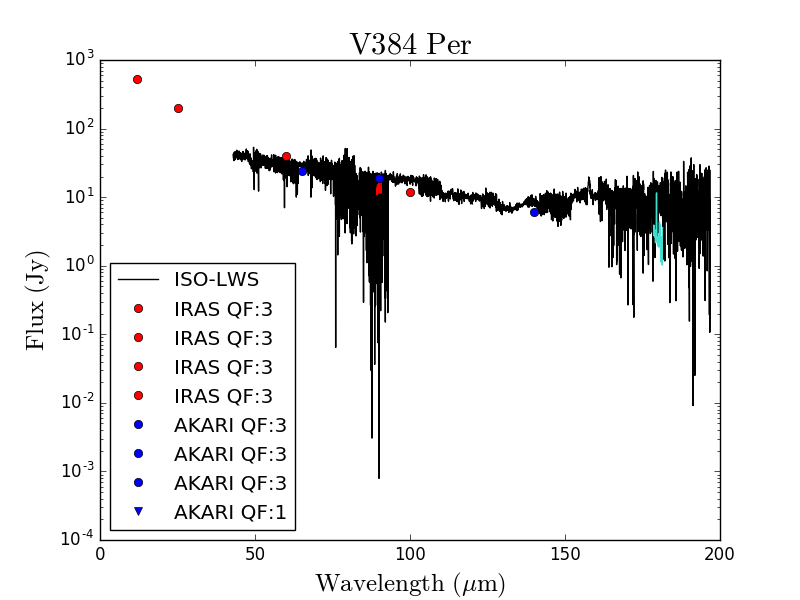}
  %\caption{1b}
  \label{fig:sfig1b}
\end{subfigure}
\begin{subfigure}{.55\textwidth}
\centering
  \includegraphics[width=0.9\linewidth,left]{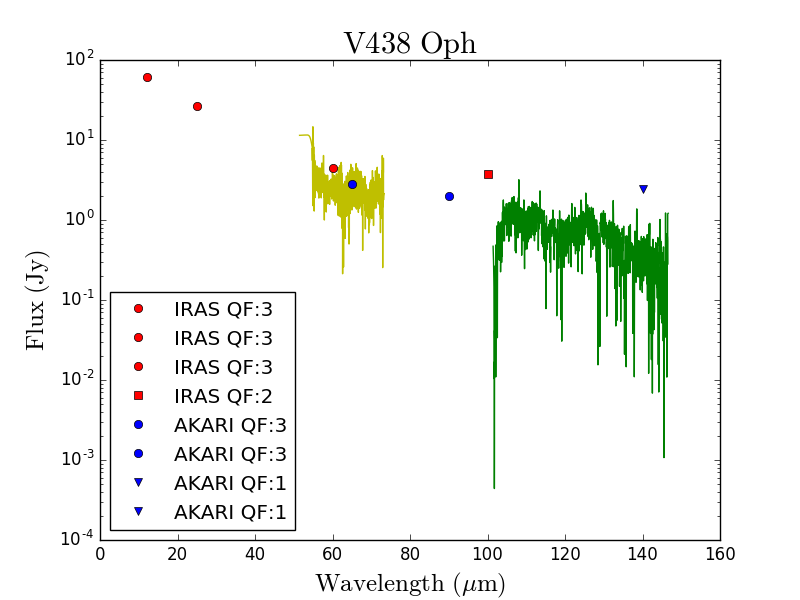}
  %\caption{1c}
  \label{fig:sfig1c}
\end{subfigure}
\begin{subfigure}{.55\textwidth}
\centering
  \includegraphics[width=0.9\linewidth,left]{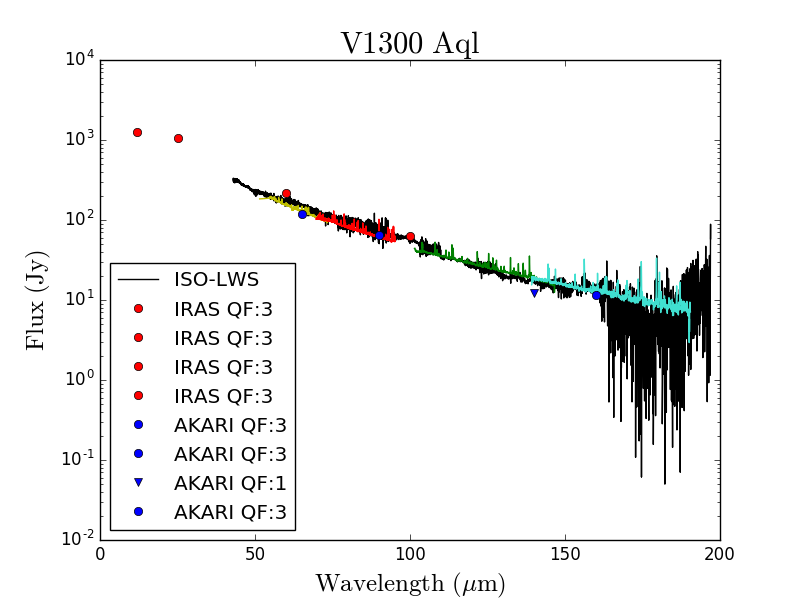}
  %\caption{1a}
  \label{fig:sfig1a}
\end{subfigure}
\begin{subfigure}{.55\textwidth}
\centering
  \includegraphics[width=0.9\linewidth,left]{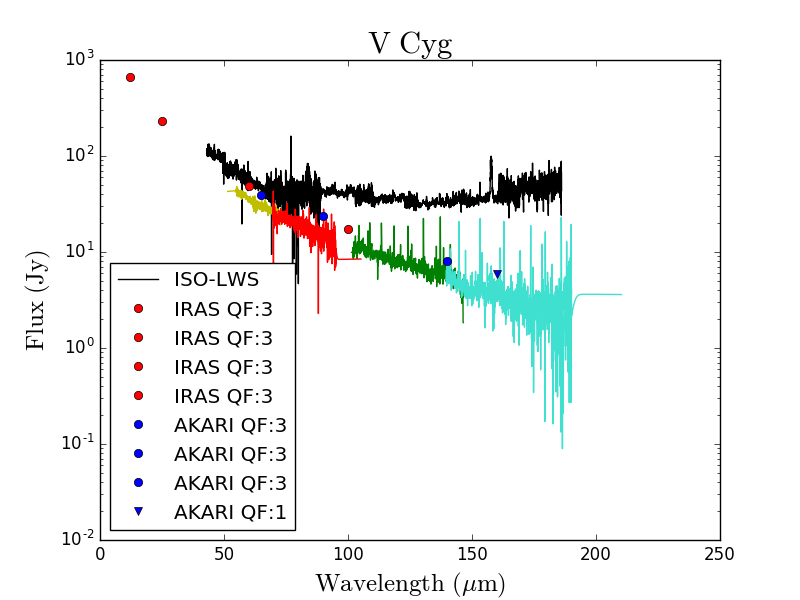}
  %\caption{1d}
  \label{fig:sfig1d}
\end{subfigure}
\caption{Continued.}
\end{figure*}

\clearpage
\addtocounter{figure}{-1}
\begin{figure*}
\begin{subfigure}{.55\textwidth}
\centering
  \includegraphics[width=0.9\linewidth,left]{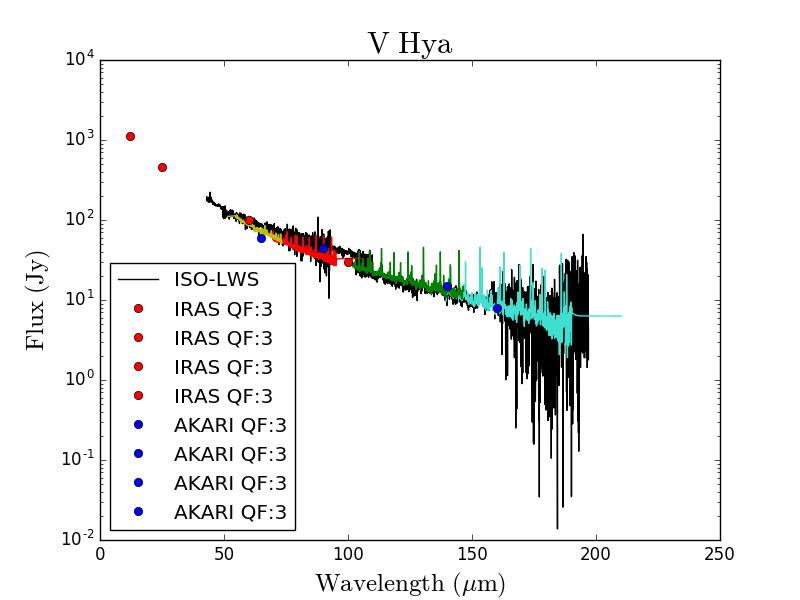}
  %\caption{1d}
  \label{fig:sfig1d}
\end{subfigure}
\begin{subfigure}{.55\textwidth}
\centering
  \includegraphics[width=0.9\linewidth,left]{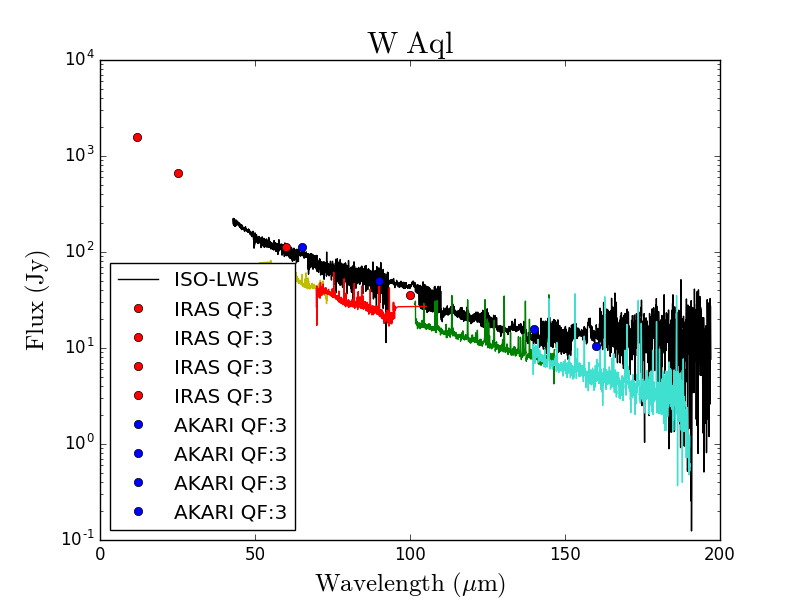}
  %\caption{1d}
  \label{fig:sfig1d}
\end{subfigure}
\begin{subfigure}{.55\textwidth}
\centering
  \includegraphics[width=0.9\linewidth,left]{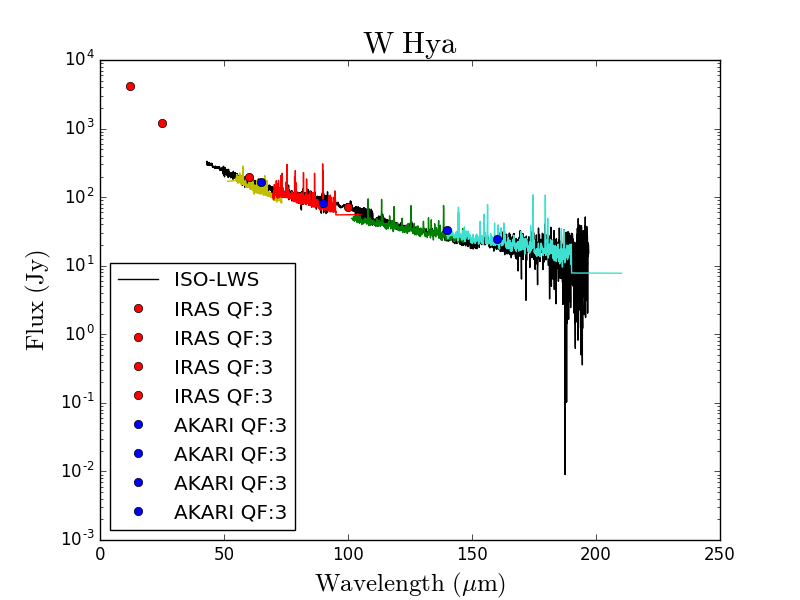}
  %\caption{1d}
  \label{fig:sfig1d}
\end{subfigure}
\begin{subfigure}{.55\textwidth}
\centering
  \includegraphics[width=0.9\linewidth,left]{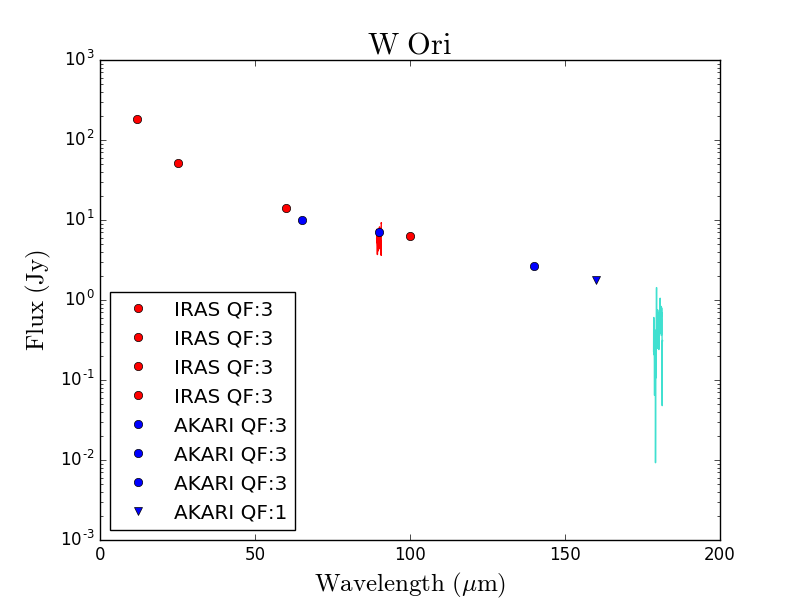}
  %\caption{1a}
  \label{fig:sfig1a}
\end{subfigure}
\begin{subfigure}{.55\textwidth}
\centering
  \includegraphics[width=0.9\linewidth,left]{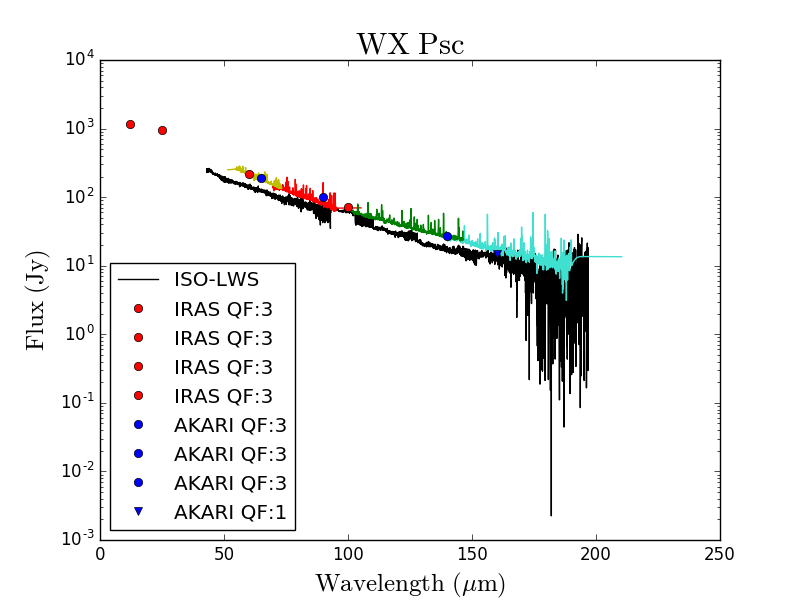}
  %\caption{1d}
  \label{fig:sfig1d}
\end{subfigure}
\begin{subfigure}{.55\textwidth}
\centering
  \includegraphics[width=0.9\linewidth,left]{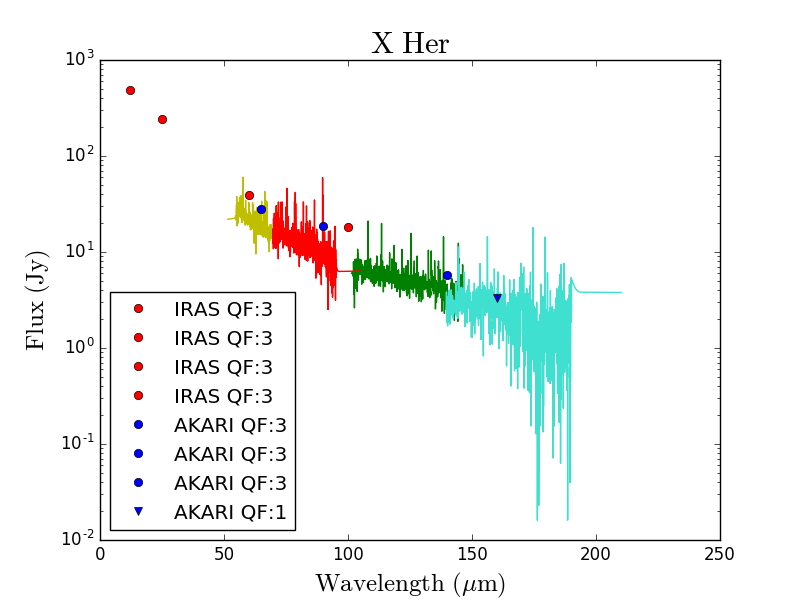}
  %\caption{1b}
  \label{fig:sfig1b}
\end{subfigure}
\caption{Continued.}
\end{figure*}

\clearpage
\addtocounter{figure}{-1}
\begin{figure*}
\begin{subfigure}{.55\textwidth}
\centering
  \includegraphics[width=0.9\linewidth,left]{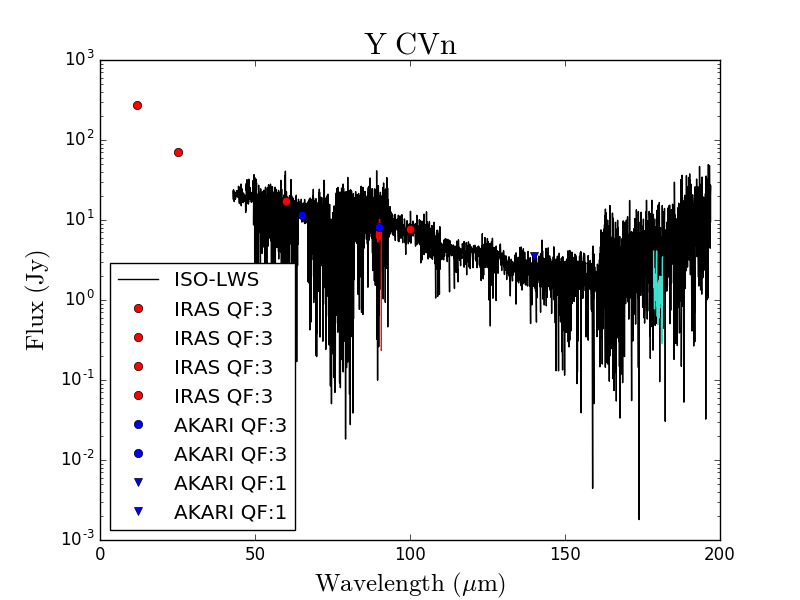}
  
  \label{fig:sfig1c}
\end{subfigure}
\caption{Continued.}

\end{figure*}

%------- Extended 5x5 corrected cases ------------

\begin{figure*}
\begin{subfigure}{.55\textwidth}
 \centering
  \includegraphics[width=0.9\linewidth,left]{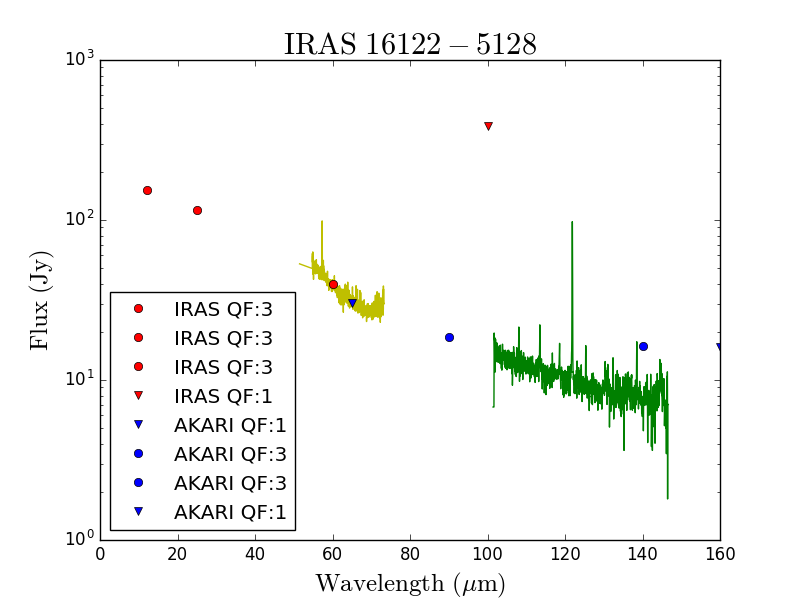}
  %\caption{1a}
  \label{fig:sfig1a}
\end{subfigure}
\begin{subfigure}{.55\textwidth}
\centering
  \includegraphics[width=0.9\linewidth,left]{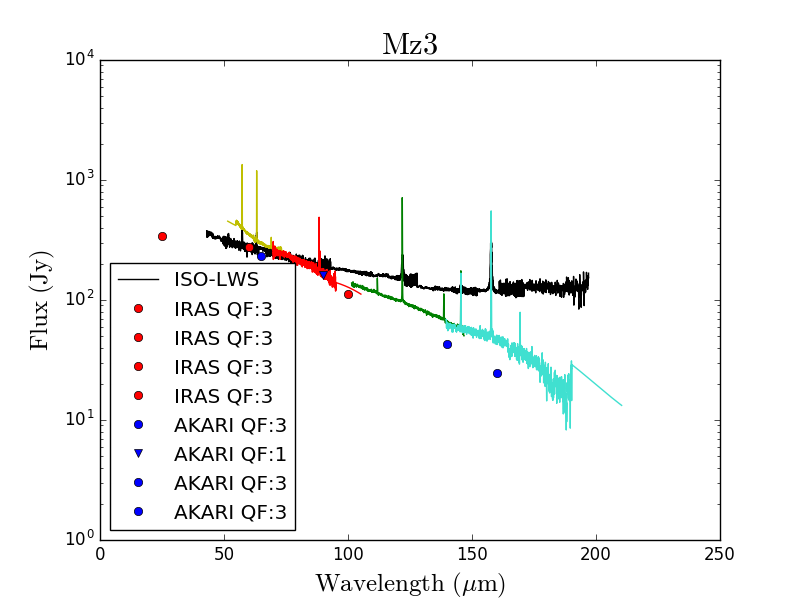}
  %\caption{1d}
  \label{fig:sfig1d}
\end{subfigure}
\begin{subfigure}{.55\textwidth}
\centering
  \includegraphics[width=0.9\linewidth,left]{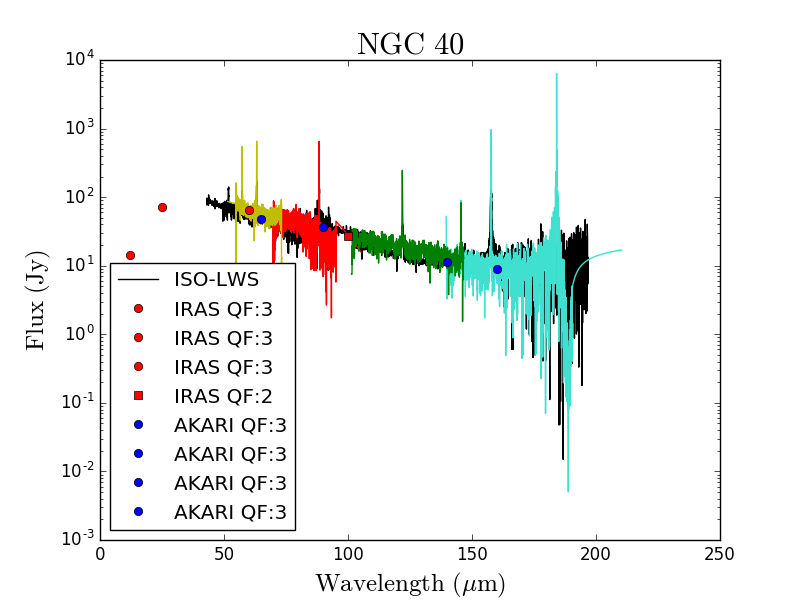}
  %\caption{1d}
  \label{fig:sfig1d}
\end{subfigure}
\begin{subfigure}{.55\textwidth}
\centering
  \includegraphics[width=0.9\linewidth,left]{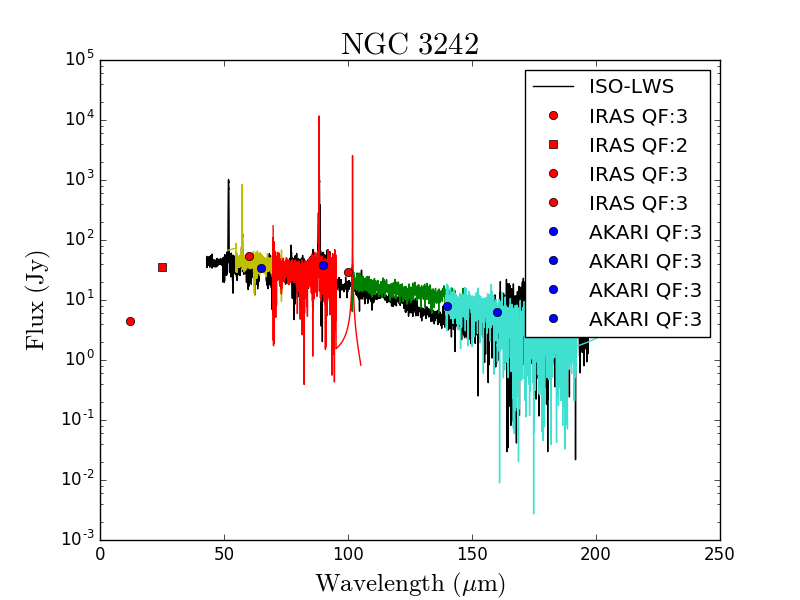}
  %\caption{1d}
  \label{fig:sfig1d}
\end{subfigure}
\begin{subfigure}{.55\textwidth}
\centering
  \includegraphics[width=0.9\linewidth,left]{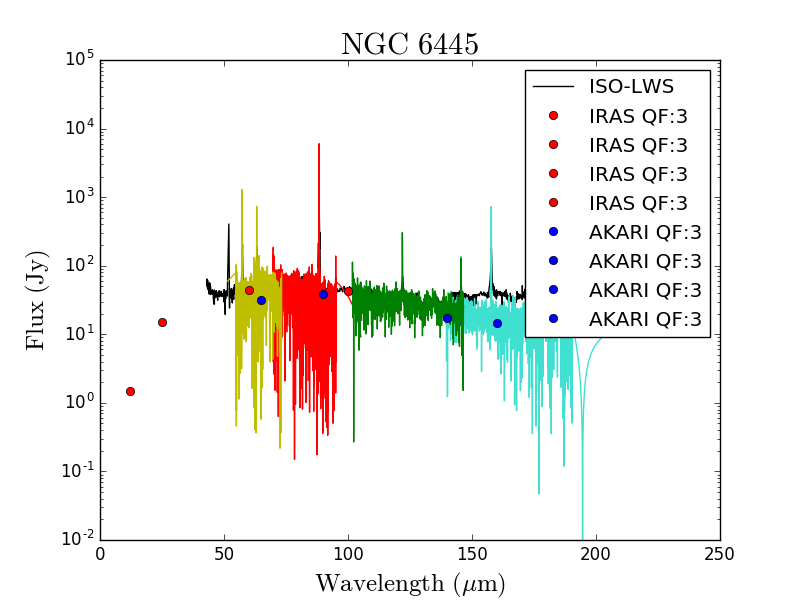}
  %\caption{1d}
  \label{fig:sfig1d}
\end{subfigure}
\begin{subfigure}{.55\textwidth}
\centering
  \includegraphics[width=0.9\linewidth,left]{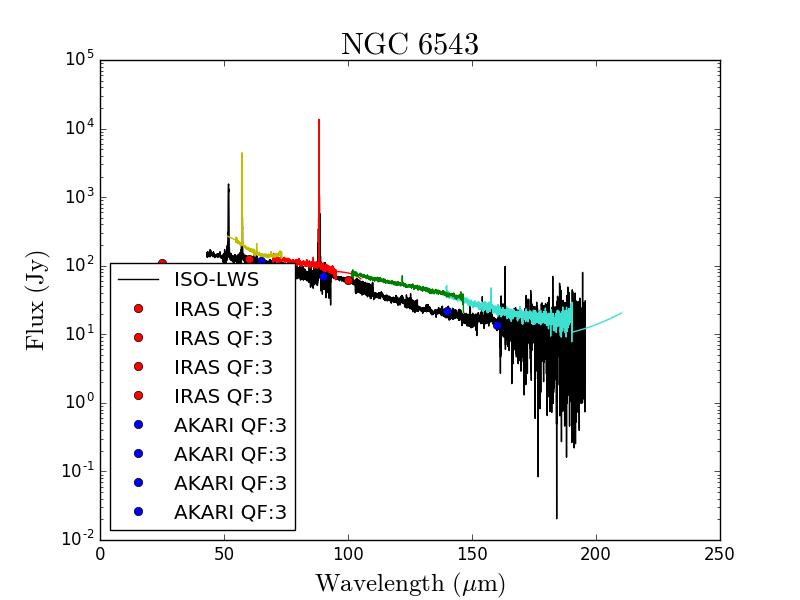}
  %\caption{1a}
  \label{fig:sfig1a}
\end{subfigure}
\caption{As in Fig. \ref{StandardSources} but for extended objects after applying point source flux loss correction 
and extended 5x5 correction. IRAS 16122-5128 appears in THROES catalogue and in HSA as MGE 4218.}
\label{Extended5x5}
\end{figure*}

\clearpage
\addtocounter{figure}{-1}
\begin{figure*}
\begin{subfigure}{.55\textwidth}
\centering
  \includegraphics[width=0.9\linewidth,left]{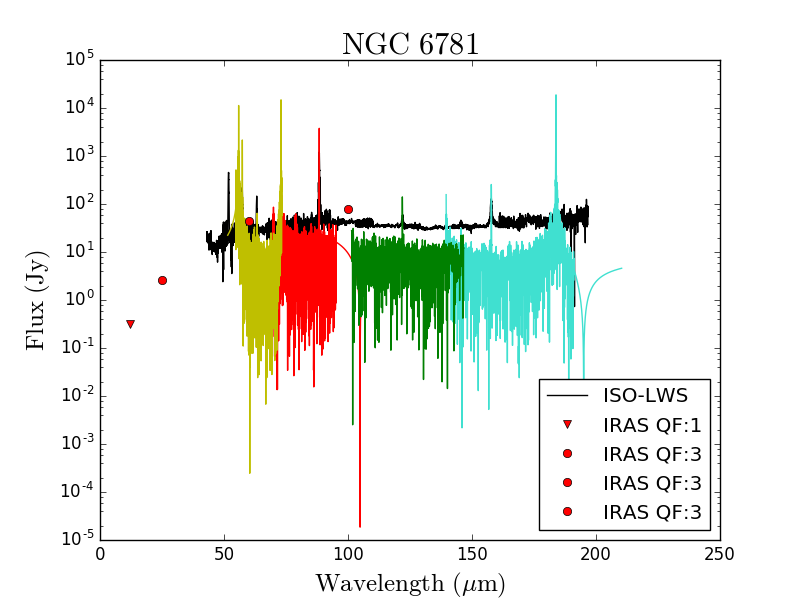}
  %\caption{1d}
  \label{fig:sfig1d}
\end{subfigure}
\begin{subfigure}{.55\textwidth}
\centering
  \includegraphics[width=0.9\linewidth,left]{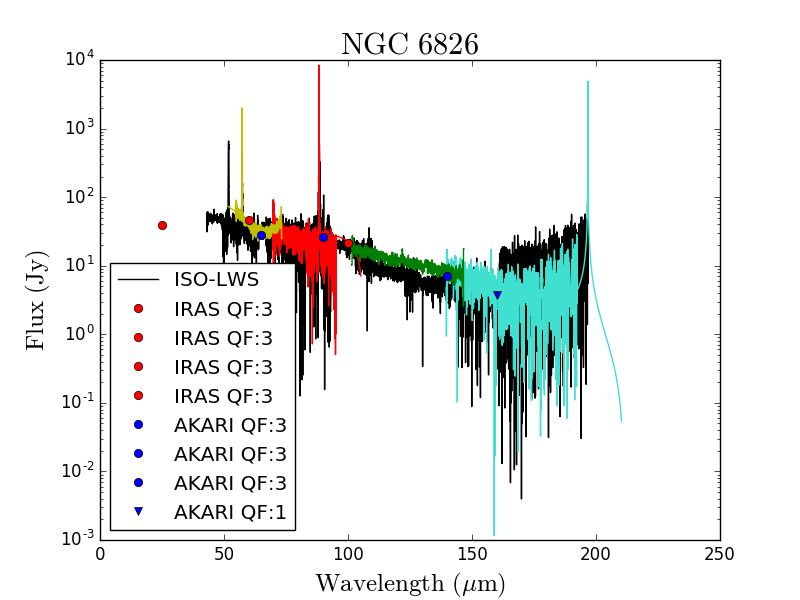}
  %\caption{1d}
  \label{fig:sfig1d}
\end{subfigure}
\begin{subfigure}{.55\textwidth}
\centering
  \includegraphics[width=0.9\linewidth,left]{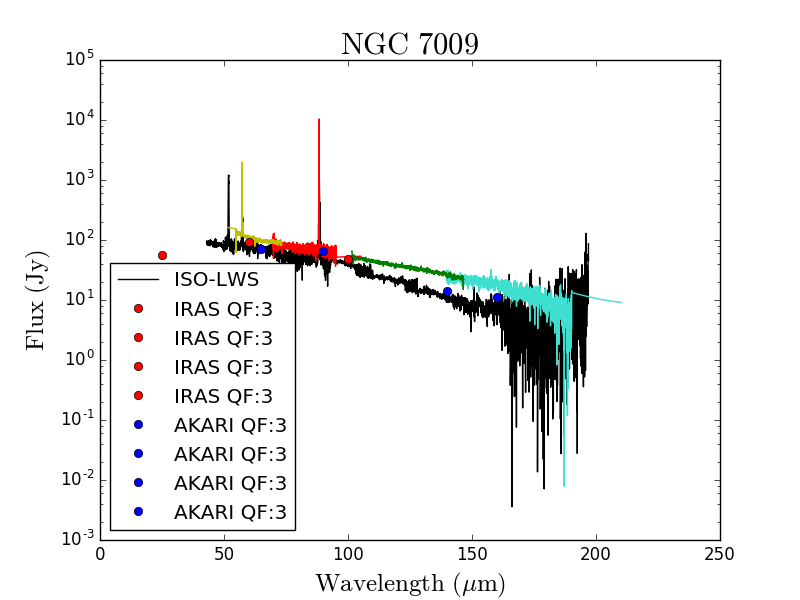}
  %\caption{1a}
  \label{fig:sfig1a}
\end{subfigure}
\begin{subfigure}{.55\textwidth}
\centering
  \includegraphics[width=0.9\linewidth,left]{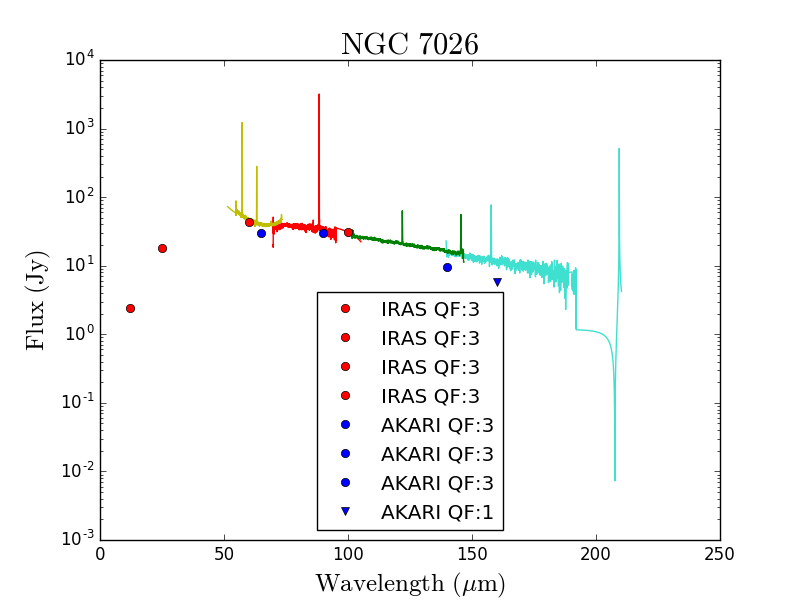}
  %\caption{1d}
  \label{fig:sfig1d}
\end{subfigure}
\caption{Continued.}
\end{figure*}

%----- Mispointing 3x3 corrected.

\begin{figure*}
\begin{subfigure}{.55\textwidth}
\centering
  \includegraphics[width=0.9\linewidth,left]{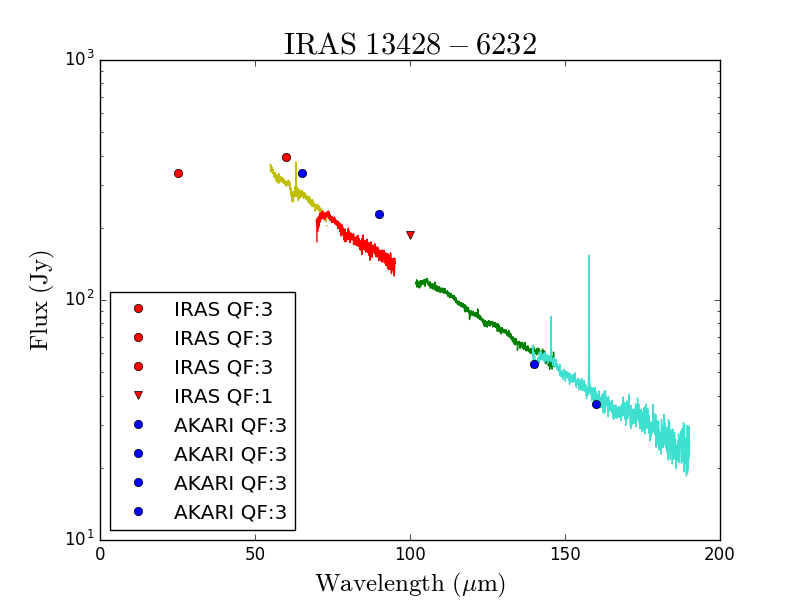}
  %\caption{1a}
  \label{fig:sfig1a}
\end{subfigure}
\begin{subfigure}{.55\textwidth}
\centering
  \includegraphics[width=0.9\linewidth,left]{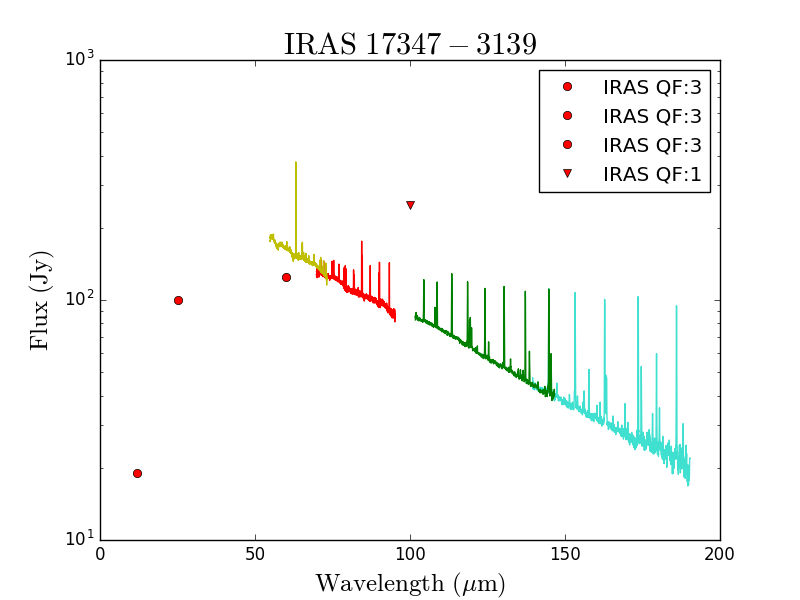}
  %\caption{1b}
  \label{fig:sfig1b}
\end{subfigure}
\begin{subfigure}{.55\textwidth}
\centering
  \includegraphics[width=0.9\linewidth,left]{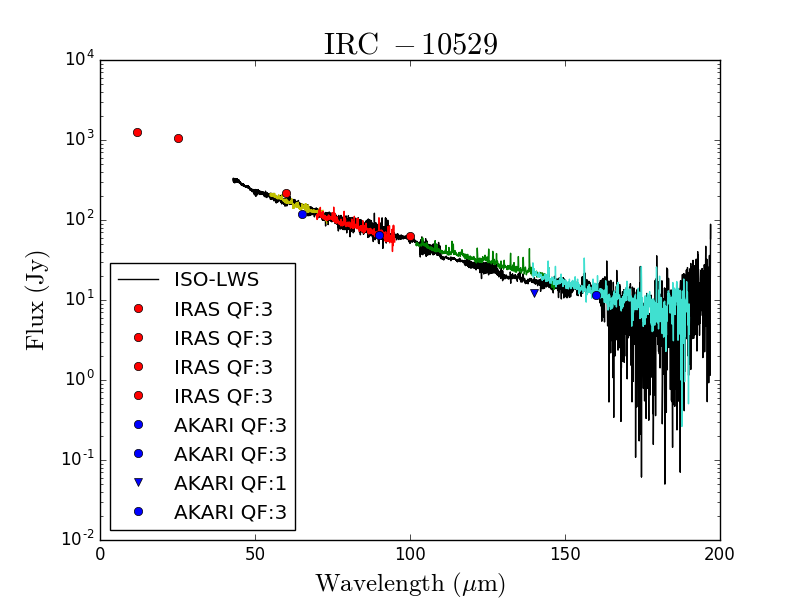}
  %\caption{1d}
  \label{fig:sfig1d}
\end{subfigure} 
\begin{subfigure}{.55\textwidth}
\centering
  \includegraphics[width=0.9\linewidth,left]{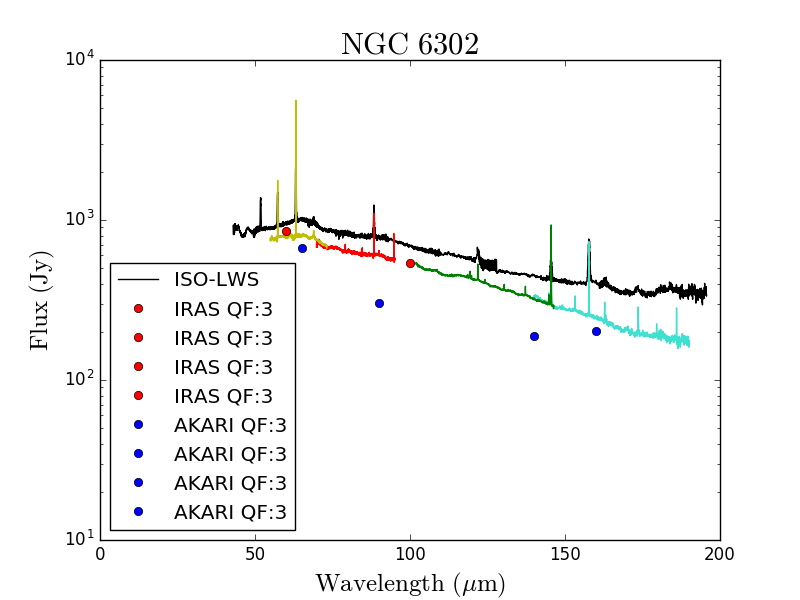}
  %\caption{1d}
  \label{fig:sfig1d}
\end{subfigure}
\begin{subfigure}{.55\textwidth}
\centering
  \includegraphics[width=0.9\linewidth,left]{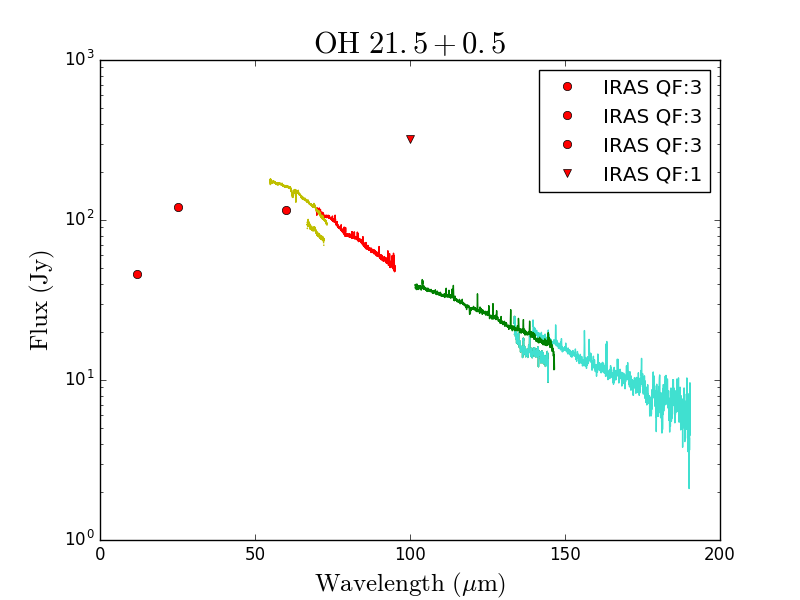}
  %\caption{1d}
  \label{fig:sfig1d}
\end{subfigure}
\caption{As in Fig. \ref{StandardSources} but for mispointed sources after applying point sources correction and 
semi-extended 3x3 correction.}
\label{Mispointed3x3}
\end{figure*}
\clearpage

%---- Mispointing No 3x3 corrected

\begin{figure*}
\begin{subfigure}{.55\textwidth}
\centering
  \includegraphics[width=0.9\linewidth,left]{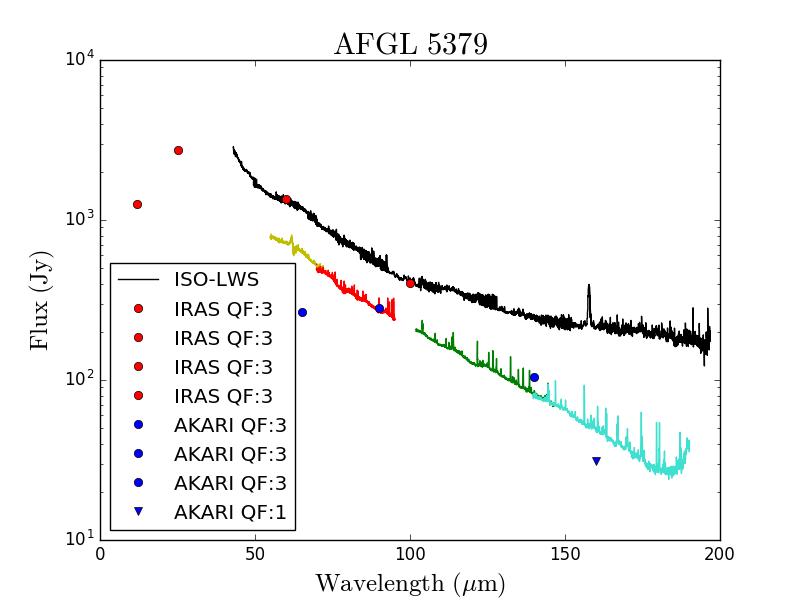}
  %\caption{1c}
  \label{fig:sfig1c}
\end{subfigure}
\begin{subfigure}{.55\textwidth}
\centering
  \includegraphics[width=0.9\linewidth,left]{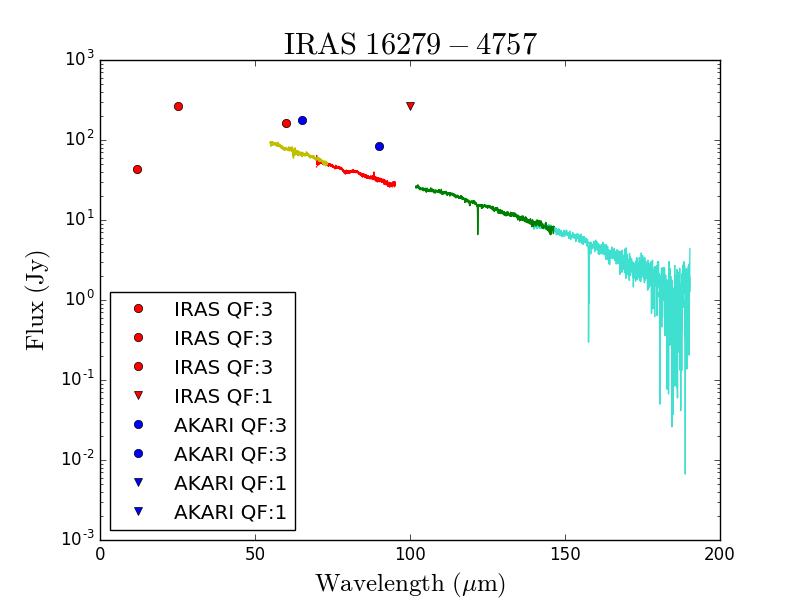}
  %\caption{1c}
  \label{fig:sfig1c}
\end{subfigure}
\begin{subfigure}{.55\textwidth}
\centering
  \includegraphics[width=0.9\linewidth,left]{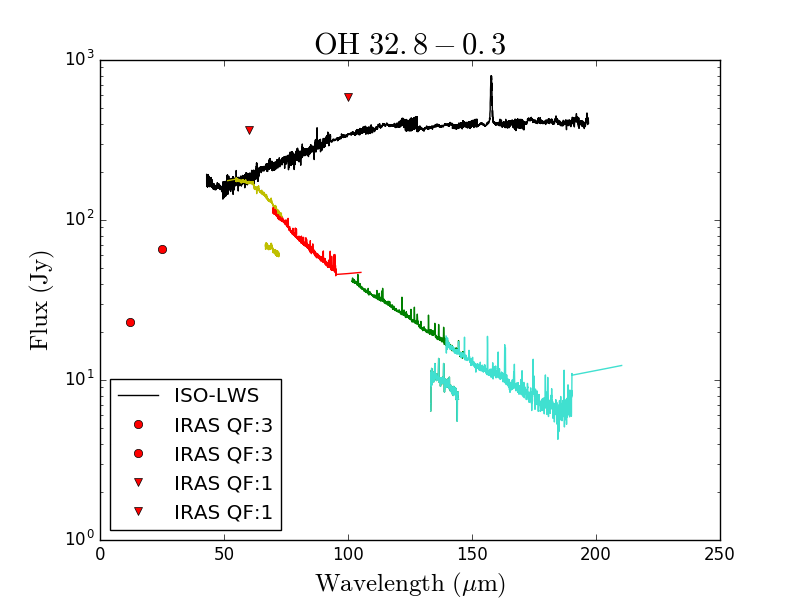}
  %\caption{1c}
  \label{fig:sfig1c}
\end{subfigure}
\caption{As in Fig. \ref{StandardSources} but for those sources corrected only by point sources correction. AFGL 5379 and 
IRAS 16279-4757 are mispointed targets for which the semi-extended 3x3 correction could not be applied as they 
are located in one of the outermost spaxels. The well-pointed object OH 32.8-0.3 shows, for unknown reasons
probably related to problems during the observations, negative fluxes in some of the 3x3 central spaxels, so the 
semi-extended 3x3 correction produced a wrong final 1D spectrum.}
\label{OnlyPSC}
\end{figure*}

%\begin{longtab}

\clearpage\onecolumn

\begin{longtable}{lcccccccc}
\caption{\label{THROESSampleInfo} Basic information about all observations reprocessed in THROES. Column 1: Target name;
Columns 2 and 3: Equatorial coordinates (J2000) in degrees; Columns 4, 5, and 6: Complementary data
for each observation in IRAS, AKARI, or ISO; Column 7: Comments on complementary information about the PACS observations 
object: 1) Incomplete PACS wavelength coverage; 2) Mispointing (target not in the central spaxel);
3) Semi-extended object (size $\sim$3x3 central spaxels); 4) Extended object (size $\sim$5x5 spaxels);
5) Contamination in ISO data due to interstellar emission; 6) Faint 
source ($\leq$10 Jy at 60 $\mu$m); 7) Multi epochs observations. If the observation presents a complete spectral 
coverage in PACS wavelength range, is well pointed, Chop/Nod and the source is point-like and intense,
there is no comment in the column (-); Column 8: ObsIds; Column 9: Date of observation.}\\
\hline
\hline
Target name & RA(deg) & Dec(deg) & IRAS & AKARI & ISO & Comments & ObsIDs & Date \\
\hline
\endfirsthead
\caption{Continued.}\\
\hline
\hline
Target name & RA(deg) & Dec(deg) & IRAS & AKARI & ISO & Comments & ObsIDs & Date \\
\hline
\endhead
\endfoot
\multirow{1}{*}{AC Her} & \multirow{1}{*}{277.5676} & \multirow{1}{*}{21.8668} & \multirow{1}{*}{Yes} & \multirow{1}{*}{Yes} & \multirow{1}{*}{Yes} & \multirow{1}{*}{1,6} & 1342208896 & 2010-11-13\\ \hline
\multirow{3}{*}{AFGL 618} & \multirow{3}{*}{70.7236} & \multirow{3}{*}{36.1147} & \multirow{3}{*}{Yes} & \multirow{3}{*}{Yes} & \multirow{3}{*}{Yes} & \multirow{3}{*}{-} & 1342225838 & 2011-08-07 \\
&&&&&&& 1342225839 & 2011-08-07 \\
&&&&&&& 1342225840 & 2011-08-07 \\ \hline
\multirow{1}{*}{AFGL 2019} & \multirow{1}{*}{268.3282} & \multirow{1}{*}{-26.9436} & \multirow{1}{*}{Yes} & \multirow{1}{*}{Yes} & \multirow{1}{*}{No} & \multirow{1}{*}{1,6} & 1342252253 & 2012-10-05 \\ \hline
\multirow{1}{*}{AFGL 2403} & \multirow{1}{*}{292.6228} & \multirow{1}{*}{19.8447} & \multirow{1}{*}{Yes} & \multirow{1}{*}{Yes} & \multirow{1}{*}{No} & \multirow{1}{*}{1} & 1342245226 & 2012-05-01 \\ \hline
\multirow{2}{*}{AFGL 2513} & \multirow{2}{*}{302.3093} & \multirow{2}{*}{31.4291} & \multirow{2}{*}{Yes} & \multirow{2}{*}{Yes} & \multirow{2}{*}{No} & \multirow{2}{*}{-} & 1342270010 & 2013-04-14 \\
&&&&&&& 1342269936 & 2013-04-12 \\ \hline
\multirow{3}{*}{AFGL 2688} & \multirow{3}{*}{315.5781} & \multirow{3}{*}{36.6938} & \multirow{3}{*}{No} & \multirow{3}{*}{No} & \multirow{3}{*}{Yes} & \multirow{3}{*}{-} & 1342199233 & 2010-06-26 \\
&&&&&&& 1342199234 & 2010-06-26 \\
&&&&&&& 1342199235 & 2010-06-27 \\ \hline
\multirow{3}{*}{AFGL 3068} & \multirow{3}{*}{349.8016} & \multirow{3}{*}{17.1931} & \multirow{3}{*}{Yes} & \multirow{3}{*}{Yes} & \multirow{3}{*}{Yes} & \multirow{3}{*}{-} & 1342199417 & 2010-06-30 \\
&&&&&&& 1342199418 & 2010-06-30 \\
&&&&&&& 1342257635 & 2012-12-21 \\ \hline
\multirow{2}{*}{AFGL 3116} & \multirow{2}{*}{353.6152} & \multirow{2}{*}{43.5506} & \multirow{2}{*}{Yes} & \multirow{2}{*}{Yes} & \multirow{2}{*}{No} & \multirow{2}{*}{-} & 1342212512 & 2011-01-11 \\
&&&&&&& 1342212513 & 2011-01-11 \\ \hline
\multirow{1}{*}{AFGL 4106} & \multirow{1}{*}{155.8311} & \multirow{1}{*}{-59.5346} & \multirow{1}{*}{Yes} & \multirow{1}{*}{Yes} & \multirow{1}{*}{Yes} & \multirow{1}{*}{1} & 1342207818 & 2010-11-03 \\ \hline
\multirow{1}{*}{AFGL 4202} & \multirow{1}{*}{223.1012} & \multirow{1}{*}{-62.0721} & \multirow{1}{*}{Yes} & \multirow{1}{*}{Yes} & \multirow{1}{*}{No} & \multirow{1}{*}{1,6} & 1342250004 & 2012-08-21 \\ \hline
\multirow{1}{*}{AFGL 4259} & \multirow{1}{*}{301.5947} & \multirow{1}{*}{27.0362} & \multirow{1}{*}{Yes} & \multirow{1}{*}{Yes} & \multirow{1}{*}{No} & \multirow{1}{*}{1,6} & 1342244918 & 2012-04-24 \\ \hline
\multirow{2}{*}{AFGL 5379} & \multirow{2}{*}{266.0942} & \multirow{2}{*}{-31.9276} & \multirow{2}{*}{Yes} & \multirow{2}{*}{Yes} & \multirow{2}{*}{Yes} & \multirow{2}{*}{2} & 1342228537 & 2011-09-13 \\
&&&&&&& 1342228538 & 2011-09-13 \\ \hline
\multirow{2}{*}{AFGL 6815} & \multirow{2}{*}{259.5830} & \multirow{2}{*}{-32.4556} & \multirow{2}{*}{Yes} & \multirow{2}{*}{Yes} & \multirow{2}{*}{No} & \multirow{2}{*}{-} & 1342216629 & 2011-03-22 \\
&&&&&&& 1342216630 & 2011-03-22 \\ \hline
\multirow{2}{*}{AQ Sgr} & \multirow{2}{*}{293.5791} & \multirow{2}{*}{-16.3741} & \multirow{2}{*}{Yes} & \multirow{2}{*}{Yes} & \multirow{2}{*}{No} & \multirow{2}{*}{-} & 1342268751 & 2013-03-28 \\
&&&&&&& 1342268752 & 2013-03-28 \\ \hline
\multirow{2}{*}{BD +30 3639} & \multirow{2}{*}{293.6884} & \multirow{2}{*}{30.5163} & \multirow{2}{*}{Yes} & \multirow{2}{*}{Yes} & \multirow{2}{*}{Yes} & \multirow{2}{*}{3} & 1342220600 & 2011-05-04 \\
&&&&&&& 1342220601 & 2011-05-04 \\ \hline
\multirow{2}{*}{CIT 6} & \multirow{2}{*}{154.0094} & \multirow{2}{*}{30.5718} & \multirow{2}{*}{Yes} & \multirow{2}{*}{No} & \multirow{2}{*}{Yes} & \multirow{2}{*}{-} & 1342197799 & 2010-06-05 \\
&&&&&&& 1342197800 & 2010-06-05 \\ \hline
\multirow{2}{*}{CPD-568032} & \multirow{2}{*}{257.2536} & \multirow{2}{*}{-56.9133} & \multirow{2}{*}{Yes} & \multirow{2}{*}{Yes} & \multirow{2}{*}{Yes} & \multirow{2}{*}{-} & 1342228201 & 2011-09-06 \\ 
&&&&&&& 1342228202 & 2011-09-06 \\ \hline
\multirow{1}{*}{CPD-642939} & \multirow{1}{*}{219.2920} & \multirow{1}{*}{-64.8013} & \multirow{1}{*}{Yes} & \multirow{1}{*}{Yes} & \multirow{1}{*}{No} & \multirow{1}{*}{1,6} & 1342250006 & 2012-08-21 \\ \hline
\multirow{2}{*}{EP Aqr} & \multirow{2}{*}{326.6327} & \multirow{2}{*}{-2.2127} & \multirow{2}{*}{Yes} & \multirow{2}{*}{Yes} & \multirow{2}{*}{No} & \multirow{2}{*}{-} & 1342270639 & 2013-04-20 \\
&&&&&&& 1342270684 & 2013-04-21 \\ \hline
\multirow{1}{*}{G Her} & \multirow{1}{*}{247.1606} & \multirow{1}{*}{41.8816} & \multirow{1}{*}{Yes} & \multirow{1}{*}{Yes} & \multirow{1}{*}{No} & \multirow{1}{*}{1} & 1342247780 & 2012-07-09 \\ \hline
\multirow{2}{*}{HD 56126} & \multirow{2}{*}{109.0427} & \multirow{2}{*}{9.9966} & \multirow{2}{*}{Yes} & \multirow{2}{*}{Yes} & \multirow{2}{*}{Yes} & \multirow{2}{*}{-} & 1342220930 & 2011-04-30\\
&&&&&&& 1342220931 & 2011-04-30 \\ \hline
\multirow{2}{*}{HD 161796} & \multirow{2}{*}{266.2311} & \multirow{2}{*}{50.0443} & \multirow{2}{*}{Yes} & \multirow{2}{*}{Yes} & \multirow{2}{*}{Yes} & \multirow{2}{*}{-} & 1342208881 & 2010-11-12 \\
&&&&&&& 1342208882 & 2010-11-12 \\ \hline
\multirow{2}{*}{HD 235858} & \multirow{2}{*}{337.2932} & \multirow{2}{*}{54.8517} & \multirow{2}{*}{Yes} & \multirow{2}{*}{Yes} & \multirow{2}{*}{No} & \multirow{2}{*}{-} & 1342196686 & 2010-05-18 \\
&&&&&&& 1342196687 & 2010-05-18 \\ \hline
\multirow{1}{*}{HD 331319} & \multirow{1}{*}{297.3731} & \multirow{1}{*}{31.4545} & \multirow{1}{*}{Yes} & \multirow{1}{*}{Yes} & \multirow{1}{*}{No} & \multirow{1}{*}{1} & 1342245232 & 2012-05-01 \\ \hline
\multirow{1}{*}{Hen 2-90} & \multirow{1}{*}{197.4009} & \multirow{1}{*}{-61.3266} & \multirow{1}{*}{Yes} & \multirow{1}{*}{Yes} & \multirow{1}{*}{Yes} & \multirow{1}{*}{1,5,6} & 1342248308 & 2012-07-16 \\ \hline
\multirow{3}{*}{Hen 2-113} & \multirow{3}{*}{224.9730} & \multirow{3}{*}{-54.3020} & \multirow{3}{*}{Yes} & \multirow{3}{*}{Yes} & \multirow{3}{*}{Yes} & \multirow{3}{*}{-} & 1342225142 & 2011-08-02 \\
&&&&&&& 1342225143 & 2011-08-02 \\ 
&&&&&&& 1342249211 & 2012-08-07 \\ \hline
\multirow{2}{*}{Hen 3-401} & \multirow{2}{*}{154.8852} & \multirow{2}{*}{-60.2248} & \multirow{2}{*}{Yes} & \multirow{2}{*}{Yes} & \multirow{2}{*}{No} & \multirow{2}{*}{-} & 1342225588 & 2011-08-03  \\
&&&&&&& 1342225589 & 2011-08-03  \\ \hline
\multirow{2}{*}{Hen 3-1475} & \multirow{2}{*}{266.3090} & \multirow{2}{*}{-17.9462} & \multirow{2}{*}{Yes} & \multirow{2}{*}{Yes} & \multirow{2}{*}{Yes} & \multirow{2}{*}{-} & 1342229719 & 2011-09-24 \\
&&&&&&& 1342229720 & 2011-09-24 \\ \hline 
\multirow{2}{*}{IC 418} & \multirow{2}{*}{81.8675} & \multirow{2}{*}{-12.6973} & \multirow{2}{*}{Yes} & \multirow{2}{*}{No} & \multirow{2}{*}{Yes} & \multirow{2}{*}{3} & 1342265942 & 2013-03-04 \\
&&&&&&& 1342265943 & 2013-03-04 \\ \hline
\multirow{1}{*}{IRAS 07027-7934} & \multirow{1}{*}{104.8599} & \multirow{1}{*}{-79.6463} & \multirow{1}{*}{Yes} & \multirow{1}{*}{Yes} & \multirow{1}{*}{Yes} & \multirow{1}{*}{1,6} & 1342245248 & 2012-05-02 \\ \hline
\multirow{3}{*}{IRAS 08011-3627} & \multirow{3}{*}{120.7568} & \multirow{3}{*}{-36.5966} & \multirow{3}{*}{Yes} & \multirow{3}{*}{Yes} & \multirow{3}{*}{No} & \multirow{3}{*}{-} & 1342210381 & 2010-11-27 \\
&&&&&&& 1342256781 & 2012-12-04 \\
&&&&&&& 1342256782 & 2012-12-04 \\ \hline
\multirow{2}{*}{IRAS 08544-4431} & \multirow{2}{*}{134.0591} & \multirow{2}{*}{-44.7196} & \multirow{2}{*}{Yes} & \multirow{2}{*}{Yes} & \multirow{2}{*}{No} & \multirow{2}{*}{-} & 1342245956 & 2012-05-21 \\
&&&&&&& 1342245957 & 2012-05-21 \\ \hline
\multirow{3}{*}{IRAS 09256-6324} & \multirow{3}{*}{141.7220} & \multirow{3}{*}{-63.6302} & \multirow{3}{*}{Yes} & \multirow{3}{*}{Yes} & \multirow{3}{*}{No} & \multirow{3}{*}{-} & 1342210386 & 2010-11-27 \\
&&&&&&& 1342248357 & 2012-07-20 \\
&&&&&&& 1342248358 & 2012-07-20 \\ \hline
\multirow{2}{*}{IRAS 09371+1212} & \multirow{2}{*}{144.9749} & \multirow{2}{*}{11.9814} & \multirow{2}{*}{Yes} & \multirow{2}{*}{Yes} & \multirow{2}{*}{No} & \multirow{2}{*}{-} & 1342245647 & 2012-05-12 \\
&&&&&&& 1342245648 & 2012-05-12 \\ \hline
\multirow{2}{*}{IRAS 09425-6040} & \multirow{2}{*}{146.0066} & \multirow{2}{*}{-60.9072} & \multirow{2}{*}{Yes} & \multirow{2}{*}{Yes} & \multirow{2}{*}{No} & \multirow{2}{*}{-} & 1342225563 & 2011-07-25 \\
&&&&&&& 1342225564 & 2011-07-25 \\ \hline
\multirow{3}{*}{IRAS 10456-5712} & \multirow{3}{*}{161.91} & \multirow{3}{*}{-57.4674} & \multirow{3}{*}{Yes} & \multirow{3}{*}{Yes} & \multirow{3}{*}{No} & \multirow{3}{*}{-} & 1342211823 & 2010-12-28 \\
&&&&&&& 1342248928 & 2012-07-31 \\
&&&&&&& 1342248929 & 2012-07-31 \\ \hline
\multirow{2}{*}{IRAS 13428-6232} & \multirow{2}{*}{206.5854} & \multirow{2}{*}{-62.8000} & \multirow{2}{*}{Yes} & \multirow{2}{*}{Yes} & \multirow{2}{*}{No} & \multirow{2}{*}{2} & 1342212212 & 2010-12-31 \\
&&&&&&& 1342212213 & 2010-12-31 \\ \hline
\multirow{2}{*}{IRAS 15194-5115} & \multirow{2}{*}{230.7704} & \multirow{2}{*}{-51.4330} & \multirow{2}{*}{Yes} & \multirow{2}{*}{Yes} & \multirow{2}{*}{Yes} & \multirow{2}{*}{-} & 1342215685 & 2011-03-10 \\
&&&&&&& 1342215686 & 2011-03-10 \\ \hline
\multirow{1}{*}{IRAS 16122-5128} & \multirow{1}{*}{244.0190} & \multirow{1}{*}{-51.5989} & \multirow{1}{*}{Yes} & \multirow{1}{*}{Yes} & \multirow{1}{*}{No} & \multirow{1}{*}{1} & 1342240167 & 2012-02-18 \\ \hline
\multirow{2}{*}{IRAS 16279-4757} & \multirow{2}{*}{247.9087} & \multirow{2}{*}{-48.0677} & \multirow{2}{*}{Yes} & \multirow{2}{*}{Yes} & \multirow{2}{*}{No} & \multirow{2}{*}{2} & 1342228203 & 2011-09-06 \\
&&&&&&& 1342228204 & 2011-09-06 \\ \hline
\multirow{2}{*}{IRAS 16342-3814} & \multirow{2}{*}{249.417} & \multirow{2}{*}{-38.3380} & \multirow{2}{*}{Yes} & \multirow{2}{*}{Yes} & \multirow{2}{*}{Yes} & \multirow{2}{*}{-} & 1342216627 & 2011-03-22 \\
&&&&&&& 1342216628 & 2011-03-22 \\ \hline
\multirow{2}{*}{IRAS 16594-4656} & \multirow{2}{*}{255.7917} & \multirow{2}{*}{-47.0076} & \multirow{2}{*}{Yes} & \multirow{2}{*}{Yes} & \multirow{2}{*}{Yes} & \multirow{2}{*}{5} & 1342228414 & 2011-09-10 \\
&&&&&&& 1342228415 & 2011-09-10 \\ \hline
\multirow{1}{*}{IRAS 17010-3840} & \multirow{1}{*}{256.1179} & \multirow{1}{*}{-38.7396} & \multirow{1}{*}{Yes} & \multirow{1}{*}{Yes} & \multirow{1}{*}{No} & \multirow{1}{*}{1,6} & 1342216176 & 2011-03-16 \\ \hline
\multirow{1}{*}{IRAS 17251-2821} & \multirow{1}{*}{262.0770} & \multirow{1}{*}{-28.3991} & \multirow{1}{*}{Yes} & \multirow{1}{*}{Yes} & \multirow{1}{*}{No} & \multirow{1}{*}{1,6} & 1342228535 & 2011-09-13 \\ \hline
\multirow{1}{*}{IRAS 17276-2846} & \multirow{1}{*}{262.7012} & \multirow{1}{*}{-28.8172} & \multirow{1}{*}{Yes} & \multirow{1}{*}{Yes} & \multirow{1}{*}{No} & \multirow{1}{*}{1,6} & 1342228536 & 2011-09-13 \\ \hline
\multirow{1}{*}{IRAS 17323-2424} & \multirow{1}{*}{263.8583} & \multirow{1}{*}{-24.4422} & \multirow{1}{*}{Yes} & \multirow{1}{*}{Yes} & \multirow{1}{*}{No} & \multirow{1}{*}{1,6} & 1342228532 & 2011-09-13 \\ \hline
\multirow{2}{*}{IRAS 17347-3139} & \multirow{2}{*}{264.5054} & \multirow{2}{*}{-31.6827} & \multirow{2}{*}{Yes} & \multirow{2}{*}{No} & \multirow{2}{*}{No} & \multirow{2}{*}{2} & 1342229696 & 2011-09-24 \\
&&&&&&& 1342229697 & 2011-09-24 \\ \hline
\multirow{1}{*}{IRAS 17521-2938} & \multirow{1}{*}{268.8404} & \multirow{1}{*}{-29.6536} & \multirow{1}{*}{Yes} & \multirow{1}{*}{No} & \multirow{1}{*}{No} & \multirow{1}{*}{1,6} & 1342229802 & 2011-09-27 \\ \hline
\multirow{1}{*}{IRAS 18433-0228} & \multirow{1}{*}{281.4800} & \multirow{1}{*}{-2.4189} & \multirow{1}{*}{Yes} & \multirow{1}{*}{No} & \multirow{1}{*}{No} & \multirow{1}{*}{1} & 1342231743 & 2011-11-01 \\ \hline
\multirow{2}{*}{IRAS 18488-0107} & \multirow{2}{*}{282.8592} & \multirow{2}{*}{-1.0645} & \multirow{2}{*}{Yes} & \multirow{2}{*}{No} & \multirow{2}{*}{No} & \multirow{2}{*}{-} & 1342268791 & 2013-03-30 \\
&&&&&&& 1342268792 & 2013-03-30\\ \hline
\multirow{2}{*}{IRAS 19067+0811} & \multirow{2}{*}{287.2845} & \multirow{2}{*}{8.2770} & \multirow{2}{*}{Yes} & \multirow{2}{*}{Yes} & \multirow{2}{*}{No} & \multirow{2}{*}{-} & 1342268797 & 2013-03-31 \\
&&&&&&& 1342268798 & 2013-03-31 \\ \hline
\multirow{1}{*}{IRAS 19306+1407} & \multirow{1}{*}{293.2295} & \multirow{1}{*}{14.2269} & \multirow{1}{*}{Yes} & \multirow{1}{*}{Yes} & \multirow{1}{*}{No} & \multirow{1}{*}{1,6} & 1342244921 & 2012-04-24 \\ \hline
\multirow{2}{*}{IRAS 19474-0744} & \multirow{2}{*}{297.5264} & \multirow{2}{*}{-7.6145} & \multirow{2}{*}{Yes} & \multirow{2}{*}{Yes} & \multirow{2}{*}{No} & \multirow{2}{*}{-} & 1342268638 & 2013-03-25 \\
&&&&&&& 1342268449 & 2013-03-26 \\ \hline
\multirow{2}{*}{IRAS 20000+3239} & \multirow{2}{*}{300.4980} & \multirow{2}{*}{32.7924} & \multirow{2}{*}{Yes} & \multirow{2}{*}{Yes} & \multirow{2}{*}{No} & \multirow{2}{*}{-} & 1342270612 & 2013-04-18 \\
&&&&&&& 1342270344 & 2013-04-17 \\ \hline
\multirow{2}{*}{IRAS 20038-2722} & \multirow{2}{*}{301.7301} & \multirow{2}{*}{-27.2249} & \multirow{2}{*}{Yes} & \multirow{2}{*}{Yes} & \multirow{2}{*}{No} & \multirow{2}{*}{-} & 1342268730 & 2013-03-29 \\
&&&&&&& 1342268569 & 2013-03-27 \\ \hline
\multirow{2}{*}{IRAS 21282+5050} & \multirow{2}{*}{322.4934} & \multirow{2}{*}{51.0666} & \multirow{2}{*}{Yes} & \multirow{2}{*}{Yes} & \multirow{2}{*}{Yes} & \multirow{2}{*}{5} & 1342223375 & 2011-06-30 \\
&&&&&&& 1342220741 & 2011-05-12 \\ \hline
\multirow{2}{*}{IRAS 21554+6204} & \multirow{2}{*}{329.2424} & \multirow{2}{*}{62.3121} & \multirow{2}{*}{Yes} & \multirow{2}{*}{Yes} & \multirow{2}{*}{No} & \multirow{2}{*}{1,6} & 1342245813 & 2012-05-15 \\
&&&&&&& 1342247150 & 2012-06-20 \\ \hline
\multirow{2}{*}{IRAS 22036+5306} & \multirow{2}{*}{331.3761} & \multirow{2}{*}{53.3591} & \multirow{2}{*}{Yes} & \multirow{2}{*}{Yes} & \multirow{2}{*}{Yes} & \multirow{2}{*}{5} & 1342221882 & 2011-05-29 \\
&&&&&&& 1342221883 & 2011-05-29 \\ \hline 
\multirow{4}{*}{IRC +10216} & \multirow{4}{*}{146.9892} & \multirow{4}{*}{13.2787} & \multirow{4}{*}{Yes} & \multirow{4}{*}{No} & \multirow{4}{*}{No} & \multirow{4}{*}{1,7} & 1342245395 & 2012-05-05 \\
&&&&&&& 1342221889 & 2011-05-29 \\
&&&&&&& 1342253754 & 2012-10-21 \\
&&&&&&& 1342256262 & 2012-11-30 \\ \hline
\multirow{2}{*}{IRC -10529\footnote{Also appears as V1300 Aql. We keep two different entries as the coordinates are different}} & \multirow{2}{*}{302.6142} & \multirow{2}{*}{-6.2710} & \multirow{2}{*}{Yes} & \multirow{2}{*}{Yes} & \multirow{2}{*}{Yes} & \multirow{2}{*}{2} & 1342208931 & 2010-11-14 \\
&&&&&&& 1342208932 & 2010-11-14 \\ \hline
\multirow{1}{*}{IRC +50137} & \multirow{1}{*}{77.831} & \multirow{1}{*}{52.8758} & \multirow{1}{*}{Yes} & \multirow{1}{*}{Yes} & \multirow{1}{*}{No} & \multirow{1}{*}{1} & 1342249315 & 2012-08-10 \\ \hline \\ \\
\multirow{2}{*}{$\chi$ Cyg} & \multirow{2}{*}{297.6413} & \multirow{2}{*}{32.9140} & \multirow{2}{*}{Yes} & \multirow{2}{*}{Yes} & \multirow{2}{*}{Yes} & \multirow{2}{*}{-} & 1342198176 & 2010-06-02 \\
&&&&&&& 1342198177 & 2010-06-02 \\ \hline
\multirow{1}{*}{MGE 4602} & \multirow{1}{*}{271.1620} & \multirow{1}{*}{-20.6242} & \multirow{1}{*}{Yes} & \multirow{1}{*}{Yes} & \multirow{1}{*}{No} & \multirow{1}{*}{1,6} & 1342243506 & 2012-03-24 \\ \hline
\multirow{2}{*}{MWC 922} & \multirow{2}{*}{275.3162} & \multirow{2}{*}{-13.0241} & \multirow{2}{*}{Yes} & \multirow{2}{*}{Yes} & \multirow{2}{*}{Yes} & \multirow{2}{*}{5} & 1342229805 & 2011-09-27 \\
&&&&&&& 1342229806 & 2011-09-27 \\ \hline
\multirow{2}{*}{Mz 3} & \multirow{2}{*}{244.3058} & \multirow{2}{*}{-51.9862} & \multirow{2}{*}{Yes} & \multirow{2}{*}{Yes} & \multirow{2}{*}{Yes} & \multirow{2}{*}{4} & 1342243109 & 2012-03-21 \\
&&&&&&& 1342243110 & 2012-03-21 \\ \hline
\multirow{2}{*}{NGC 40} & \multirow{2}{*}{3.2542} & \multirow{2}{*}{72.5219} & \multirow{2}{*}{Yes} & \multirow{2}{*}{Yes} & \multirow{2}{*}{Yes} & \multirow{2}{*}{4,6} & 1342236879 & 2012-01-08 \\
&&&&&&& 1342236880 & 2012-01-08 \\ \hline
\multirow{2}{*}{NGC 2392} & \multirow{2}{*}{112.2948} & \multirow{2}{*}{20.9118} & \multirow{2}{*}{Yes} & \multirow{2}{*}{Yes} & \multirow{2}{*}{No} & \multirow{2}{*}{6} & 1342229792 & 2011-09-26  \\
&&&&&&& 1342229816 & 2011-09-27 \\ \hline
\multirow{2}{*}{NGC 3242} & \multirow{2}{*}{156.1920} & \multirow{2}{*}{-18.6411} & \multirow{2}{*}{Yes} & \multirow{2}{*}{Yes} & \multirow{2}{*}{Yes} & \multirow{2}{*}{4,6} & 1342232278 & 2011-11-12 \\
&&&&&&& 1342232279 & 2011-11-12 \\ \hline
\multirow{2}{*}{NGC 6302} & \multirow{2}{*}{258.4342} & \multirow{2}{*}{-37.1044} & \multirow{2}{*}{Yes} & \multirow{2}{*}{Yes} & \multirow{2}{*}{Yes} & \multirow{2}{*}{2} & 1342230150 & 2011-10-05 \\
&&&&&&& 1342230151 & 2011-10-05 \\ \hline
\multirow{2}{*}{NGC 6445} & \multirow{2}{*}{267.3133} & \multirow{2}{*}{-20.0095} & \multirow{2}{*}{Yes} & \multirow{2}{*}{Yes} & \multirow{2}{*}{Yes} & \multirow{2}{*}{4,6} & 1342242440 & 2012-03-26 \\
&&&&&&& 1342242441 & 2012-03-26 \\ \hline
\multirow{2}{*}{NGC 6537} & \multirow{2}{*}{271.3045} & \multirow{2}{*}{-19.8430} & \multirow{2}{*}{No} & \multirow{2}{*}{Yes} & \multirow{2}{*}{Yes} & \multirow{2}{*}{3,5} & 1342231322 & 2011-10-22 \\
&&&&&&& 1342231323 & 2011-10-22 \\ \hline
\multirow{3}{*}{NGC 6543} & \multirow{3}{*}{269.6385} & \multirow{3}{*}{66.6330} & \multirow{3}{*}{No} & \multirow{3}{*}{Yes} & \multirow{3}{*}{Yes} & \multirow{3}{*}{4} & 1342238388 & 2012-01-29 \\
&&&&&&& 1342238389 & 2012-01-29 \\
&&&&&&& 1342212264 & 2011-01-02 \\ \hline
\multirow{2}{*}{NGC 6543 WKnot} & \multirow{2}{*}{269.5722} & \multirow{2}{*}{66.6356} & \multirow{2}{*}{No} & \multirow{2}{*}{No} & \multirow{2}{*}{No} & \multirow{2}{*}{-} & 1342235679 & 2011-12-27 \\
&&&&&&& 1342235680 & 2011-12-27 \\ \hline
\multirow{1}{*}{NGC 6720} & \multirow{1}{*}{283.3961} & \multirow{1}{*}{33.0291} & \multirow{1}{*}{Yes} & \multirow{1}{*}{Yes} & \multirow{1}{*}{Yes} & \multirow{1}{*}{1,6} & 1342208920 & 2010-11-14 \\ \hline
\multirow{2}{*}{NGC 6720 OFFCenter} & \multirow{2}{*}{283.3937} & \multirow{2}{*}{33.0326} & \multirow{2}{*}{Yes} & \multirow{2}{*}{Yes} & \multirow{2}{*}{Yes} & \multirow{2}{*}{-} & 1342233716 & 2011-12-07 \\
&&&&&&& 1342233717 & 2011-12-07 \\ \hline
\multirow{2}{*}{NGC 6781} & \multirow{2}{*}{289.6170} & \multirow{2}{*}{6.5386} & \multirow{2}{*}{Yes} & \multirow{2}{*}{No} & \multirow{2}{*}{Yes} & \multirow{2}{*}{4,6} & 1342230999 & 2011-10-14 \\
&&&&&&& 1342231000 & 2011-10-14 \\ \hline
\multirow{2}{*}{NGC 6781 Rim} & \multirow{2}{*}{289.6313} & \multirow{2}{*}{6.5388} & \multirow{2}{*}{No} & \multirow{2}{*}{No} & \multirow{2}{*}{No} & \multirow{2}{*}{-} & 1342231001 & 2011-10-15 \\
&&&&&&& 1342231002 & 2011-10-15 \\ \hline
\multirow{2}{*}{NGC 6826} & \multirow{2}{*}{296.2006} & \multirow{2}{*}{50.5250} & \multirow{2}{*}{Yes} & \multirow{2}{*}{Yes} & \multirow{2}{*}{Yes} & \multirow{2}{*}{3,6} & 1342238926 & 2012-02-10 \\
&&&&&&& 1342238927 & 2012-02-10 \\ \hline
\multirow{2}{*}{NGC 6826 Rim} & \multirow{2}{*}{296.2207} & \multirow{2}{*}{50.5290} & \multirow{2}{*}{No} & \multirow{2}{*}{No} & \multirow{2}{*}{No} & \multirow{2}{*}{-} & 1342235850 & 2011-12-31 \\
&&&&&&& 1342235851 & 2011-12-31 \\ \hline
\multirow{2}{*}{NGC 7009} & \multirow{2}{*}{316.0450} & \multirow{2}{*}{-11.3635} & \multirow{2}{*}{Yes} & \multirow{2}{*}{Yes} & \multirow{2}{*}{Yes} & \multirow{2}{*}{4} & 1342232300 & 2011-11-13 \\
&&&&&&& 1342232301 & 2011-11-13 \\ \hline
\multirow{2}{*}{NGC 7026} & \multirow{2}{*}{316.5773} & \multirow{2}{*}{47.8519} & \multirow{2}{*}{Yes} & \multirow{2}{*}{Yes} & \multirow{2}{*}{No} & \multirow{2}{*}{3} & 1342234268 & 2011-12-06 \\
&&&&&&& 1342234269 & 2011-12-06 \\ \hline
\multirow{2}{*}{NML Cyg} & \multirow{2}{*}{311.606} & \multirow{2}{*}{40.1165} & \multirow{2}{*}{No} & \multirow{2}{*}{Yes} & \multirow{2}{*}{Yes} & \multirow{2}{*}{-} & 1342198174 & 2010-06-02 \\
&&&&&&& 1342198175 & 2010-06-02 \\ \hline
\multirow{3}{*}{NML Tau} & \multirow{3}{*}{58.3701} & \multirow{3}{*}{11.4062} & \multirow{3}{*}{Yes} & \multirow{3}{*}{Yes} & \multirow{3}{*}{Yes} & \multirow{3}{*}{-} & 1342203679 & 2010-08-28 \\
&&&&&&& 1342203680 & 2010-08-28 \\
&&&&&&& 1342203681 & 2010-08-28 \\ \hline
\multirow{3}{*}{OH 21.5+0.5} & \multirow{3}{*}{277.1316} & \multirow{3}{*}{-9.9696} & \multirow{3}{*}{No} & \multirow{3}{*}{No} & \multirow{3}{*}{No} & \multirow{3}{*}{2,7} & 1342242438 & 2012-03-26 \\
&&&&&&& 1342268778 & 2013-03-29 \\
&&&&&&& 1342268748 & 2013-03-30 \\ \hline
\multirow{2}{*}{OH 26.5+0.6} & \multirow{2}{*}{279.3852} & \multirow{2}{*}{-5.3998} & \multirow{2}{*}{Yes} & \multirow{2}{*}{Yes} & \multirow{2}{*}{Yes} & \multirow{2}{*}{5} & 1342207776 & 2010-10-31 \\
&&&&&&& 1342207777 & 2010-10-31 \\ \hline
\multirow{3}{*}{OH 30.1-0.7} & \multirow{3}{*}{282.1746} & \multirow{3}{*}{-2.8411} & \multirow{3}{*}{Yes} & \multirow{3}{*}{Yes} & \multirow{3}{*}{No} & \multirow{3}{*}{-} & 1342216207 & 2011-03-04 \\
&&&&&&& 1342269304 & 2013-04-03 \\
&&&&&&& 1342269305 & 2013-04-03 \\ \hline 
\multirow{2}{*}{OH 30.7+0.4} & \multirow{2}{*}{281.5241} & \multirow{2}{*}{-1.9881} & \multirow{2}{*}{Yes} & \multirow{2}{*}{No} & \multirow{2}{*}{No} & \multirow{2}{*}{6} & 1342268789 & 2013-03-30 \\
&&&&&&& 1342268790 & 2013-03-30 \\ \hline 
\multirow{3}{*}{OH 32.8-0.3} & \multirow{3}{*}{283.0924} & \multirow{3}{*}{-0.2371} & \multirow{3}{*}{Yes} & \multirow{3}{*}{No} & \multirow{3}{*}{Yes} & \multirow{3}{*}{5,7} & 1342209738 & 2010-11-07 \\
&&&&&&& 1342268793 & 2013-03-30 \\
&&&&&&& 1342268794 & 2013-03-30 \\ \hline
\multirow{1}{*}{OH 104.91+2.41} & \multirow{1}{*}{334.8645} & \multirow{1}{*}{59.8560} & \multirow{1}{*}{Yes} & \multirow{1}{*}{Yes} & \multirow{1}{*}{Yes} & \multirow{1}{*}{1} & 1342212261 & 2011-01-01 \\ \hline \\ \\
\multirow{2}{*}{OH 231.8+4.2} & \multirow{2}{*}{115.5701} & \multirow{2}{*}{-14.7144} & \multirow{2}{*}{Yes} & \multirow{2}{*}{Yes} & \multirow{2}{*}{No} & \multirow{2}{*}{-} & 1342196694 & 2010-05-19 \\
&&&&&&& 1342196695 & 2010-05-19 \\ \hline
\multirow{2}{*}{$o$ Cet} & \multirow{2}{*}{34.8366} & \multirow{2}{*}{-2.9776} & \multirow{2}{*}{Yes} & \multirow{2}{*}{Yes} & \multirow{2}{*}{No} & \multirow{2}{*}{-} & 1342213286 & 2011-01-25 \\
&&&&&&& 1342213287 & 2011-01-25 \\ \hline
\multirow{2}{*}{$\pi$ Gru} & \multirow{2}{*}{335.6842} & \multirow{2}{*}{-45.9479} & \multirow{2}{*}{Yes} & \multirow{2}{*}{Yes} & \multirow{2}{*}{Yes} & \multirow{2}{*}{-} & 1342210397 & 2010-11-28 \\
&&&&&&& 1342210398 & 2010-11-28 \\ \hline
\multirow{1}{*}{R Aql} & \multirow{1}{*}{286.5927} & \multirow{1}{*}{8.2300} & \multirow{1}{*}{Yes} & \multirow{1}{*}{Yes} & \multirow{1}{*}{Yes} & \multirow{1}{*}{1,5} & 1342243900 & 2012-04-08 \\ \hline
\multirow{2}{*}{R Cas} & \multirow{2}{*}{359.6036} & \multirow{2}{*}{51.3888} & \multirow{2}{*}{Yes} & \multirow{2}{*}{Yes} & \multirow{2}{*}{Yes} & \multirow{2}{*}{-} & 1342212576 & 2011-01-12 \\
&&&&&&& 1342212577 & 2011-01-12 \\ \hline
\multirow{2}{*}{R Dor} & \multirow{2}{*}{69.1899} & \multirow{2}{*}{-62.0771} & \multirow{2}{*}{Yes} & \multirow{2}{*}{Yes} & \multirow{2}{*}{Yes} & \multirow{2}{*}{3} & 1342197794 & 2010-06-05 \\
&&&&&&& 1342197795 & 2010-06-05 \\ \hline
\multirow{1}{*}{R Lep} & \multirow{1}{*}{74.9014} & \multirow{1}{*}{-14.8062} & \multirow{1}{*}{Yes} & \multirow{1}{*}{Yes} & \multirow{1}{*}{No} & \multirow{1}{*}{1,6} & 1342249509 & 2012-08-14 \\ \hline
\multirow{1}{*}{RAFGL 2374} & \multirow{1}{*}{290.4021} & \multirow{1}{*}{9.4656} & \multirow{1}{*}{Yes} & \multirow{1}{*}{Yes} & \multirow{1}{*}{No} & \multirow{1}{*}{1,6} & 1342244920 & 2012-04-24 \\ \hline
\multirow{2}{*}{Red Rectangle} & \multirow{2}{*}{94.9925} & \multirow{2}{*}{-10.6374} & \multirow{2}{*}{Yes} & \multirow{2}{*}{Yes} & \multirow{2}{*}{Yes} & \multirow{2}{*}{-} & 1342220928 & 2011-04-30 \\
&&&&&&& 1342220929 & 2011-04-30 \\ \hline
\multirow{1}{*}{RR Aql} & \multirow{1}{*}{299.4002} & \multirow{1}{*}{-1.8864} & \multirow{1}{*}{Yes} & \multirow{1}{*}{Yes} & \multirow{1}{*}{No} & \multirow{1}{*}{1} & 1342269414 & 2013-04-05 \\ \hline
\multirow{2}{*}{RT Cap} & \multirow{2}{*}{304.2772} & \multirow{2}{*}{-21.3179} & \multirow{2}{*}{Yes} & \multirow{2}{*}{Yes} & \multirow{2}{*}{No} & \multirow{2}{*}{6} & 1342269308 & 2013-04-03 \\
&&&&&&& 1342269355 & 2013-04-04 \\ \hline
\multirow{1}{*}{S Aur} & \multirow{1}{*}{81.7810} & \multirow{1}{*}{34.1496} & \multirow{1}{*}{Yes} & \multirow{1}{*}{Yes} & \multirow{1}{*}{No} & \multirow{1}{*}{1,6} & 1342250896 & 2012-09-11 \\ \hline
\multirow{1}{*}{ST Her} & \multirow{1}{*}{237.6942} & \multirow{1}{*}{48.4830} & \multirow{1}{*}{Yes} & \multirow{1}{*}{Yes} & \multirow{1}{*}{No} & \multirow{1}{*}{1,6} & 1342247537 & 2012-06-30 \\ \hline
\multirow{1}{*}{T Cep} & \multirow{1}{*}{317.3824} & \multirow{1}{*}{68.4908} & \multirow{1}{*}{Yes} & \multirow{1}{*}{Yes} & \multirow{1}{*}{No} & \multirow{1}{*}{1} & 1342246557 & 2012-06-01 \\ \hline
\multirow{3}{*}{T Mic} & \multirow{3}{*}{306.9799} & \multirow{3}{*}{-28.2610} & \multirow{3}{*}{Yes} & \multirow{3}{*}{Yes} & \multirow{3}{*}{No} & \multirow{3}{*}{-} & 1342268729 & 2013-03-29 \\
&&&&&&& 1342268788 & 2013-03-30 \\
&&&&&&& 1342269458 & 2013-04-06 \\ \hline
\multirow{2}{*}{Tc 1} & \multirow{2}{*}{266.3970} & \multirow{2}{*}{-46.0899} & \multirow{2}{*}{Yes} & \multirow{2}{*}{Yes} & \multirow{2}{*}{No} & \multirow{2}{*}{6} & 1342231319 & 2011-10-22 \\
&&&&&&& 1342231320 & 2011-10-22 \\ \hline
\multirow{2}{*}{TX Cam} & \multirow{2}{*}{75.2099} & \multirow{2}{*}{56.1812} & \multirow{2}{*}{Yes} & \multirow{2}{*}{Yes} & \multirow{2}{*}{Yes} & \multirow{2}{*}{3} & 1342225855 & 2011-08-08 \\
&&&&&&& 1342225856 & 2011-08-08 \\ \hline
\multirow{1}{*}{U Hya} & \multirow{1}{*}{159.3886} & \multirow{1}{*}{-13.3845} & \multirow{1}{*}{Yes} & \multirow{1}{*}{Yes} & \multirow{1}{*}{No} & \multirow{1}{*}{1,6} & 1342256947 & 2012-12-11 \\ \hline
\multirow{3}{*}{U Mon} & \multirow{3}{*}{112.6977} & \multirow{3}{*}{-9.7768} & \multirow{3}{*}{Yes} & \multirow{3}{*}{Yes} & \multirow{3}{*}{No} & \multirow{3}{*}{-} & 1342206993 & 2010-10-23 \\
&&&&&&& 1342245243 & 2012-05-02 \\
&&&&&&& 1342245244 & 2012-05-02 \\ \hline
\multirow{1}{*}{V384 Per} & \multirow{1}{*}{51.6229} & \multirow{1}{*}{47.5301} & \multirow{1}{*}{Yes} & \multirow{1}{*}{Yes} & \multirow{1}{*}{Yes} & \multirow{1}{*}{1,6} & 1342250572 & 2012-09-04 \\ \hline
\multirow{1}{*}{V438 Oph} & \multirow{1}{*}{258.6657} & \multirow{1}{*}{11.0694} & \multirow{1}{*}{Yes} & \multirow{1}{*}{Yes} & \multirow{1}{*}{No} & \multirow{1}{*}{1,6} & 1342252326 & 2012-10-07 \\ \hline
\multirow{2}{*}{V1300 Aql} & \multirow{2}{*}{302.6161} & \multirow{2}{*}{-6.2704} & \multirow{2}{*}{Yes} & \multirow{2}{*}{Yes} & \multirow{2}{*}{Yes} & \multirow{2}{*}{-} & 1342269516 & 2013-04-07 \\
&&&&&&& 1342269910 & 2013-04-11 \\ \hline
\multirow{2}{*}{V Cyg} & \multirow{2}{*}{310.3261} & \multirow{2}{*}{48.1413} & \multirow{2}{*}{Yes} & \multirow{2}{*}{Yes} & \multirow{2}{*}{Yes} & \multirow{2}{*}{5} & 1342208939 & 2010-11-15 \\
&&&&&&& 1342208940 & 2010-11-15 \\ \hline
\multirow{2}{*}{V Hya} & \multirow{2}{*}{162.9052} & \multirow{2}{*}{-21.2500} & \multirow{2}{*}{Yes} & \multirow{2}{*}{Yes} & \multirow{2}{*}{Yes} & \multirow{2}{*}{-} & 1342197790 & 2010-06-05 \\
&&&&&&& 1342197791 & 2010-06-05 \\ \hline
\multirow{2}{*}{W Aql} & \multirow{2}{*}{288.8476} & \multirow{2}{*}{-7.0471} & \multirow{2}{*}{Yes} & \multirow{2}{*}{Yes} & \multirow{2}{*}{Yes} & \multirow{2}{*}{-} & 1342209731 & 2010-11-06 \\
&&&&&&& 1342209732 & 2010-11-07 \\ \hline
\multirow{2}{*}{W Hya} & \multirow{2}{*}{207.2583} & \multirow{2}{*}{-28.3676} & \multirow{2}{*}{Yes} & \multirow{2}{*}{Yes} & \multirow{2}{*}{Yes} & \multirow{2}{*}{-} & 1342212604 & 2011-01-14 \\
&&&&&&& 1342223808 & 2011-07-09  \\ \hline
\multirow{1}{*}{W Ori} & \multirow{1}{*}{76.3488} & \multirow{1}{*}{1.1776} & \multirow{1}{*}{Yes} & \multirow{1}{*}{Yes} & \multirow{1}{*}{No} & \multirow{1}{*}{1,6} & 1342249503 & 2012-08-14 \\ \hline
\multirow{2}{*}{WX Psc} & \multirow{2}{*}{16.6082} & \multirow{2}{*}{12.5980} & \multirow{2}{*}{Yes} & \multirow{2}{*}{Yes} & \multirow{2}{*}{Yes} & \multirow{2}{*}{-} & 1342202121 & 2010-07-28 \\
&&&&&&& 1342202122 & 2010-07-28 \\ \hline
\multirow{3}{*}{X Her} & \multirow{3}{*}{240.6632} & \multirow{3}{*}{47.2403} & \multirow{3}{*}{Yes} & \multirow{3}{*}{Yes} & \multirow{3}{*}{No} & \multirow{3}{*}{-} & 1342197802 & 2010-06-05 \\
&&&&&&& 1342197803 & 2010-06-05 \\
&&&&&&& 1342202120 & 2010-07-28 \\ \hline
\multirow{1}{*}{Y CVn} & \multirow{1}{*}{191.2826} & \multirow{1}{*}{45.4402} & \multirow{1}{*}{Yes} & \multirow{1}{*}{Yes} & \multirow{1}{*}{Yes} & \multirow{1}{*}{1,6} & 1342254305 & 2012-11-02 \\ \hline
%\footnote{}

\end{longtable}

%\end{longtab}

%\begin{longtab}
\begin{landscape}
%\resizebox{\textwidth}{!}{
\begin{longtable}{l r r c c c c}% c c}
%\label{table:3}
\caption{\label{THROESSamplePhoto}. Photometric data of THROES targets. Target name (Col. 1), as it appears in HSA, 
galactic coordinates (Cols. 2 and 3), IRAS (100 $\mu$m) and AKARI (160 $\mu$m) photometric data  (Cols. 4 and 5), 
and PACS synthetic photometry at 100 and 160 $\mu$m (Cols. 6 and 7)}\\
\hline\hline
Target name & b(deg) & l(deg) & IRAS$_{100}$ (Jy) & AKARI$_{160}$ (Jy) & PACS$_{IRAS100}$ (Jy) & PACS$_{AKARI160}$ (Jy) \\%& PACS$_{100}$ (Jy) & PACS$_{160}$ (Jy) \\ %PACS$_{60}$ (Jy) 
\hline
\endfirsthead
\caption{Continued.}\\
\hline\hline
Target name & b(deg) & l(deg) & IRAS$_{100}$ (Jy) & AKARI$_{160}$ (Jy) & PACS$_{IRAS100}$ (Jy) & PACS$_{AKARI160}$ (Jy) \\%& PACS$_{100}$ (Jy) & PACS$_{160}$ (Jy) \\ %PACS$_{60}$ (Jy) & 
\hline
\endhead
\hline
\endfoot

AC Her & 14.2415 & 50.4928 & 8.0 & <3.2 & - & - \\%& 9.8[1] & 11.6[1]\\
AFGL 618 & -6.5275 & 166.4460 & 340.0 & 111.7 & 859.0 & 195.9 \\%& 538.5[2] & 163.5[3]\\
AFGL 2019 & -0.4324 & 2.5826 & <227.0 & 48.6 & - & - \\%& 27.8[1] & 11.7[1]\\
AFGL 2403 & 0.7353 & 54.9494 & <31.3 & <10.1 & - & - \\%& 13.6[1] & -23.7[1]\\
AFGL 2513 & -0.8653 & 69.3522 & 20.8 & <5.8 & 10.7 & 3.9 \\%& 7.3[2] & 2.6[3]\\
AFGL 2688 & -6.5032 & 80.1675 & <863.0 & - & 1969.1 & 467.8 \\%& 1196.2[2] & 408.6[3]\\
AFGL 3068 & -40.3540 & 93.5266 & 73.7 & 24.4 & 102.0 & 34.3 \\%& 83.3[2] & 30.5[3]\\
AFGL 3116 & -17.1467 & 108.4549 & 35.5 & 10.9 & 46.8 & 14.9 \\%& 36.0[2] & 12.7[3] \\
AFGL 4106 & -1.8789 & 285.1439 & 181.0 & 60.1 & - & - \\%& 228.2[1] & 56.4[1] \\
AFGL 4202 & -2.4514 & 316.5887 & <75.3 & <6.5 & - & - \\%& 12.8[1] & 6.2[1]\\
AFGL 4259 & -2.7141 & 65.3167 & 7.5 & <2.6 & - & - \\%& 11.3[1] & 19.8[1] \\
AFGL 5379 & -1.3330 & 357.3084 & 406.0 & <31.2 & - & - \\%& 208.7[2] & 49.9[2]\\
AFGL 6815 & 2.9842 & 353.8442 & 82.4 & <18.37 & 120.0 & 40.3 \\%& 99.4[2] & 32.5[3]\\
AQ Sgr & -16.7007 & 22.7377 & 5.9 & <0.5 & 1.3 & 0.4 \\%& 1.1[2] & 0.4[3]\\
BD +30 3639 & 5.0194 & 64.7853 & 70.1 & <23.1 & 70.6 & 21.0 \\%& 64.2[2] & 24.7[3]\\
CIT 6 & 55.9641 & 197.7148 & 86.1 & - & 83.5 & 24.2 \\%& 69.2[2] & 25.1[3]\\
CPD-568032 & -9.9082 & 332.9152 & 91.7 & 29.4 & 106.2 & 26.7 \\%& 98.5[2] & 32.1[3]\\
CPD-642939 & -4.2026 & 313.8867 & 20.6 & <4.7 & - & - \\%& 9.4[1] & -29.6[1]\\
EP Aqr & -39.2603 & 54.2000 & 16.4 & <4.9 & 15.0 & 4.7 \\%& 9.7[2] & 3.6[3]\\
G Her & 43.7146 & 66.1579 & 6.6 & <0.1 & - & - \\%& 5.7[1] & -10.4[1]\\
HD 56126 & 9.9947 & 206.7457 & 18.7 & <3.8 & 20.5 & 6.0 \\%& 15.7[2] & 4.6[3]\\
HD 161796 & 30.8696 & 77.1331 & 48.7 & 10.7 & 56.1 & 8.0 \\%& 47.8[2] & 13.0[3]\\
HD 235858 & -2.5181 & 103.3488 & 41.0 & 8.5 & 34.0 & 10.0 \\%& 26.3[2] & 8.5[3]\\
HD 331319 & 2.7327 & 67.1633 & 14.8 & <4.1 & - & - \\%& 18.3[1] & 20.8[1]\\
Hen 2-90 & 1.4683 & 305.1099 & <50.2 & <11.1 & - & - \\%& -5.7[1] & -26.0[1]\\
Hen 2-113 & 3.9885 & 321.0483 & 71.3 & 20.9 & 90.7 & 26.0 \\%& 82.9[2] & 25.8[3]\\
Hen 3-401 & -2.7162 & 285.1194 & 41.5 & 16.6 & 45.2 & 15.7 \\%& 44.5[2] & 16.8[3]\\
Hen 3-1475 & 5.7788 & 9.3644 & 33.4 & 12.1 & 36.2 & 9.7 \\%& 34.5[2] & 12.6[3]\\
IC 418 & -24.2837 & 215.2120 & 31.2 & - & 29.1 & 9.5 \\%& 19.4[2] & 8.8[3]\\
IRAS 07027-7934 & -26.2939 & 291.3749 & 13.6 & <4.4 & - & - \\%& 16.4[1] & 12.2[1]\\
IRAS 08011-3627 & -2.9959 & 253.0207 & 12.0 & <4.7 & 13.6 & 4.6 \\%& 11.4[2] & 3.9[3]\\
IRAS 08544-4431 & 0.3864 & 265.5011 & <28.4 & <3.2 & 29.3 & 11.5 \\%& 25.2[2] & 9.4[3]\\
IRAS 09256-6324 & -9.2362 & 282.4206 & 13.5 & <5.3 & 15.6 & 7.1 \\%& 13.2[2] & 6.3[3]\\
IRAS 09371+1212 & 42.7271 & 221.8893 & 28.2 & <11.6 & 32.7 & 15.5 \\%& 29.3[2] & 12.8[3]\\
IRAS 09425-6040 & -5.8842 & 282.0386 & 5.1 & <1.3 & 7.5 & 1.8 \\%& 6.3[2] & 2.1[3]\\
IRAS 10456-5712 & 1.4921 & 286.8707 & 10.9 & <5.1 & 15.6 & 5.8 \\%& 13.4[2] & 5.1[3]\\
IRAS 13428-6232 & -0.5939 & 309.1588 & <186.0 & 36.9 & - & - \\%& 47.2[2] & 17.8[2]\\
IRAS 15194-5115 & 4.6593 & 325.5339 & 51.0 & 15.0 & 37.0 & 11.4 \\%& 31.8[2] & 12.2[3]\\
IRAS 16122-5128 & -0.6095 & 331.8689 & <387.0 & <16.1 & - & - \\%& 6.5[1] & -8.3[1]\\
IRAS 16279-4757 & 0.0851 & 336.1433 & <265.0 & - & - & - \\%& 24.9[2] & 4.1[2]\\
IRAS 16342-3814 & 5.8464 & 344.0739 & 139.0 & 43.2 & 193.9 & 65.4 \\%& 183.0[2] & 66.8[3]\\
IRAS 16594-4656 & -3.2887 & 340.3924 & 34.4 & <10.6 & 45.7 & 10.8 \\%& 40.2[2] & 12.0[3]\\
IRAS 17010-3840 & 1.5481 & 347.1050 & <50.1 & <6.3 & - & - \\%& 48.7[1] & 82.8[1]\\
IRAS 17251-2821 & 3.4922 & 358.4142 & <19.9 & - & - & - \\%& -4.6[1] & -14.0[1]\\
IRAS 17276-2846 & 2.8046 & 358.3664 & <30.0 & - & - & - \\%& 1.9[1] & 0.9[1]\\
IRAS 17323-2424 & 4.3084 & 2.6112 & <13.5 & <0.7 & - & - \\%& -5.9[1] & -15.6[1]\\
IRAS 17347-3139 & -0.0557 & 356.8018 & <249.0 & - & - & - \\%& 66.9[2] & 23.9[2]\\
IRAS 17521-2938 & -2.1916 & 0.4718 & <110.0 & - & - & - \\%& 0.3[1] & -10.8[1]\\
IRAS 18433-0228 & 0.1239 & 30.1503 & <240.0 & - & - & - \\%& 7.6[1] & 1.4[1]\\
IRAS 18488-0107 & -0.4851 & 31.9844 & 72.8 & - & 20.9 & 6.9 \\%& 17.0[2] & 5.8[3]\\
IRAS 19067+0811 & -0.1323 & 42.3093 & <91.1 & - & 12.4 & 2.4 \\%& 9.3[2] & 1.7[3]\\
IRAS 19306+1407 & -2.4774 & 50.3034 & 10.0 & <7.6 & - & - \\%& 11.6[1] & 17.0[1]\\
IRAS 19474-0744 & -16.4859 & 32.7232 & 17.2 & <3.2 & 20.8 & 6.5 \\%& 12.6[2] & 3.9[3]\\
%IRAS 19312+1950 & 0.1853 & 55.3706 & 423.0 (24) [3] & 275.1 (31.8) [3] & 402.5 & 265.7 & 412.3[2] & 285.0[3]\\
IRAS 20000+3239 & 1.1609 & 69.6792 & <43.1 & - & 12.0 & 3.4 \\%& 8.1[2] & 2.6[3]\\
IRAS 20038-2722 & -27.6822 & 14.7708 & 13.6 & <0.8 & 9.8 & 3.3 \\%& 5.8[2] & 2.0[3]\\
IRAS 21282+5050 & -0.1178 & 93.9871 & 15.0 & <3.7 & 16.6 & 7.2 \\%& 14.4[2] & 5.8[3]\\
IRAS 21554+6204 & 5.9916 & 104.1301 & 15.6 & <3.3 & - & - \\%& 12.6[1] & 2.4[1]\\
IRAS 22036+5306 & -1.8353 & 99.6333 & 50.7 & 20.3 & 66.1 & 21.2 \\%& 61.0[2] & 24.9][3]\\
IRC +10216 & 45.0604 & 221.4467 & 922.0 & - & - & - \\%& 818.5[1] & 249.7[1]\\
IRC -10529\footnote{Also appears as V1300 Aql. We keep two different entries as the coordinates are different.} & -20.4151 & 36.3565 & 63.7 & 11.4 & - & - \\%& 27.0[2] & 6.1[3]\\
IRC +50137 & 7.8349 & 156.4379 & 23.0 & 6.6 & - & - \\%& 27.6[1] & 30.3[1]\\
$\chi$ Cyg & 3.2762 & 68.5386 & 17.7 & <6.9 & 24.2 & 6.8 \\%& 20.9[2] & 6.7[3]\\
MGE 4602 & 0.4734 & 9.3523 & <50.0 & <14.0 & - & - \\%& 3.7[1] & 0.2[1]\\MWC 922 & 0.6341 & 17.9269 & <436.0 & 28.4 & 70.7 & 22.3 \\%& 56.8[2] & 18.1[3]\\
Mz 3 & -1.0113 & 331.7275 & 113.0 & 24.7 & 77.5 & 28.3 \\%& 65.1[2] & 22.6[3]\\
NGC 40 & 9.8680 & 120.0162 & 27.5 & 9.1 & 11.5 & 4.8 \\%& 11.0[2] & 4.5[3]\\
NGC 2392 & 17.4002 & 197.8784 & 16.1 & <3.3 & - & - \\%& 6.5[2] & 2.6[3]\\
NGC 3242 & 32.0509 & 261.0499 & 29.6 & 6.4 & 15.4 & 3.6 \\%& 13.2[2] & 1.8[3]\\
NGC 6302 & 1.0558 & 349.5075 & 537.0 & 201.7 & - & - \\%& 240.2[2] & 118.5[2]\\
NGC 6445 & 3.9051 & 8.0758 & 43.2 & 14.6 & 13.4 & 3.5 \\%& 14.8[2] & 6.8[3]\\
NGC 6537 & 0.7395 & 10.0989 & 166.0 & <85.5 & 155.4 & 70.9 \\%& 156.9[2] & 79.7[3]\\
NGC 6543 & 29.9548 & 96.4680 & 60.9 & 13.9 & 48.5 & 13.0 \\%& 48.2[2] & 12.1[3]\\
NGC 6543 WKnot & 29.9810 & 96.4712 & - & - & - & - \\%& -0.2[2] & 0.2[3]\\
NGC 6720 & 13.9782 & 63.1700 & 54.6 & 20.6 & - & - \\%& -20.1[1] & -66.1[1]\\
NGC 6720 OFFcenter & 13.9814 & 63.1725 & 54.6 & 20.6 & - & - \\%& 5.7[2] & 2.7[3]\\
%NGC 2022 & -10.9425 & 196.6847 & 5.9 (11) [3] &  () [1] & - & - & -14.1[1] & 5.7[1]\\
%NGC 2440 & 2.4197 & 234.8363 & 26.3 (14) [2] &  () [] & - & - & -20.2[1] & -8.8[1]\\
NGC 6781 & -2.9878 & 41.8407 & 77.0 & - & 0.6 & 1.6 \\%& 0.7[2] & 1.3[3]\\
NGC 6781 Rim & -3.0000 & 41.8475 & 77.0 & - & - & - \\%& 6.4[2] & 3.9[3]\\
NGC 6826 & 12.7923 & 83.5616 & 21.3 & <3.7 & 11.7 & 5.8 \\%& 13.1[2] & 1.3[3]\\
NGC 6826 Rim & 12.7828 & 83.5713 & - & - & - & - \\%& 0.5[1] & -0.1[1]\\
NGC 7009 & -34.5713 & 37.7621 & 48.1 & 11.3 & 29.0 & 8.6 \\%& 29.6[2] & 7.8[3]\\
NGC 7026 & 0.3753 & 89.0020 & 30.9 & <5.8 & 17.8 & 1.7 \\%& 18.0[2] & 6.4[3]\\
NML Cyg & -1.9208 & 80.7982 & <334.0 & 100.1 & 400.0 & 109.2 \\%& 343.6[2] & 111.3[3]\\
NML Tau & -31.4120 & 177.9545 & 103.0 & 27.4 & 131.1 & 22.9 \\%& 69.2[2] & 23.2[3]\\
OH 21.5+0.5 & 0.4893 & 21.4588 & <319.0 & - & - & - \\%& 35.5[2] & 7.7[2]\\
OH 26.5+0.6 & 0.6179 & 26.5434 & <311.0 & <30.25 & 181.0 & 33.5 \\%& 142.9[2] & 39.5[3]\\
OH 30.1-0.7 & -0.6862 & 30.0911 & <807.0 & - & 93.6 & 15.0 \\%& 63.4[2] & 12.6[3]\\
OH 30.7+0.4 & 0.2813 & 30.5537 & 290.0 & - & 6.9 & 12.7 \\%& 8.1[1] & 12.6[1]\\
OH 32.8-0.3 & -0.3154 & 32.8271 & <583.0 & - & 68.6 & 14.0 \\%& 50.2[2] & 6.7[3]\\
OH 104.91+2.41 & 2.4134 & 104.9083 & 35.0 & 7.8 & - & - \\%& 52.4[1] & 54.6[1]\\
OH 231.8+4.2 & 4.2196 & 231.8354 & 294.0 & 98.9 & 586.2 & 198.0 \\%& 549.8[2] & 199.8[3]\\
$o$ Cet & -57.9827 & 167.7549 & 88.4 & 16.1 & 74.3 & 18.5 \\%& 61.0[2] & 21.8[3]\\
$\pi$ Gru & -55.1638 & 350.2822 & 23.3 & <5.5 & 18.4 & 6.0 \\%& 14.4[2] & 5.0[3]\\
R Aql & 0.4539 & 41.9524 & <83.1 & - & - & - \\%& 13.0[1] & 11.8[1]\\
R Cas & -10.6191 & 114.5608 & 38.8 & 9.5 & 29.6 & 10.5 \\%& 23.5[2] & 8.7[3]\\
R Dor & -39.3424 & 272.6713 & 83.4 & 25.5 & 92.8 & 33.2 \\%& 74.8[2] & 28.1[3]\\
R Lep & -31.3270 & 214.3244 & 9.1 & <0.5 & - & - \\%& 7.3[1] & 3.8[1]\\
RAFGL 2374 & -2.3064 & 44.7947 & 9.8 & <4.5 & - & - \\%& 6.6[1] & -8.6[1]\\
Red Rectangle & -11.7648 & 218.9680 & 66.2 & <28.0 & 88.4 & 35.0 \\%& 75.7[2] & 29.4[3]\\
RR Aql & -15.5601 & 38.9171 & 10.1 & <1.4 & - & - \\%& 7.2[1] & 3.7[1]\\
RT Cap & -27.9386 & 21.9411 & 3.6 & <3.8 & 2.2 & 0.6 \\%& 1.6[2] & 0.8[3]\\
S Aur & -0.5112 & 173.4866 & 11.6 & <0.7 & - & - \\%& 2.4[1] & 1.4[1]\\
ST Her & 49.4432 & 76.9788 & 6.0 & <1.4 & - & - \\%& 3.0[1] & -3.3[1]\\
T Cep & 13.8449 & 104.8050 & 15.3 & <3.0 & - & - \\%& 11.6[1] & 2.8[1]\\
T Mic & -32.4287 & 15.1857 & 12.7 & <3.1 & 10.4 & 3.2 \\%& 7.0[2] & 2.6[3]\\
Tc 1 & -8.8349 & 345.2375 & 4.7 & <0.5 & 2.0 & 1.0 \\%& 1.7[2] & 0.8[3]\\TX Cam & 8.5669 & 152.8375 & 38.6 & 8.0 & 31.9 & 7.4 \\%& 25.7[2] & 10.8[3]\\
U Hya & 38.0741 & 259.9663 & 14.5 & <0.8 & - & - \\%& 3.7[1] & 2.1[1]\\
U Mon & 4.1536 & 226.1413 & 9.5 & - & 12.0 & 4.9 \\%& 10.5[2] & 4.3[3]\\
V384 Per & -7.5975 & 148.1771 & 11.9 & - & - & -  \\%& 9.7[1] & 4.6[1] \\ 
V438 Oph & 26.4554 & 32.1306 & 3.7 & - & - & - \\%& 1.6[1] & 9.6[1]\\ 
V1300 Aql & -20.4165 & 36.3580 & 63.7 & 11.4 & 72.7 & 17.5 \\%& 46.4[2] & 12.9[3]  \\ 
V Cyg & 3.7667 & 86.5361 & 17.2 & <6.0 & 13.3 & 5.2 \\%& 10.6[2] & 4.4[3] \\ 
V Hya & 33.6014 & 268.9648 & 29.9 & 8.0 & 33.9 & 9.6 \\%& 28.4[2] & 10.3[3] \\ 
W Aql & -8.5161 & 29.3389 & 36.0 & 10.6 & 26.6 & 7.6 \\%& 20.7[2] & 6.5[3] \\ 
W Hya & 32.8108 & 318.0223 & 72.2 & 24.3 & 72.3 & 26.6  \\%& 57.9[2] & 22.8[3]\\ 
W Ori & -22.8179 & 199.0092 & 6.27 & <1.8 & - & -  \\%& 4.4[1] & 0.9[1]\\ 
WX Psc & -50.1074 & 128.6416 & 72.1 & <15.0 & 77.5 & 19.8 \\%& 64.3[2] & 20.0[3] \\ 
X Her & 47.7855 & 74.4639 & 18.3 & <3.3 & 9.5 & 3.6 \\%& 7.7[2] & 2.9[3]\\ 
Y CVn & 71.6450 & 126.4472 & 7.8 & - & - & -  \\%& 6.4[1] & 2.0[1]\\

%\footnote{}
         
\end{longtable}
%}
\end{landscape}
%\end{longtab}

\end{document}